\documentclass[aps,prb,reprint,twocolumn,english,superscriptaddress,longbibliography,floatfix]{revtex4-1}
\usepackage{amsmath}
\usepackage{amssymb,scalerel}
\usepackage{graphicx}
\usepackage[colorlinks,bookmarks=false,citecolor=red,linkcolor=blue,urlcolor=blue]{hyperref}
\usepackage{lipsum, color}
\usepackage{soulutf8}
\usepackage{wasysym}

\def\Hhat{\hat{H}}

\def\i{{\boldsymbol i}}

\def\k{{\boldsymbol k}}
\def\q{{\boldsymbol q}}

\def\Q{{\boldsymbol Q}}

\newcommand{\ve}[1]{\boldsymbol{#1}}

\graphicspath{ {./FIGURES/}{./SM/} } 

\setcitestyle{square,numbers,compress}

\begin{document}
\title{SU($\boldsymbol N$) Kondo-Heisenberg chain: Phase diagram, Ising criticality, and the 
coexistence of heavy quasiparticles and  valence bond solid order}
\author{Marcin Raczkowski}
\affiliation{Institut f\"ur Theoretische Physik und Astrophysik, Universit\"at W\"urzburg, 97074 W\"urzburg, Germany}
\author{Fakher F. Assaad}
\affiliation{Institut f\"ur Theoretische Physik und Astrophysik and W\"urzburg-Dresden Cluster of Excellence ct.qmat, Universit\"at W\"urzburg, 97074 W\"urzburg, Germany}
\date{\today}
\begin{abstract}
We map out the ground state phase diagram of a one-dimensional SU($N$) Kondo-Heisenberg lattice model 
at half filling and in the fully antisymmetric self-adjoint representation  as a function of $\tfrac {1}{N}$ 
and Kondo coupling $J_k/t$. 
On the basis of auxiliary field  quantum Monte Carlo (QMC)  simulations with even $N$ 
up to 8,  we show that the enlarged SU($N\ge 4$) symmetry realizes a quantum phase transition separating 
a valence bond solid (VBS) phase occupying a weak coupling part of the phase diagram and the Kondo insulator (KI) 
state dominating in the strong $J_k/t$ limit. Along the phase boundary, we always observe critical exponents that 
belong to a two-dimensional classical Ising universality class. 
We next trace the evolution of the composite  fermion  and spin  spectral  functions across the 
phase boundary and conclude that  VBS order triggers a bond order wave state of conduction electrons, 
both coexisting with Kondo screening. 
Upon further reducing $J_k/t$ we observe that composite quasiparticles lose their  spectral weight indicating 
that  VBS order gradually liberates localized $f$ spins from forming Kondo singlets with conduction electron spins. 
We contrast the QMC results with a static large-$N$ approximation, and we show that in the limit of 
infinite degeneracy $N\to\infty$  the order-disorder transition becomes of first order and is accompanied by 
a full decoupling of conduction electrons and localized $f$ spins.
We complete our analysis by considering the limit of a vanishing Heisenberg coupling $J_h=0$, and we provide 
evidence that the enlarged SU($N$) symmetry and low dimensionality of the RKKY exchange interaction are  
not sufficient to induce VBS order in the conventional  SU($N$) Kondo chain which hosts solely 
a KI phase.
Finally, we quantify the role of increasing $N$ in reducing the separation of charge and spin times scales specific 
to the KI state, and we show that the latter is adiabatically connected to a trivial band insulator captured by 
the large-$N$ approximation.

\end{abstract}

\maketitle

\section{\label{sec:intro}Introduction}

The complexity of solving the Kondo lattice model arises from many body correlation effects involving 
both the localized spin and the itinerant electron degrees of freedom~\cite{Doniach1977}. 
A coherent Bloch-like screening of impurity spins by conduction electrons 
leads to the formation of a heavy Fermi liquid state~\cite{Tsukahara11,Raczkowski18,Morr19,Wan23}. 
The volume of the resultant large Fermi surface includes both the localized spins and conduction electrons~\cite{Oshikawa00a}. 
Depending on the strength of Kondo coupling $J_k/t$, the Kondo quenching might be incomplete or even gets 
broken down by the competing Rudermann-Kittel-Kasuya-Yosida (RKKY) interaction. The latter corresponds to an indirect coupling between 
magnetic impurities mediated by the conduction electrons and promotes a magnetically ordered state. 
The breakdown of the Kondo effect~\cite{Coleman01,Si01,PhysRevLett.125.206602,Raczkowski22,PhysRevLett.130.246402} 
or equivalently the orbital selective  Mott transition~\cite{Vojta10,gleis2023emergent,eickhoff2024kondo} 
involves a reconstruction of the Fermi surface such that its volume is determined solely by the conduction electrons. 

These two underlying physical mechanisms, lattice Kondo screening and RKKY interaction,  are believed to determine 
the physical properties of heavy fermion compounds with $f$ electron orbitals~\cite{RMP07,Gegenwart08,Si10}. 
Following a phenomenological approach by Doniach~\cite{Doniach1977}, substantial analytical and numerical efforts have been 
undertaken to deepen our understanding of a resultant  magnetic order-disorder transition in a two-dimensional (2D) 
Kondo lattice model~\cite{PhysRevB.51.15630,Assaad99a,Capponi00,PhysRevB.66.045103,PhysRevB.67.214406,PhysRevB.69.035111,
Watanabe07,PhysRevLett.101.066404,Fabrizio08,Vojta08,Otsuki09,PhysRevB.82.245105,Becca13,PhysRevB.92.075103,PhysRevB.96.155119,
PhysRevB.97.235151,PhysRevB.98.245125,Raczkowski20,Danu21,PhysRevB.66.045103,PhysRevResearch.5.L032014}. 


While long range antiferromagnetic order does not appear in a strictly one-dimensional (1D) geometry, the physics of a dense 
Kondo chain at half filling turns out to be equally exciting.  
While a semiclassical analysis based on the  nonlinear $\sigma$ model~\cite{PhysRevLett.72.1048}, 
bosonization approach~\cite{PhysRevLett.77.1342}, exact diagonalization~\cite{PhysRevB.46.3175},  
and  density-matrix renormalization group (DMRG) calculations~\cite{PhysRevLett.71.3866} yield  an exponentially 
small spin excitation gap, the charge gap was shown to track $J_k$ for small Kondo couplings 
instead~\cite{PhysRevLett.71.3866,PhysRevB.53.R8828}.
Thus strong antiferromagnetic spin correlations in a Kondo insulator lead to the separation of spin and charge  
energy scales~\cite{Ueda97,PhysRevB.105.155134,PhysRevB.108.195116,SciPostPhys.14.6.166} 
absent in a trivial band insulator~\cite{PhysRevB.80.155116,PhysRevB.106.125116}.
The precise nature of the spin excitation spectrum in a Kondo insulator has been the subject of recent 
debate~\cite{chen2023matrix,PhysRevB.108.195120}.
Furthermore, a generalized Luttinger theorem  guarantees that, in a translationally invariant case, the large filled 
Fermi sea picture  holds for an arbitrarily small value of $J_k/t$~\cite{PhysRevLett.79.1110}. 
Tracing spectral signatures of the heavy Tomonaga-Luttinger liquid constitutes, however, an ongoing challenge for numerical 
simulations  carried out on finite size systems~\cite{PhysRevB.52.R15723,PhysRevB.54.12212,PhysRevB.54.13495,PhysRevB.56.330,PhysRevB.65.214406,PhysRevB.91.195116,Xie2017,pnas.1719374115,nikolaenko2023numerical}.


In comparison, much less is known about the occurrence of dimer order in the Kondo lattice model beyond the 
static large-$N$ approximation~\cite{PhysRevB.83.214427,PhysRevLett.113.176402,Rey15,Li16,PhysRevB.92.165431,PhysRevB.98.085110}.
A frequently used route to address numerically the interplay of dimerization and Kondo screening is to include 
geometrical frustration~\cite{Coleman2010}. For example, DMRG studies of 
a Kondo lattice model on the zigzag ladder discovered a spontaneous dimerization as an easy way out to  alleviate 
the magnetic frustration~\cite{PhysRevB.54.9862,PhysRevB.97.115124,PhysRevB.99.085140}.
Another possibility is to consider a spin-1/2 Heisenberg 1D chain with competing nearest- and next-nearest-neighbor interactions 
$J_h$ and $J_h'$, respectively, tuned to the dimerized Majumdar-Ghosh point~\cite{1.1664978,PhysRevB.25.4925} and to explore 
the fate of the spin-dimerized phase when Kondo-coupled to the Fermi sea~\cite{PhysRevB.84.014413,PhysRevB.87.245102,PhysRevB.109.014103}.

A key idea of this paper is to consider a generalized Kondo-Heisenberg chain characterized by a spin symmetry 
group extended from SU(2) to SU($N$). On the one hand, it is known that already an SU(4) Heisenberg chain has a dimerized 
ground state~\cite{Paramekanti07,PhysRevB.100.085103}. The two-fold degeneracy of the latter comes from two choices of 
forming a singlet dimer with either a left or right nearest neighbor spin.
On the other hand, the ground state of the SU(4) Kondo-Heisenberg model in the strong Kondo coupling $J_k/J_h \to\infty$ 
shall be given by a direct product of Kondo singlets. 
Thus, it is plausible that increasing Kondo coupling $J_k/t$ in the SU(4) Kondo-Heisenberg chain
will drive a phase transition from a dimerized ground  state to the Kondo insulating phase.
Since both phases possess a finite spin gap, 
one can assume that the essential dynamics of the model is dominated by short range valence bond configurations 
and the problem should be effectively captured by an Ising gauge theory, 
known to be dual to the 1D transverse field Ising model~\cite{PhysRevB.65.024504}. 
Thus based on quantum-classical mapping, we foresee that the phase transition belongs 
to the 2D classical Ising universality class.

A related issue concerns the existence of a spontaneously  dimerized state of the localized spins driven solely by 
the RKKY interaction. 
On the one hand,  our detailed study of the phase diagram and the low energy properties of the 2D SU($N$) Kondo lattice model 
in the fully antisymmetric self-adjoint representation led us to conclude that no other phases  aside from the Kondo 
insulator and N\'eel state intervene at half filling~\cite{Raczkowski20}. On the other hand,  
early DMRG evidence of an insulating dimerized ground state in the weak-coupling regime $J_k/t\lesssim 1.2$ 
of the  quarter-filled 1D Kondo lattice model~\cite{PhysRevLett.90.247204,PhysRevB.78.144406}  has recently been 
confirmed, along with the coexisting bond order wave state of conduction electrons, by unbiased infinite DMRG 
calculations~\cite{PhysRevB.102.245143}. 
Hence, the low dimensionality of the RKKY interaction could possibly enhance dimer fluctuations thus motivating 
studies of a pure SU($N$) Kondo chain as well.

In fact, quantum simulations with alkali-earth-like atoms in optical lattices  have already made the SU($N$) quantum 
magnetism tangible~\cite{Kaden22}.   Steady progress in this domain allows one to envisage a highly controllable 
implementation of the Kondo singlet state with an enlarged SU($N$) symmetry in single impurity and lattice situations 
in the near future~\cite{Gorshkov10,Zhang14,Scazza14,Riegger18}.

In addition to a general interest in exploring novel quantum many-body phenomena driven by SU($N$) symmetric interactions in low 
dimensions, an additional motivation to study 1D Kondo chains is provided by real materials, e.g.,  
a recently discovered Kondo lattice compound ${\mathrm{CeCo}}_{2}{\mathrm{Ga}}_{8}$~\cite{Wang2017}. 
Upon cooling, resistivity measurements~\cite{PhysRevMaterials.3.021402}  and optical spectra~\cite{PhysRevB.105.035112} 
of ${\mathrm{CeCo}}_{2}{\mathrm{Ga}}_{8}$ indicate the occurrence of coherent Kondo screening along one single crystal 
direction. Thus, ${\mathrm{CeCo}}_{2}{\mathrm{Ga}}_{8}$ represents a rare example of quasi-1D heavy fermion behavior 
and numerical results obtained in the 1D geometry can serve as a starting point to develop at least qualitatively theories 
of such a state.

The rest of the paper is organized as follows. 
In Sec.~\ref{sec:Model} we introduce an SU($N$) Kondo-Heisenberg Hamiltonian, we discuss its mapping onto 
the U(1) lattice gauge theory which sets the basis for our auxiliary field quantum Monte Carlo (QMC) simulations, and 
we describe the saddle-point approximation to the auxiliary field action. 
Next, we present an extended discussion of the numerical results by considering in Sec.~\ref{sec:KHLM} 
the SU$(N)$ Kondo-Heisenberg model and in Sec.~\ref{sec:KLM} a pure SU($N$) Kondo chain. 
Finally, Sec.~\ref{sec:summary} gives a summary and outlook.

\section{\label{sec:Model} Model and methods}

Here we  first  define the model.  We  then  use  a  numerically exact approach,   quantum Monte Carlo (QMC),  to  solve  the  problem.    In the 
presence  of  particle-hole symmetry  this  approach can be formulated   without  confronting the  so-called  negative sign problem such  that   we  can 
obtain   numerically  exact    zero  temperature results    with  computational costs  scaling  as  the  cube of the number of lattice sites  
times the  so-called   projection parameter $\Theta$.   Since the particle-hole  symmetry  is not  broken   for  even  values of fermion flavors  $N$,  
this  method   can also  be  used,   at  no extra  computational cost,   for  the  SU($N$)  model.     The  limitations  occur at  finite  doping  
where  we  encounter the negative  sign problem  and  an  exponential  scaling of  the  computational costs  with  lattice size.   
We  will compare  our  results with  large-$N$ mean-field  theories.   Strictly  speaking   these  methods  are  not  controlled,   
and they  can  produce  wrong  results,  if  e.g.  there  is  a  phase  transition between  the  finite  $N$ case  of  interest  and  $N=  \infty$. 

\subsection{Model }

The  Kondo-Heisenberg chain Hamiltonian  
reads~\cite{PhysRevLett.79.929,PhysRevB.64.033103,PhysRevLett.105.146403,PhysRevB.94.165114,PhysRevB.101.165133,PhysRevB.106.014411}:
   \begin{align}
	   \hat{H}       &= \hat{H}_{t} + \hat{H}_{J_k} + \hat{H}_{J_h}, \label{su2}  \\   
	   \hat{H}_{t}   &= -t\sum_{\langle\ve{i},\ve{j}\rangle} 
	  \bigl ( \ve{\hat{c}}^{\dagger}_{\ve{i}} \ve{\hat{c}}^{\phantom\dagger}_{\ve{j}} + h.c. \bigr),  \\
	   \hat{H}_{J_k} &= \frac{J_k}{2}   \sum_{\ve{i}=1}^L  
	\ve{\hat{c}}^{\dagger}_{\ve{i}} \ve{\sigma} \ve{\hat{c}}^{\phantom\dagger}_{\ve{i}} \cdot \ve{\hat{S}}^{}_{\ve{i}},  \\
	    \hat{H}_{J_h} &= J_h  \sum_{\ve{i}=1}^L \ve{\hat{S}}^{}_{\ve{i}} \cdot \ve{\hat{S}}^{}_{\ve{i}+\ve{1}}, 
   \end{align}
where 
$\ve{\hat{c}}^{\dagger}_{\ve{i}}  = \left( \hat{c}^{\dagger}_{\ve{i},\uparrow}, \hat{c}^{\dagger}_{\ve{i},\downarrow}  \right) $  
is a spinor  where ${\hat c}^{\dagger}_{\i,\sigma}$ creates an electron  in Wannier  state  centered  around lattice site $\ve{i}$ 
and $z$ component of spin $\sigma = \uparrow,\downarrow$; 
$t$ is the nearest-neighbor hopping amplitude,  $J_k$ is the Kondo exchange coupling between the spins of conduction electrons 
and  the localized spin $s=1/2$ degrees of freedom, $\ve{\hat{S}}_{\ve{i}}$, with $\ve{\sigma}$  being   a vector of Pauli 
spin matrices, $J_h$ is the nearest-neighbor Heisenberg exchange interaction, and $L$ is the length of the chain.

To incorporate a general flavor symmetry SU($N$), we generalize the above model  with the help of a fermionic 
representation of the SU($N$) generators, 
\begin{equation}
\hat{S} ^{\mu}_{\ve{i},\nu}   = \hat{f}^{\dagger}_{\ve{i},\nu} \hat{f}^{\phantom\dagger}_{\ve{i},\mu}   
 - \frac{\delta_{\mu,\nu}}{N}  \sum_{\sigma=1}^{N} \hat{f}^{\dagger}_{\ve{i},\sigma} \hat{f}^{\phantom\dagger}_{\ve{i},\sigma},
\label{generator}
\end{equation}
subject to the local  constraint,
\begin{equation}
        \sum_{\sigma=1}^N  \hat{f}^{\dag}_{{\ve i},\sigma} \hat{f}^{\phantom\dagger}_{{\ve i},\sigma} = \frac{N}{2}.
\label{constraint}
\end{equation}
It selects the fully antisymmetric self-adjoint representation corresponding to a Young tableau with a single column and 
$N/2$ rows~\cite{Raczkowski20,hazra2021luttinger,PhysRevA.107.033317}.

\subsection{QMC formulation}

In the  above   representation, the model Eq.~(\ref{su2}) reads,
\begin{align}
\Hhat&=\Hhat_t-\frac{J_k}{4N}\sum_{\ve{i}}  
	\Big\{ \big(\ve{\hat{c}}^{\dagger}_{\ve{i} }\ve{\hat{f}}^{\phantom\dagger}_{\ve{i} }+h.c.\big)^2 
      + \big(i\ve{\hat{c}}^{\dagger}_{\ve{i} }\ve{\hat{f}}^{\phantom\dagger}_{\ve{i}}+h.c.\big)^2 \Big\}\nonumber\\
&-\frac{J_h}{4N}\sum_{\ve{i}}  \Big\{ \big(\ve{\hat{f}}^{\dagger}_{\ve{i} }\ve{\hat{f}}^{\phantom\dagger}_{\ve{i}+\ve{1}} +h.c.\big)^2 
      + \big(i \ve{\hat{f}}^{\dagger}_{\ve{i} }\ve{\hat{f}}^{\phantom\dagger}_{\ve{i}+\ve{1}} +h.c.\big)^2 \Big\} \nonumber \\
	&+\frac{U}{N}\sum_{\ve{i}}   \Big(\ve{\hat{f}}^\dagger_{\ve{i}} \ve{\hat{f}}^{\phantom\dagger}_{\ve{i} }-\frac{N}{2}\Big)^2,
\label{sun}
\end{align}
where $\ve{\hat{c}}^{\dagger}_{\ve{i}}$  and  $\ve{\hat{f}}^{\dagger}_{\ve{i}}$ are both $N$ flavor spinor operators.
Note that the extra Hubbard-$U$  interaction  term commutes with the Hamiltonian and suppresses charge fluctuations 
on the $f$ orbitals by the Boltzmann factor $\propto e^{-\Theta U/N}$. This guarantees that the constraint Eq.~(\ref{constraint}) will be automatically 
imposed in simulations with any finite positive $U$ and when the projection parameter $\Theta$  is chosen to be sufficiently large.

To   proceed we  use the  Trotter   decomposition   and  Hubbard Stratonovich   transformation    to    decouple  the   perfect  
square terms  in Eq.~(\ref{sun}). To this end, we introduce complex  bond  fields $\mathcal{V( \ve{i},\tau)}$ and $\mathcal{\chi( \ve{i},\tau)}$  
to handle the dynamics of Kondo and Heisenberg terms as well as the scalar Lagrange multiplier $\mathcal{\lambda}( \ve{i},\tau)$  to enforce 
the constraint.  Since all  the three fields couple to SU($N$) symmetric operators, the number of flavors  $N$ can be   pulled out 
in front of the action and one arrives at the  following  form  for the  grand-canonical  partition function
\begin{equation}
Z~\equiv~\int \mathcal{D}\{\mathcal{V},\mathcal{\chi}, \mathcal{\lambda} \} ~e^{- N \mathcal{S}\{\mathcal{V},\mathcal{\chi},\mathcal{\lambda}\}}
\label{z}
\end{equation}
with the  action
\begin{align}
	\mathcal{S}\{\mathcal{V},\mathcal{\chi},\lambda\}&=
	-\ln \Big[~\mbox{Tr} ~ \mathcal{T} e^{-\int^\beta_0 d\tau~ \Hhat \{\mathcal{V},\mathcal{\chi},\lambda\}}\Big] \nonumber \\ 
	&+\int^\beta_0 d \tau \sum_{\ve{i}} \Big\{\frac{J_k}{4}|\mathcal{V}( \ve{i},\tau)|^2 \nonumber \\
	&+\frac{J_h}{4} | \mathcal{\chi}( \ve{i},\tau)|^2+\frac{ U}{4} |\lambda( \ve{i},\tau)|^2\Big\}
	\label{action}
\end{align}
and the time dependent Hamiltonian
\begin{widetext}
\begin{equation}
 \Hhat \{\mathcal{V},\mathcal{\chi},\lambda\}=
	\Hhat_t+ \sum_{\ve{i}} \Big\{-\frac{J_k}{2}\Big(\mathcal{V}( \ve{i},\tau) 
	\hat{c}^{\dagger}_{\ve{i} }\hat{f}^{\phantom\dagger}_{\ve{i} }+h.c.\Big)
	-\frac{J_h}{2}\Big(\mathcal{\chi}(\ve{i},\tau) \hat{f}^{\dagger}_{\ve{i} }\hat{f}^{\phantom\dagger}_{\ve{i}+\ve{1}}+h.c.\Big)
	-i U\lambda(\ve{i},\tau) \Big(\hat{f}^\dagger_{\ve{i}} \hat{f}^{\phantom\dagger}_{\ve{i} }-\frac{1}{2}\Big) \Big\}.
\end{equation}
\end{widetext}
Note that in the above the spinor operators $\ve{\hat{c}}^{\dagger}_{\ve{i}}$  and  $\ve{\hat{f}}^{\dagger}_{\ve{i}}$  have  lost their 
flavor index $N$ since it merely  comes in to Eq.~(\ref{z}) as a parameter. 

As a matter of fact, by using Grassmann variables~\cite{Negele} $\ve{c}(\ve{i},\tau)$  and $\ve{f}(\ve{i},\tau)$ and taking the limit 
$U \rightarrow \infty $, one can show that the constraint leads to a local  U(1) gauge invariance of the 
action~\cite{Read83,Auerbach86,Saremi07,Raczkowski22}.
In that case, the   canonical transformation,   $\ve{f}(\ve{i},\tau)   \rightarrow  e^{i \varphi_{\ve{i}}(\tau)} \ve{f}(\ve{i},\tau) $
amounts  to  redefining the  fields 
$ \mathcal{V}(\ve{i},\tau)  \rightarrow e^{-i \varphi_{\ve{i}}(\tau)} \mathcal{V}(\ve{i},\tau)   $, 
$ \mathcal{\chi}(\ve{i},\tau)  \rightarrow   e^{i \varphi_{\ve{i}}(\tau)}  \mathcal{\chi}(\ve{i},\tau)  e^{-i \varphi_{\ve{i}+\ve{1}}(\tau)} $, 
and $  \lambda(\ve{i},\tau) \rightarrow \lambda(\ve{i},\tau)  + \partial_\tau \varphi_{\ve{i}}(\tau) $,  
such  that the  partition function remains  invariant.
As is apparent, fermion fields $\ve{f}(\ve{i},\tau)$ carry the unit gauge charge. 
It is, however,   possible to define a gauge neutral field    
\begin{equation}
\label{eq:tildef}
	 \ve{\tilde{f}}(\ve{i},\tau)   =   e^{i \varphi_{\ve{i}}(\tau)}  \ve{f}(\ve{i},\tau),   \text{ with }   e^{i \varphi_{\ve{i}}
(\tau)}  =  \frac{\mathcal{V}(\ve{i},\tau)} {|\mathcal{V}(\ve{i},\tau)|}
\end{equation}
which  has quantum numbers of  a   physical  fermion.  In  the  heavy  fermion  phase,  
 $  \ve{\tilde{f}}(\ve{i},\tau) $   emerges then as  a new  particle  excitation that acquires  coherence and contributes to the Luttinger  volume. 

The  above  can  be  understood  in terms of a   Higgs~\cite{Fradkin79}  mechanism    in  which  the phase  fluctuations of  
$\varphi_{\ve{i}}(\tau)$    become very slow  such   that   $\varphi_{\ve{i}}(\tau)$    can be  set to  a  constant.    In this  case   
there is  no  distinction  between  the  fields $ \ve{\tilde{f}}(\ve{i},\tau)   $  and $\ve{f}(\ve{i},\tau)$   or,  in other  words,   
$\ve{f}(\ve{i},\tau)$   has  lost its gauge  charge  and  has    acquired  a unit   electric charge.  This  Higgs  mechanism  is  
captured  in  mean-field large-$N$  approaches  of  the Kondo  lattice    where  Kondo  screening  corresponds  to  
$ \left< \mathcal{V}(\ve{i},\tau)     \right> \neq 0 $~\cite{Burdin00,Morr17}.  Moreover, as  shown in  Ref.~\cite{Danu21}  
and in  the large-$N$ limit,  $  \ve{\tilde{f}}(\ve{i},\tau) $   is   nothing  but  the so-called composite  fermion  field~\cite{Costi00,Maltseva09}: 
\begin{equation}
	\ve{\tilde{f}}(\ve{i},\tau)  \propto   \ve{\psi}(\ve{i},\tau) =    \ve{S}(\ve{i},\tau) \cdot   \ve{\sigma}  \ve{c}(\ve{i},\tau). 
\end{equation}
More specifically, for Kondo impurity problems the  Green's function of   $ \hat{\ve{\psi}}^{\dagger}_{\i} $ corresponds  to the 
$T$-matrix~\cite{Borda07} while $ \hat{\ve{\psi}}^{\dagger}_{\i}  $  itself    corresponds  to the    Schrieffer-Wolff   transformation 
of the localized electron operator  in the  realm of the Anderson model~\cite{Raczkowski18}.

The mapping of the model Eq.~(\ref{sun}) onto the U(1) lattice gauge theory sets the basis for  our auxiliary  field  QMC 
simulations~\cite{Blankenbecler81,Sorella89,Assaad08_rev}. 
The  integration over  the   Grassmann  variables  yields  the  fermion determinant which, for  a particle-hole symmetric  
conduction electron band  and  even values of $N$ used  in our study,  is positive semi-definite such that no negative sign problem 
occurs~\cite{Wu04}. 
The  integration over  the  Hubbard-Stratonovich   fields  is then conveniently carried out  with Monte  Carlo  importance sampling. 
Specifically for the calculation presented  here,  
we  have  used  the  implementation of the Kondo  lattice  model  of the Algorithms for Lattice  Fermions (ALF-2.0) library~\cite{ALF_v2} 
extended by the Heisenberg term, and we adopted a projective (zero-temperature) version of the algorithm.  It is based on the 
imaginary time evolution of a trial wave function $| \Psi_\text{T}\rangle$, with $ \langle\Psi_\text{T}  |\Psi_0 \rangle  \neq 0 $,  
to the ground state $|\Psi_0 \rangle$:              %
\begin{equation}
  \frac{ \langle  \Psi_0 | \hat{O} |  \Psi_0 \rangle  }{ \langle  \Psi_0  |  \Psi_0 \rangle  }  =  
  \lim_{\Theta \rightarrow \infty} 
  \frac{ \langle  \Psi_\text{T} | e^{-\Theta \hat{\mathcal{H}}  } \hat{O} e^{-\Theta \hat{\mathcal{H}}   } |  \Psi_\text{T}\rangle  }
  { \langle  \Psi_\text{T}  | e^{-2\Theta \hat{\mathcal{H}} }  |  \Psi_\text{T} \rangle  }.
\end{equation}
To assess the hermiticity of the imaginary time propagation in the presence of current operators in Eq.~(\ref{sun}), 
we have opted for a symmetric Trotter decomposition~\cite{ALF_v2} with  a finite imaginary time step $\Delta\tau t=0.1$.  
Furthermore, we used a $J_k$ and $N$ dependent projection parameter $20 \leqslant 2\Theta t \leqslant 200 $ chosen large enough 
to obtain converged ground-state quantities in test run simulations.  
As for the constraint, the choice $2\Theta U /N \geqslant 10$ was found sufficient to suppress charge fluctuations 
on the $f$ sites within statistical uncertainty.   
Finally, for  the analytical continuation of the single and two particle imaginary time QMC data,  we have made  use of the  
stochastic maximum entropy  method~\cite{Sandvik98,Beach04a}  implemented in the ALF-2.0 library. 

Let us emphasize that the formulation of  the auxiliary field QMC  approach 
maps  the Kondo lattice model  onto a compact  U(1) lattice  gauge theory,  where the local U(1) symmetry  
reflects  the infinitely strong bare  coupling. The Monte Carlo  simulations  that we  carry out make  no  
approximations, and  correspond  to an exact  integration over  the  gauge  fields.   By  systematically 
carrying out  simulations at larger and larger  values of  $N$  we  can actually  assess,  as was done in 
Ref.~\cite{Raczkowski20},  that  there is no  phase  transition between the  $N=2$ Kondo  insulating state  
and  the one at  $N= \infty$  for  the 2D case.    This  adiabatic path between $N=2$ and  $N = \infty$ is  in our  view  an 
important numerical result  that can  justify  the large-$N$ approximation, which we  discuss in the next section.

\subsection{Large-${\boldsymbol N}$ approach}

 In the large-$N$ limit,  the saddle-point approximation to the action Eq.~(\ref{action}):
\begin{align}
\frac{\partial\mathcal{S}\{\mathcal{V},\mathcal{\chi},\mathcal{\lambda}\}}{\partial \mathcal{V}( \ve{i},\tau)},
	\frac{\partial\mathcal{S}\{\mathcal{V^*},\mathcal{\chi^*},\lambda\}}{\partial \mathcal{V}^*( \ve{i},\tau)}&=0, \\ 
\frac{\partial \mathcal{S}\{\mathcal{V},\mathcal{\chi}, \mathcal{\lambda}\}}{\partial \mathcal{\chi}( \ve{i},\tau)},\frac{\partial\mathcal{S}\{\mathcal{V^*},
	\mathcal{\chi^*},\mathcal{\lambda}\}}{\partial \mathcal{\chi}^*( \ve{i},\tau)}&=0,   
\quad \mbox{and} \\
	\quad\frac{\partial\mathcal{S}\{\mathcal{V},\mathcal{\chi},\mathcal{\lambda}\}}{\partial \mathcal{\lambda}( \ve{i},\tau)}&=0,
\end{align}
becomes exact. 
In the mean-field approximation carried out below,   we consider only saddle points with time independent fields. We furthermore 
assume translational invariance of the field $ \mathcal{V}( \ve{i},\tau)=\mathcal{V}^*( \ve{i},\tau)=V(\in \mathbb{R})$, 
introduced to handle  the  Kondo term, but since we are interested in a solution with dimer order and the resultant doubling 
of the unit cell, we allow for a  site dependent field, 
$\mathcal{\chi}( \ve{i},\tau)=\mathcal{\chi}^*( \ve{i},\tau)=\chi_{\ve{i}}(\in \mathbb{R})$, used to decouple the Heisenberg term. 
Furthermore, since the 1D geometry of the Kondo-Heisenberg chain does not sustain solutions involving 
spontaneous breaking  of a continuous spin rotational symmetry group, we leave out in our analysis long range antiferromagnetic 
order.  Finally, in the large-$N$ approach, the constraint is enforced only on average by means of a Lagrange multiplier $\lambda$.

\begin{figure*}[t!]
\includegraphics[width=0.23\textwidth]{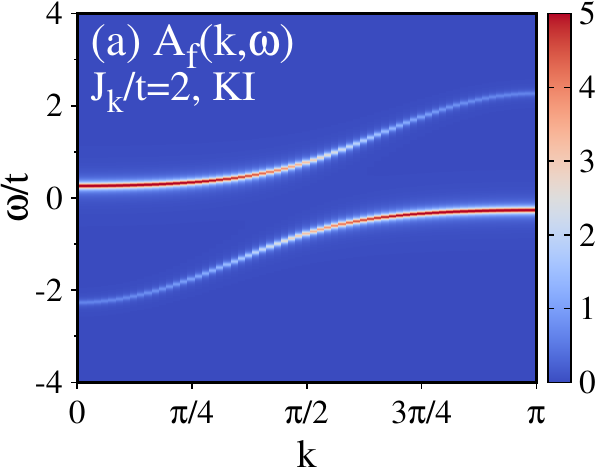}
\includegraphics[width=0.23\textwidth]{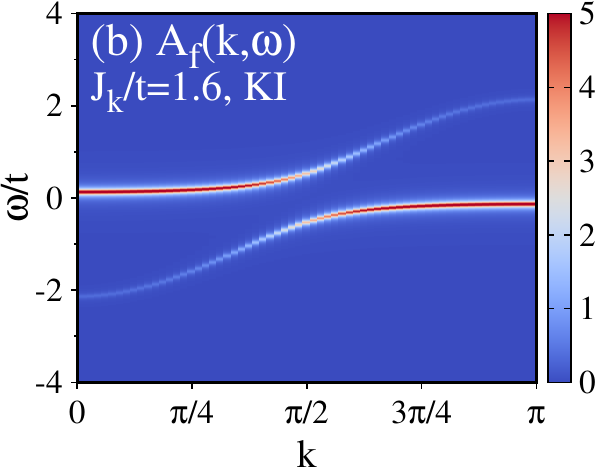}
\includegraphics[width=0.23\textwidth]{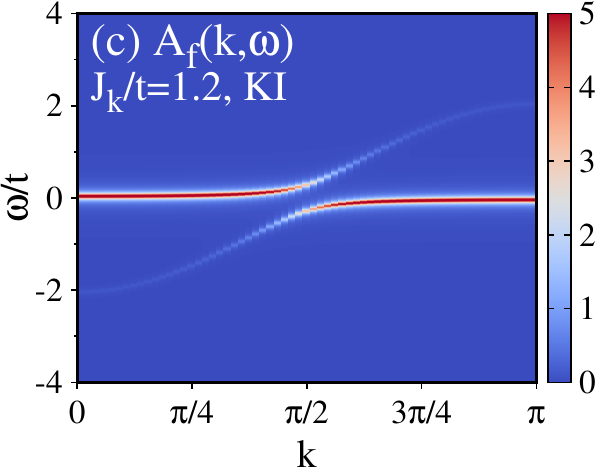}
\includegraphics[width=0.23\textwidth]{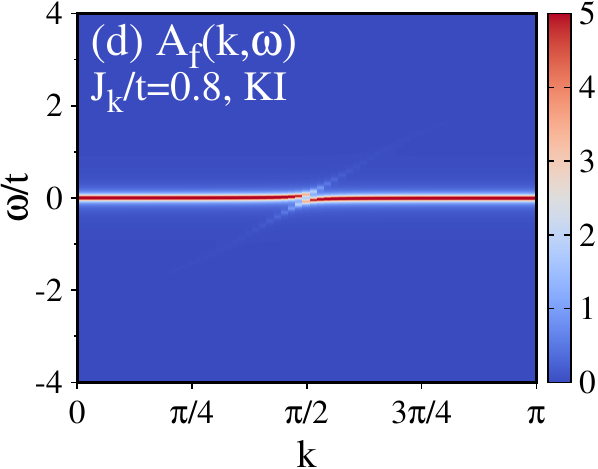}\\
\includegraphics[width=0.23\textwidth]{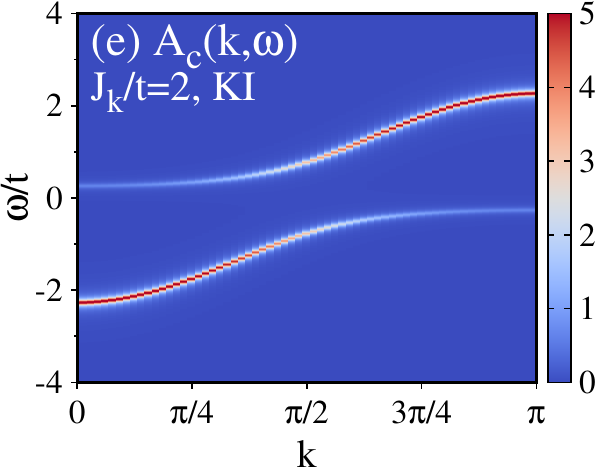}
\includegraphics[width=0.23\textwidth]{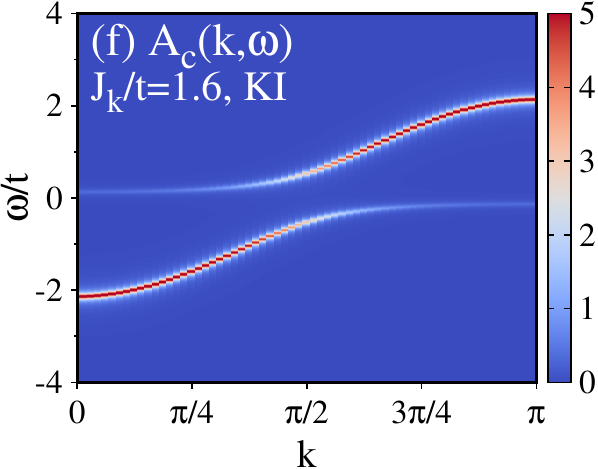}
\includegraphics[width=0.23\textwidth]{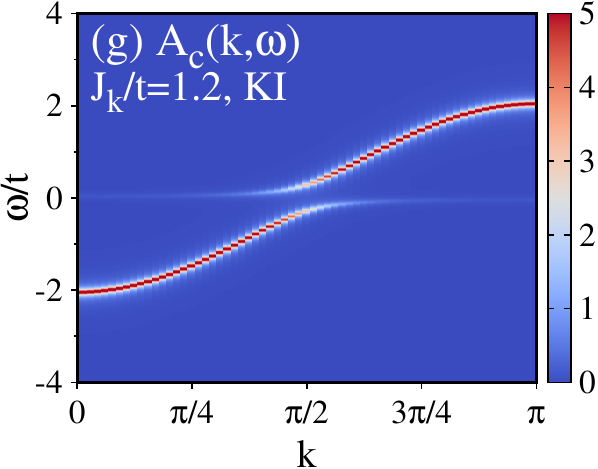}
\includegraphics[width=0.23\textwidth]{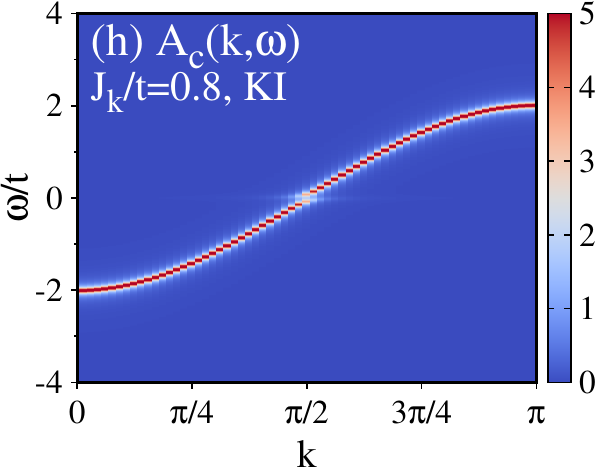}
        \caption{(a)-(d) $f$ fermion $A_{f}(\ve{k},\omega)$  and (e)-(h) conduction electron $A_{c}(\ve{k},\omega)$
        spectral functions as obtained from the large-$N$ approach accounting for Kondo screening in the translationally 
	invariant Kondo chain with $J_h=0$ for representative values of $J_k/t$.}
\label{Fig:Ak_klm}
\end{figure*}

We perform a standard mean-field decoupling~\cite{PhysRevB.37.3774,PhysRevB.39.11538,Saremi07}  based on the hybridization order parameter 
$ V   =  \langle \hat{f}^{\dagger}_{\ve{i}} \hat{c}^{\phantom\dagger}_{\ve{i}} + h.c. \rangle $ 
measuring the strength of Kondo screening and the valence bond order parameter  
$ \chi^{}_{\ve{i}} = \langle \hat{f}^{\dagger}_{\ve{i}} \hat{f}^{\phantom\dagger}_{\ve{i}+\ve{1}} + h.c. \rangle $ 
accounting for a spinon description of the spin-1/2 Heisenberg  antiferromagnetic chain~\cite{Raczkowski13,PhysRevB.99.205102}.
Given that the smallest unit cell capturing the valence bond solid (VBS) phase consists of two sites, we split the lattice into sublattices 
$A$ and $B$, and assume a perfect dimerization of the chain, $\chi_A=\chi_0+\delta$ and $\chi_B=\chi_0-\delta$,  
such that $\chi_0=\tfrac{1}{2}(\chi_A+\chi_B)$ [$\delta=\tfrac{1}{2}(\chi_A-\chi_B)$] correspond to the uniform (staggered) 
part of the valence bond order parameter, respectively.  Within the above Ansatz, the mean-field (MF) Hamiltonian takes the following Fourier 
transformed form:
\begin{align}
\label{LargeN_vbs.eq}
        \hat{H}_{MF}   & =  \sum_{\ve{k} \in RBZ}  
\left(
\hat{c}^{\dagger}_{\ve{k}},
	\hat{c}^{\dagger}_{\ve{k} + \ve{\pi}},
\hat{f}^{\dagger}_{\ve{k}},
	\hat{f}^{\dagger}_{\ve{k} + \ve{\pi}} 
	\right)
\nonumber\\
&  \times
\begin{pmatrix} 
\epsilon(\ve{k})     &   0                   &    -\frac{J_k V}{2}            &   0                  \\
     0               & -\epsilon(\ve{k})     &       0                        & -\frac{J_k V}{2}      \\
-\frac{J_k V}{2}     &   0                   &   \epsilon_f(\ve{k})           &  i\tilde{\epsilon}_f(\ve{k}) \\
     0               & -\frac{J_k V}{2}      & -i\tilde{\epsilon}_f(\ve{k})   &  -\epsilon_f(\ve{k})
\end{pmatrix} 
\nonumber  \\
&  \times \begin{pmatrix} 
\hat{c}^{}_{\ve{k}} \\
\hat{c}^{}_{\ve{k} + \ve{\pi}} \\
\hat{f}^{}_{\ve{k}} \\
\hat{f}^{}_{\ve{k} + \ve{\pi}} 
\end{pmatrix}  
	+ \frac{N_u}{2} \left[ J_h(\chi_0^2+\delta_{}^2)  + J_k V^2 \right], 
\end{align}
where $\epsilon(\ve{k})  = -2t\cos k$ and $\epsilon_f(\ve{k})  = -J_h\chi_0\cos k$ such that 
$ \epsilon(\ve{k} + \ve{\pi}) = -  \epsilon(\ve{k}) $ and $ \epsilon_f(\ve{k} + \ve{\pi}) = -  \epsilon_f(\ve{k})$; 
$\tilde{\epsilon}_f(\ve{k})  = -J_h\delta\sin k$,
the $\ve{k}$ sum runs over a reduced Brillouin zone (RBZ), and $N_u$ denotes the number of unit cells. 
Note that since the underlying particle hole-symmetry pins automatically the $f$ occupation to half filling,  
in the above we have left  out the  Lagrange multiplier $\lambda$. The saddle-point equations then read: 
\begin{equation}
    \frac{\partial F} {\partial \chi_{0} } =  \frac{\partial F} {\partial \delta } =  \frac{\partial F} {\partial V }   = 0,
\label{saddle}
\end{equation}
with $F = -\frac{1}{\beta} \ln{ \text{Tr}   e^{- \beta  \hat{H}_{MF}}} $.

\begin{figure*}[t!]
\includegraphics[width=0.23\textwidth]{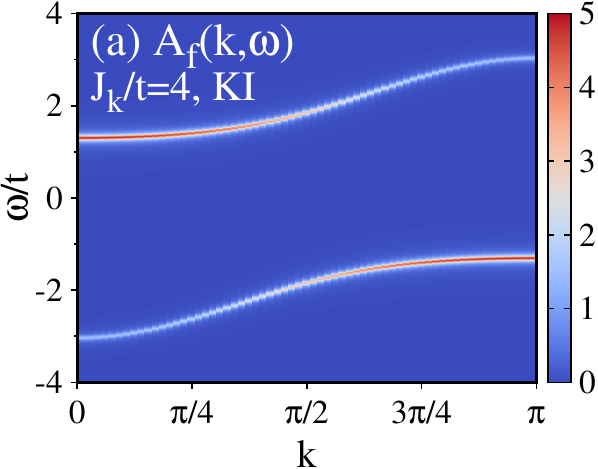}
\includegraphics[width=0.23\textwidth]{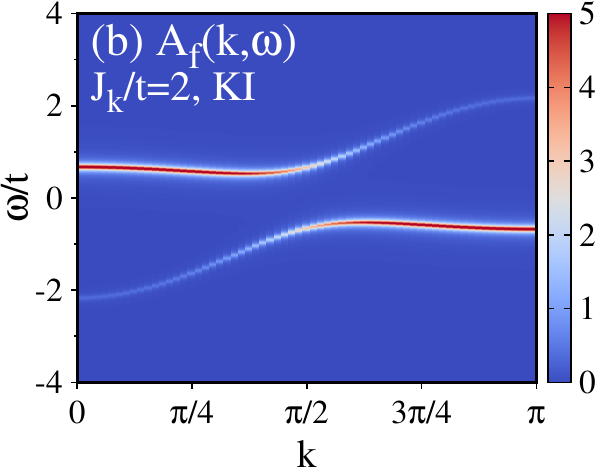}
\includegraphics[width=0.23\textwidth]{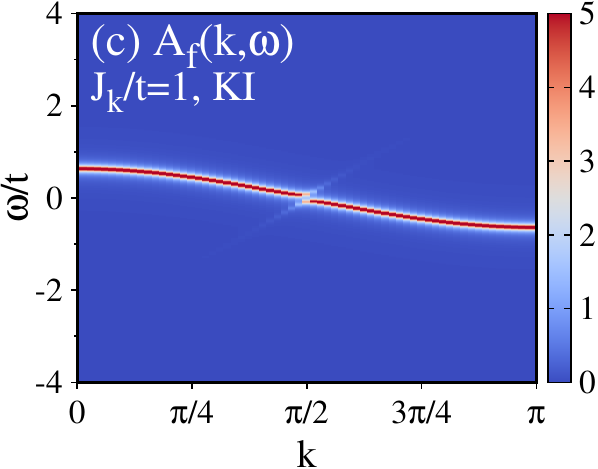}
\includegraphics[width=0.23\textwidth]{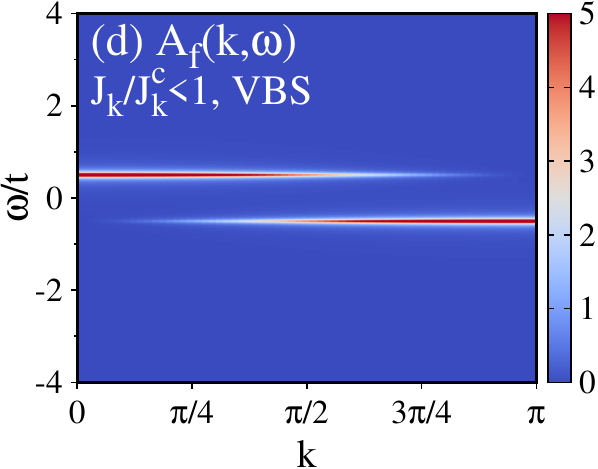}\\
\includegraphics[width=0.23\textwidth]{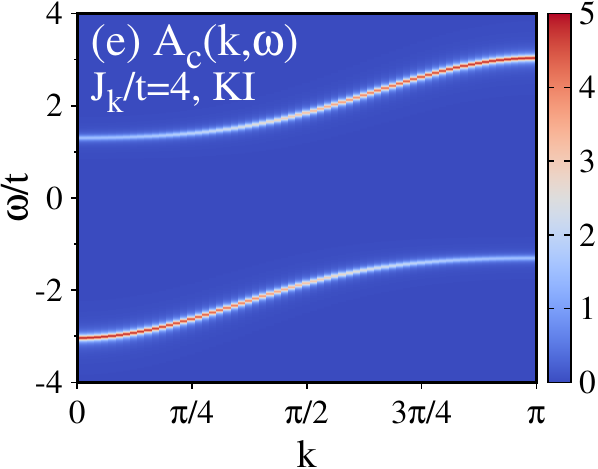}
\includegraphics[width=0.23\textwidth]{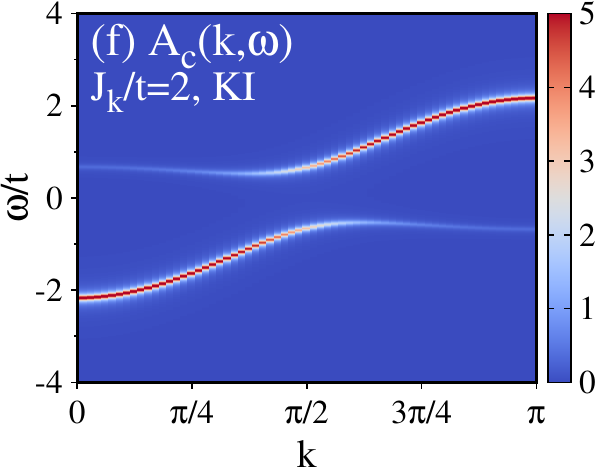}
\includegraphics[width=0.23\textwidth]{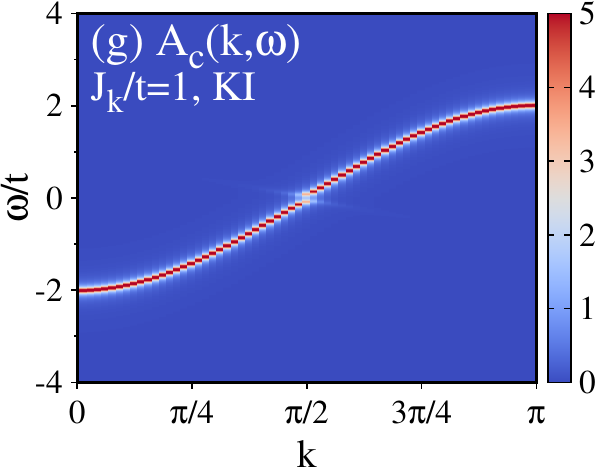}
\includegraphics[width=0.23\textwidth]{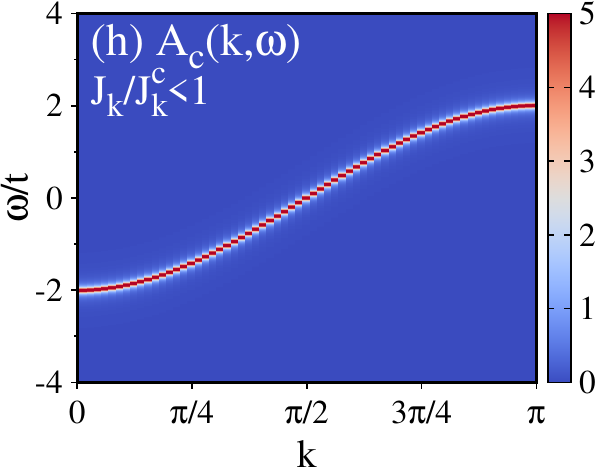}
        \caption{(a)-(d) $f$ fermion $A_{f}(\ve{k},\omega)$  and (e)-(h) conduction electron $A_{c}(\ve{k},\omega)$
        spectral functions as obtained from the large-$N$ approach accounting for both Kondo screening and spinon description 
	of the Kondo-Heisenberg chain with $J_h/t=1$ for representative values of $J_k/t$. 
        Panels (a)-(c) and (e)-(g) correspond to self-consistent solutions with  a uniform valence bond order parameter 
	Ansatz, i.e., $\chi_{ij}=\chi_0$. For $J_k\leqslant J_k^c=1.67t$ the ground state corresponds to the VBS phase 
	where each $f$ spin forms a dimer with only one of its two neighbors, i.e., $\chi_A=-1$ and $\chi_B=0$, and no Kondo 
	screening occurs, $V=0$, see panels (d) and (h). 
	}
\label{Fig:Ak_kh}
\end{figure*}

In principle, a spin dimerization on the $f$ spin layer can trigger, via Kondo coupling $J_k$, a bond order wave (BOW) state
with a dimerized kinetic energy of the conduction electrons such that $\xi_A\ne\xi_B$ where 
$\xi_{\ve{i}}=\langle \hat{c}^{\dagger}_{\ve{i}} \hat{c}^{\phantom\dagger}_{\ve{i}+\ve{1}} + h.c. \rangle$
~\cite{PhysRevB.102.245143,PhysRevB.109.014103}. 
Before exploring this issue by numerically solving the saddle-point equations in Sec.~\ref{sec:LargeN},  let us first discuss 
the following three limiting cases, each of them reducing the $4\times 4$ matrix in Eq.~(\ref{LargeN_vbs.eq}) to a $2\times 2$ 
matrix problem:

\begin{itemize}
\item
Vanishing Heisenberg interaction  $J_h=0$, i.e., conventional Kondo chain with the RKKY energy scale being the only 
intersite exchange interaction between the localized $f$ spins. Assuming  a finite hybridization order parameter  $V\ne 0$  
as appropriate for the KI phase, one arrives at  the single particle  dispersion relation,
\begin{equation}
	E_{\ve{k}}^{\pm}  = \frac{ \epsilon(\ve{k}) \pm \sqrt{ \epsilon^2(\ve{k})+ (J_k V)^2}}{2}, 
\end{equation}
and the corresponding quasiparticle residues
\begin{equation}
	Z_{\ve{k}}^{f,c}  =  \frac{1}{2} \left[ 1 \pm \frac{\epsilon(\ve{k}) }{ \sqrt{\epsilon^2(\ve{k})+ (J_k V)^2 }}\right],
	\label{QP_largeN}
\end{equation} 
with $+$ ($-$) sign denoting a doped hole away from half filling into the $f$ ($c$) state, respectively.
As illustrated in Fig.~\ref{Fig:Ak_klm}, the resultant low energy hybridized bands are nearly dispersionless, extend up to 
$\k=\pi$ momentum, and have predominantly $f$ character. While decreasing $J_k/t$ reduces both the slope of 
		the hybridized bands and the $c$ electron spectral weight at $\k=\pi$, the quasiparticle (QP) gap remains 
pinned to $\k=\pi$, $\Delta_{qp}  \equiv {\rm min}_{\ve{k}} \Delta_{qp}({\ve{k}}) = - E^{-}_{\ve{k}=\pi}$. 

\item
Vanishing dimerization order parameter $\delta=0$; consequently $\tilde{\epsilon}_f(\ve{k})=0$ for each $\ve{k}$. 
The delocalization of spinons via $\epsilon_f(\ve{k})$ on top of that through 
a finite hybridization order parameter $V$ modifies the dispersion relation of the hybridized bands,
\begin{equation}
	E_{\ve{k}}^{\pm}  = \frac{ \epsilon(\ve{k}) + \epsilon_f(\ve{k}) 
			 \pm \sqrt{ [ \epsilon(\ve{k}) - \epsilon_f(\ve{k}) ]^2 +(J_k V)^2}}{2},
\end{equation}
as well as the QP weights
\begin{equation}
        Z_{\ve{k}}^{f,c}  =  \frac{1}{2} \left[ 1 \pm \frac{\epsilon(\ve{k}) - \epsilon_f(\ve{k}) }
        { \sqrt{ [\epsilon(\ve{k}) - \epsilon_f(\ve{k}) ]^2 + (J_k V)^2 }}\right].
\end{equation}
Figure~\ref{Fig:Ak_kh} plots the evolution of $f$ and $c$ electron spectra upon varying $J_k/t$ at  $J_h/t=1$. 
As can be seen,  the nearly flat hybridized bands for $J_k/J_h\gg 1$  acquire appreciable dispersion 
upon reducing $J_k/t$. Consequently, the location of the QP gap $\Delta_{qp}$ becomes $J_k$ dependent drifting from 
$\k=\pi$ in the large $J_k/t=4$ limit to $\k=\pi/2$ at $J_k/t=1$, see Figs.~\ref{Fig:Ak_kh}(a)-\ref{Fig:Ak_kh}(c) 
and Figs.~\ref{Fig:Ak_kh}(e)-\ref{Fig:Ak_kh}(g).

\item	
Vanishing hybridization order parameter $V=0$ such that the $f$ and $c$ layers are fully decoupled. 
The self-consistency procedure for the $f$ subsystem yields a saddle-point solution where each spin forms a dimer 
with only one of its two neighbors, i.e., $\chi_A\ne 0$ and $\chi_B=0$~\cite{PhysRevB.37.3774,PhysRevB.39.11538}. 
As a result, $\chi_0=\delta\equiv\chi$ and thus the dispersion relation of the $f$ states 
\begin{equation}
	E_{\ve{k}}^{\pm}  = \pm J_h \sqrt{ \chi_0^2\cos^2 k + \delta_{\phantom 0}^2\sin^2 k }
\end{equation}
becomes completely localized with quasiparticle energies $\pm J_h\chi$, see Fig.~\ref{Fig:Ak_kh}(d). 
On the other hand,  the conduction electron layer corresponds to a free electron gas whose energy spectrum 
is given by a tight-binding dispersion relation without a gap, see Fig.~\ref{Fig:Ak_kh}(h). 
As we show it in Sec.~\ref{sec:LargeN}  by explicitly performing self-consistent large-$N$ calculations using 
the full $4\times 4$ matrix in Eq.~(\ref{LargeN_vbs.eq}), the $J_k=0$ (or equivalently $V=0$) VBS ground state 
extends up to $J_k^c/t=1.67$, i.e., at the mean-field level, VBS order is generically accompanied by vanishing Kondo 
screening~\cite{PhysRevB.83.214427,PhysRevLett.113.176402,Rey15,PhysRevB.92.165431,Li16,PhysRevB.98.085110}.

\end{itemize}

\section{\label{sec:results} Numerical results }


\subsection{\label{sec:KHLM} SU($\pmb{N}$) Kondo-Heisenberg chain}

In this section we explore the ground state as well as spin and single particle excitation spectra  
of the SU(${N}$) Kondo-Heisenberg model. Throughout, we consider the value $J_h/t=1$. 

\subsubsection{\label{sec:PD} Phase diagram}

\begin{figure}[t!]
\includegraphics[width=0.46\textwidth]{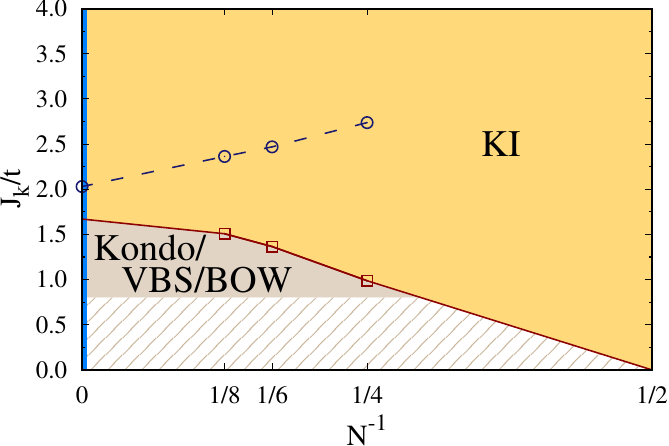}
\caption{The conjectured zero temperature phase diagram of the SU($N$) Kondo-Heisenberg chain with $J_h/t=1$ 
	hosts: (i) Kondo insulator (KI) state and (ii) valence bond solid (VBS) phase of $f$ spins coexisting with both 
	Kondo screening and a bond order wave (BOW) state of itinerant electrons. 
	Available system sizes in QMC simulations determine  the reachable energy scales of the model 
	and restrict reliable studies to values of $J_k/t\ge 0.8$ above the stripy region.
	The transition between the KI and VBS/BOW phases for $N\ge 4$ (squares) is shown to belong to the 2D classical 
	Ising universality class.
	The limit of infinite degeneracy $\tfrac{1}{N}=0$ represents a singular line (solid blue): 
        The phase transition is of first order, the $f$ and $c$ layers become fully decoupled in the VBS phase,  and 
	the BOW state does not appear.
	Circles indicate the change in the position of the minimal quasiparticle (QP) gap in the KI state 
	from  $\ve{k}=\pi/2$ at intermediate values of $J_k/t$ to $\ve{k}=\pi$ in the large $J_k/t$ region 
	reminiscent of the physics of the bare Kondo chain. 
        For $N=2$, the critical nature of the spin-1/2 Heisenberg antiferromagnetic chain~\cite{PhysRevB.36.5291}
        implies the instability towards the KI phase at any finite value of $J_k/t$.
        The $N=2$  data quality does not allow us to meaningfully estimate the QP gap at the $\k=\pi$ point. 
	Given a tiny  $\sim 10^{-4}$ QP residue at $\k=\pi$, we believe that the low energy physics will 
	be dominated by QP excitations at the $\k=\pi/2$ point. }
	\label{Fig:PD}
\end{figure}

\begin{figure*}[t!]
\includegraphics[width=0.32\textwidth]{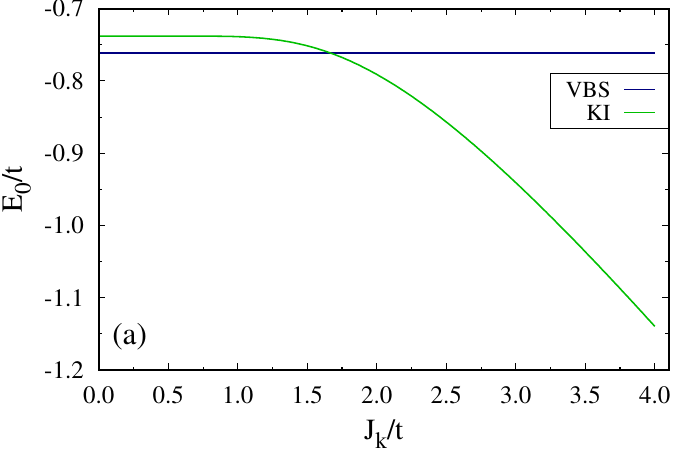}  
\includegraphics[width=0.32\textwidth]{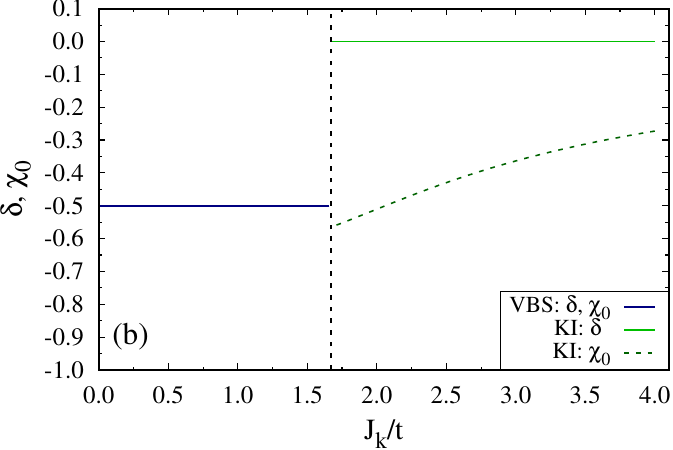}
\includegraphics[width=0.32\textwidth]{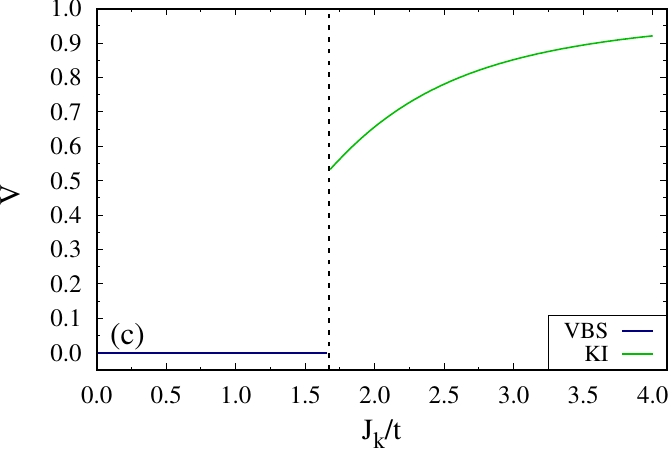} \\
\includegraphics[width=0.32\textwidth]{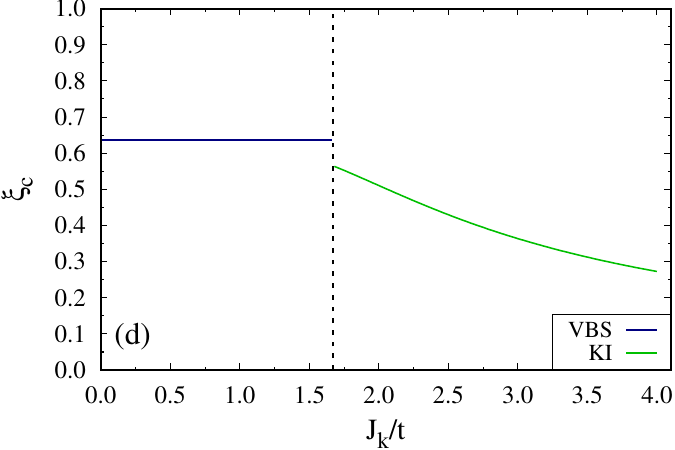}
\includegraphics[width=0.32\textwidth]{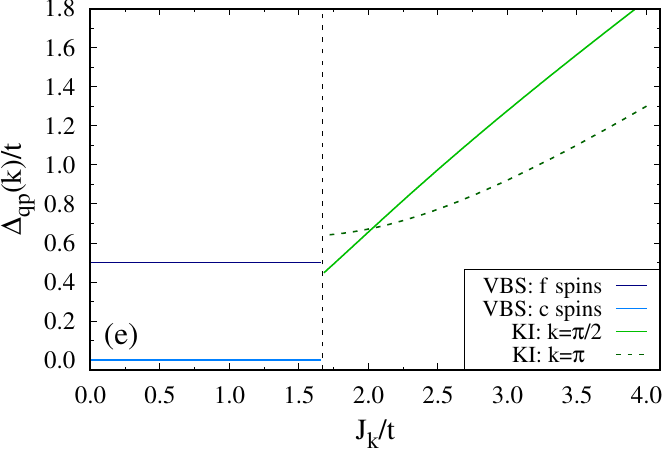} 
	\caption {Large-$N$ analysis of the Kondo-Heisenberg chain:  
        (a) Ground state energy $E_0/t$, (b) valence bond $\delta$, $\chi_0$, and (c) Kondo singlet $V$ order parameters 
	as a function of $J_k/t$ reveal a direct first order phase transition 
	at $J_k^c/t=1.67$ separating a VBS phase with fully decoupled $f$ spin dimers ($\delta=\chi_0\ne 0$ and $V=0$) 
	and a uniform KI state ($\delta = 0$, $\chi_0\ne 0$, and $V\ne 0$).
	For completeness, we also show the evolution of (d) bond kinetic energy of the conduction electrons  $\xi_c$ 
	and (e) quasiparticle gap $\Delta_{qp}(\ve{k})$. 
	The latter is momentum independent in the VBS phase of the $f$ layer and since $V=0$ it vanishes in the conduction electron layer. 
	In the KI phase, the gap displays a crossover between the $\ve{k}=\pi/2$ and $\ve{k}=\pi$ points.}
\label{Fig:MF}
\end{figure*}

Figure~\ref{Fig:PD} maps out the ground state phase diagram as a function of $\tfrac {1}{N}$ and Kondo coupling $J_k/t$
and provides a concise summary of our main findings.  In the large Kondo coupling limit  $J_k/J_h\gg 1$, 
the effect of the Heisenberg interaction can be neglected and the low energy physics should correspond 
to that of the bare Kondo chain with $J_h=0$. We thus find a homogeneous KI phase for each $N$. 

Upon decreasing the Kondo coupling $J_k/t$, the Heisenberg interaction progressively dominates over the Kondo scale.  
As indicated by circles, it gives rise to a change in the position of the minimal quasiparticle gap in the KI state  
from $\ve{k}=\pi$ consistent with a bare Kondo chain to the $\ve{k}=\pi/2$ point.  
This effect is equally captured by the large-$N$ approach accounting for both Kondo screening and the spinon description
of the Kondo-Heisenberg chain.  

Given a dimerized ground state of the bare SU($N\ge 4$) Heisenberg antiferromagnetic  chain~\cite{Paramekanti07,PhysRevB.100.085103}, 
further reduction of $J_k/t$  triggers a phase transition to the VBS phase of the $f$ spins. 
While the finite size scaling of the QMC data is consistent with the 2D classical Ising universality class of the transition, 
the large-$N$ approach erroneously produces a direct first order phase transition.

Another difference between the QMC and large-$N$ methods concerns the nature of the ordered phase: 
The QMC data are consistent with VBS order coexisting with both Kondo screening and a BOW state
displaying a dimerized kinetic energy of itinerant electrons.
In contrast, at the mean-field level, the KI-VBS phase transition represents an example of the Kondo breakdown  such that 
the $f$ and $c$ layers are fully decoupled, and thus the BOW phase does not appear.

Since both the relevant energy scales, i.e., dimer order and Kondo screening are generated fully dynamically in QMC simulations, 
reaching the genuine ground state becomes less tractable for small Kondo couplings and requires us 
to simulate  exponentially long chains. Within the available system sizes $L\le 130$, we are hence limited to values of 
$J_k/t\ge 0.8$ above the stripy region in Fig.~\ref{Fig:PD}.

\subsubsection{\label{sec:LargeN}  Large-$N$ analysis}

To account for the possible coexistence of dimer order and Kondo screening,  we solve the saddle-point equations (\ref{saddle}) 
for a unit cell with two sites $A$ and $B$  using the large-$N$ Hamiltonian matrix in Eq.~(\ref{LargeN_vbs.eq}). 
Figures \ref{Fig:MF}(a)-\ref{Fig:MF}(c) show the ground state energy of the solutions we have found 
and the corresponding mean-field order parameters as a function of Kondo coupling $J_k/t$.

Despite initializing our large-$N$ calculations using different sets of the hybridization order
$ V   =  \langle \hat{f}^{\dagger}_{\ve{i}} \hat{c}^{\phantom\dagger}_{\ve{i}} + h.c. \rangle $, 
uniform $\chi_0=\tfrac{1}{2}(\chi_A+\chi_B)$, 
and staggered $\delta=\tfrac{1}{2}(\chi_A-\chi_B)$ valence bond order parameters,  
the self-consistent procedure yields only two distinct phases: 
(i) VBS state with fully decoupled dimers characterized by $\delta=\chi_0\ne 0$ and $V=0$ and 
(ii) homogeneous KI state with $\delta = 0$, $\chi_0\ne 0$, and $V\ne 0$.
By comparing the ground state energies of these solutions, we find that the $J_k=0$ (or equivalently $V=0$) 
VBS phase persists as the ground state for a finite range of Kondo coupling $J_k/t\leqslant 1.67$.   
At this point,  Kondo screening becomes competitive enough to compensate both the energy gain due to the formation of 
strong valence bond singlets in the VBS phase and a loss of the bond kinetic energy of conduction electrons 
$\xi_c=\langle \hat{c}^{\dagger}_{\ve{i}} \hat{c}^{\phantom\dagger}_{\ve{i}+\ve{1}} + h.c.\rangle $
due to the formation of Kondo singlets with local moments, see Fig.~\ref{Fig:MF}(d). Consequently, 
a uniform KI phase becomes the ground state. 
As is apparent,  the transition is signaled by a jump in each order parameter and thus is of first order.   
Given existing evidence for the first order nature of the VBS-KI transition in frustrated 2D 
lattices~\cite{PhysRevB.83.214427,PhysRevLett.113.176402,Rey15,Li16,PhysRevB.92.165431,PhysRevB.98.085110}, 
it appears that frozen dimers in the realm of the large-$N$ approximation are generically accompanied by 
a vanishing of Kondo screening.

\begin{figure*}[t!]
\includegraphics[width=0.32\textwidth]{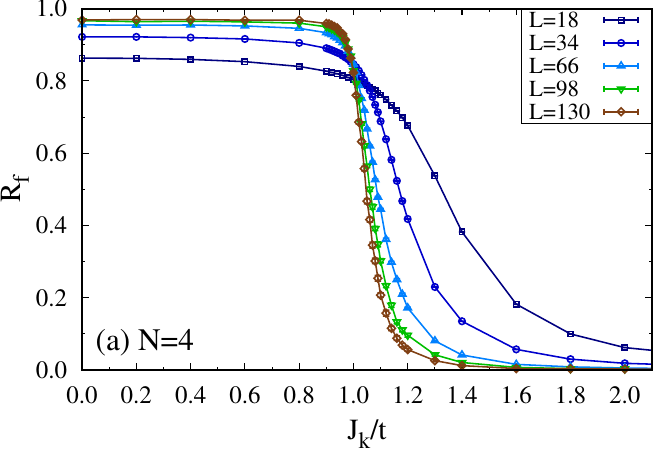}
\includegraphics[width=0.32\textwidth]{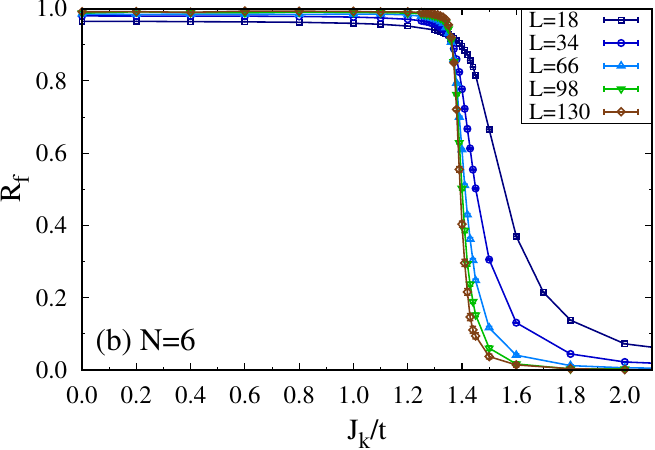}
\includegraphics[width=0.32\textwidth]{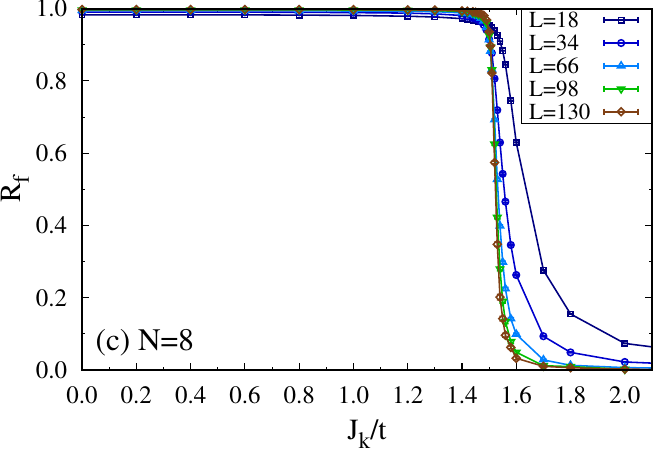}
\caption{Correlation ratio $R_f$ from the structure factor of $f$ spin dimer correlations in the SU($N$) Kondo-Heisenberg chain:  
	(a) $N=4$; (b) $N=6$, and (c) $N=8$.}
\label{Fig:Rf}
\end{figure*}

\begin{figure*}[t!]
\includegraphics[width=0.23\textwidth]{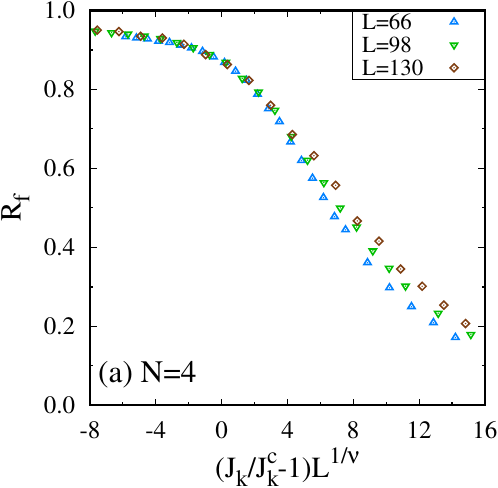}  
\includegraphics[width=0.23\textwidth]{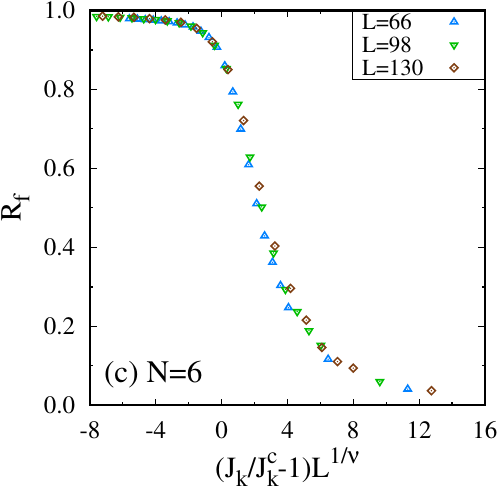}
\includegraphics[width=0.23\textwidth]{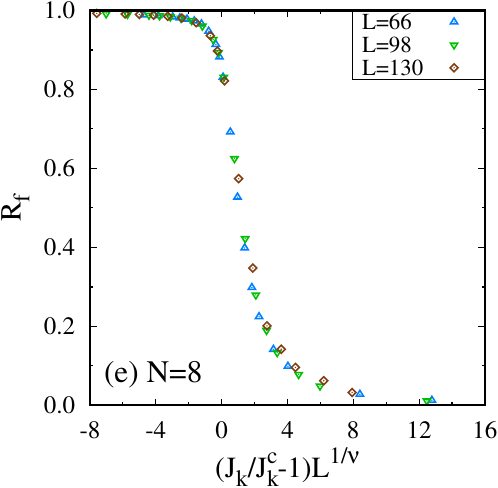}\\
\includegraphics[width=0.23\textwidth]{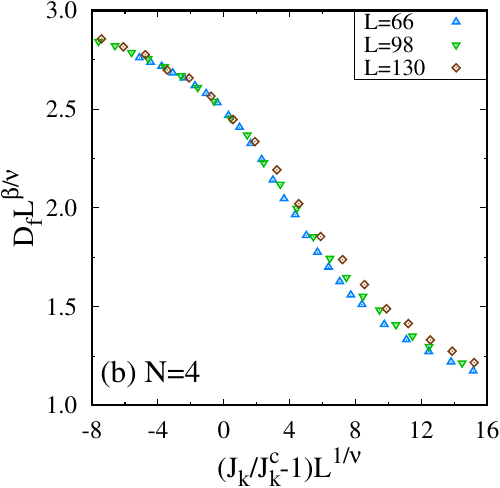}
\includegraphics[width=0.23\textwidth]{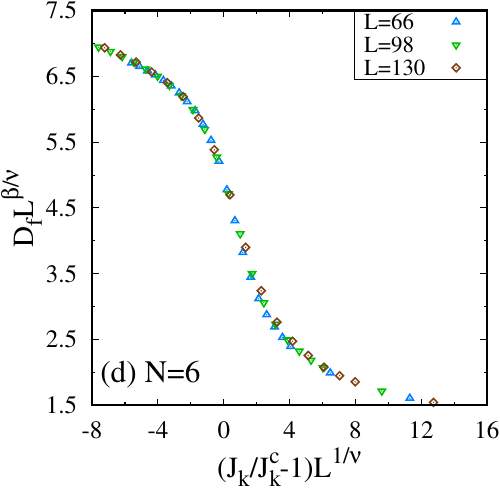}
\includegraphics[width=0.23\textwidth]{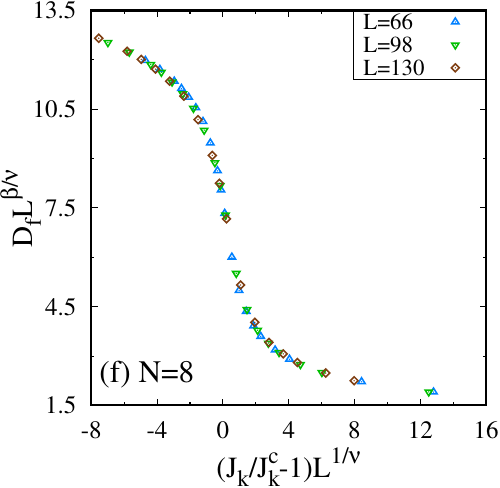}
\caption{Data collapse for the $L\ge 66$ SU($N$) Kondo-Heisenberg chains using the 2D Ising critical exponents 
	$\nu=1$ and $\beta = 1/8$  of the: 
	(a), (c), and (e) correlation ratio $R_f$ from the structure factor of $f$ spin dimer
	correlations  and (b), (d), and (f) the dimer order parameter $D_f=\sqrt{\frac{\mathcal{D}_f(\ve{q}=\pi)}{L}}$ 
	for $N=4$ (left), $N=6$ (middle), and $N=8$ (right).}    
\label{Fig:Ising}
\end{figure*}

Finally, Fig.~\ref{Fig:MF}(e) illustrates the evolution of the quasiparticle gap. 
In the VBS phase, the dispersion relation of the $f$ states is completely localized with momentum independent 
quasiparticle energies $\pm J_h\chi_0$ such that $\Delta_{qp}/t=0.5$. 
In contrast, the interplay between Kondo screening $V$ and valence bond singlets $\chi_0$  in the KI state determines 
the position of the minimal quasiparticle gap which shifts from  $\ve{k}=\pi/2$ for small $J_k/t$ 
to the $\ve{k}=\pi$ point in the large $J_k/t$ region.

\begin{figure*}[t!]
\includegraphics[width=0.32\textwidth]{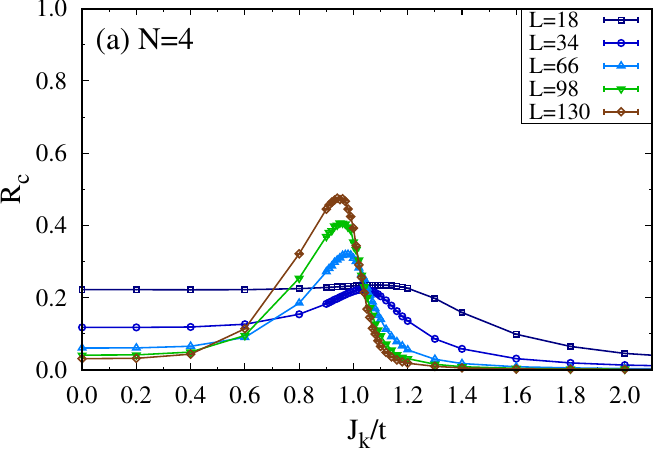}
\includegraphics[width=0.32\textwidth]{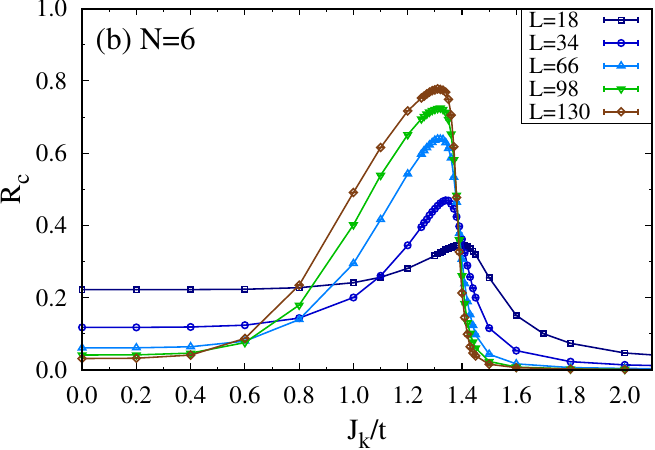}
\includegraphics[width=0.32\textwidth]{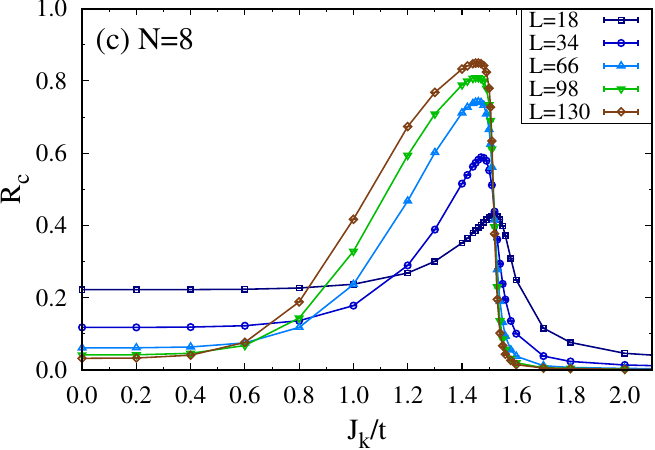}
	\caption{Correlation ratio $R_c$ from the structure factor of kinetic dimer correlations for the conduction electrons 
	in the SU($N$) Kondo-Heisenberg chain:
	(a) $N=4$; (b) $N=6$, and (c) $N=8$. }
\label{Fig:Rc}
\end{figure*}

\begin{figure*}[t!]
\includegraphics[width=0.32\textwidth]{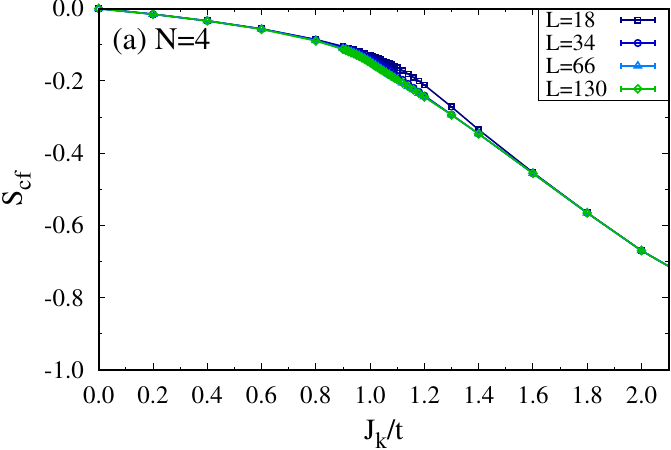}
\includegraphics[width=0.32\textwidth]{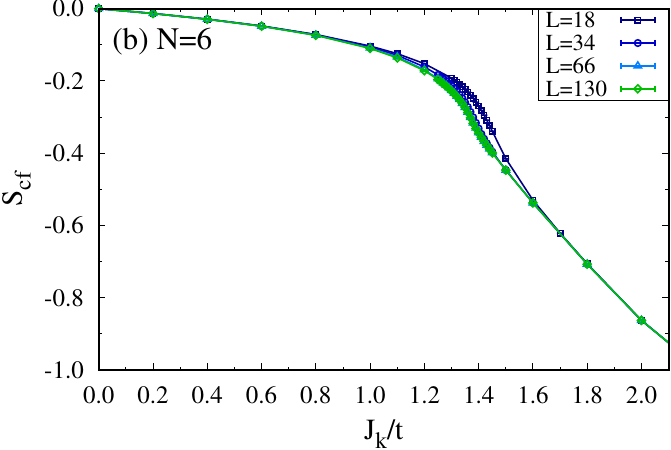}
\includegraphics[width=0.32\textwidth]{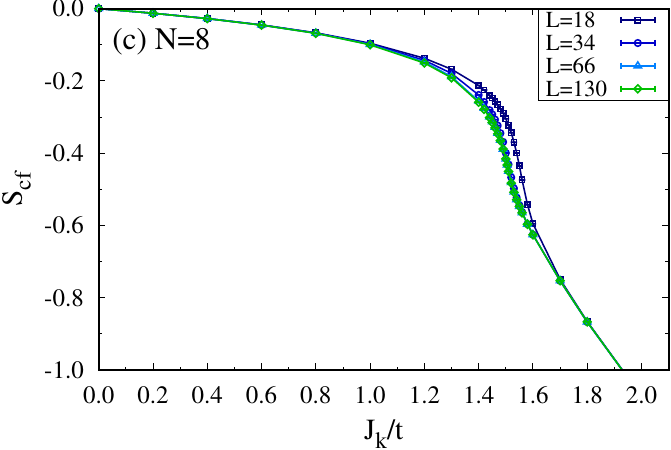}
	\caption{Local spin-spin correlation function $S_{cf}$ as a function of $J_k/t$ in the SU($N$) 
	Kondo-Heisenberg chain with a different length $L$: 
	(a) $N=4$; (b) $N=6$, and (c) $N=8$.
	}
\label{Fig:Scf}
\end{figure*}

\subsubsection{\label{sec:Ising} Critical behavior}

In order to locate the phase boundary in Fig.~\ref{Fig:PD}  we carry out a finite size scaling analysis of the
spin dimer structure factor for the $f$ spins: 
\begin{align}
	\mathcal{D}_f(\ve{q})  &  = 
    \frac{1}{L} \sum_{\ve{i},\ve{j}}  e^{i \ve{q} \cdot \left( \ve{i} - \ve{j} \right) }  \times  \nonumber \\   
  &   \left( \langle \hat{\Delta}_{\ve{i},\ve{i}+\ve{1}} \hat{\Delta}_{\ve{j},\ve{j}+\ve{1}} \rangle  
     - \langle \hat{\Delta}_{\ve{i},\ve{i}+\ve{1}} \rangle \langle \hat{\Delta}_{\ve{j},\ve{j}+\ve{1} } 
	\rangle \right),
\end{align}
 with $\hat{\Delta}_{\ve{i},\ve{i}+\ve{1}} = 
     \sum_{\mu,\nu}  \hat{S}^{f,\mu}_{\ve{i},\nu} \hat{S}^{f,\nu}_{\ve{i}+\ve{1},\mu}$.
Next, we calculate the renormalization group invariant correlation ratio
\begin{equation}
	R_{f} =  1 - \frac{\mathcal{D}^{f}(\ve{Q} - \delta\ve{q}) }{\mathcal{D}^{f}(\ve{Q}) },
\label{eq:R}
\end{equation}
where $\ve{Q}=\pi$ is the ordering wavevector and $\delta\ve{q}=2\pi/L$ is the smallest finite wavevector 
for a given system size $L$. As can be seen in Fig.~\ref{Fig:Rf}, $R_f$ scales to unity (zero) for ordered (disordered) 
states and shows a crossing point as a function of system size at the critical Kondo coupling $J_k^c$. 
The latter can be obtained accurately by noting that a two-fold degeneracy of the dimerized ground state implies
the same critical behavior as the 2D Ising model, i.e., the correlation length exponent $\nu=1$ and the order parameter 
exponent $\beta = 1/8$. Indeed, using the scaling assumption
\begin{equation}
	R_f=\mathcal{F}[(J_k/J_k^c-1)L^{1/\nu}],
\end{equation}
we obtain for $L\geq 66$ and in the vicinity of the critical point, a decent data collapse of $R_f$ shown 
in Figs.~\ref{Fig:Ising}(a) ($N=4$), \ref{Fig:Ising}(c) ($N=6$), and \ref{Fig:Ising}(e) ($N=8$). 
It allows us to estimate critical couplings $J_k^c/t = 0.987(2)$ for $N=4$, $J_k^c/t =1.366(1)$ for $N=6$, 
and $J_k^c/t =1.508(1)$ for $N=8$. 
Clearly, an enhanced SU($N$) symmetry favors the dimerized ground state and shifts the critical coupling
to larger values of $J_k$ in accordance with the large-$N$ calculation predicting $J_k^c/t=1.67$. 

To further assess the consistency of our numerical data  with the 2D Ising universality class, we consider a dimer order parameter 
$D_f=\sqrt{\frac{\mathcal{D}_f(\ve{q}=\pi)}{L}}$ which  shall satisfy the finite size scaling relation  
\begin{equation}
	D_fL^{\beta/\nu}=\mathcal{F}[(J_k/J_k^c-1)L^{1/\nu}],
\end{equation}
involving both the critical exponents $\nu$ and $\beta$. The corresponding scaling collapses are shown 
in Figs.~\ref{Fig:Ising}(b) ($N=4$), \ref{Fig:Ising}(d) ($N=6$), and \ref{Fig:Ising}(f) ($N=8$) and lead 
to critical Kondo couplings consistent within error bars with the previous estimates.

Interestingly, even though it is not the SU($N$) spin symmetry that gets broken at the phase transition, changing $N$ modifies the shape of 
the scaling function $\mathcal{F}$ such that it progressively develops a step-like behavior.
Since a correlation ratio  $R$  generically  scales as $(\xi/L)^2$ where $\xi$ is the correlation length \cite{CARACCIOLO1993475}, 
a constant value of $R$ for $J_k>J_k^c$ implies no divergence of the correlation length and thus a first order phase transition. 
This line of arguing is consistent with the large-$N$ analysis predicting the first order nature of transition 
in the limit of infinite degeneracy $N\to\infty$. We believe, however, that it is a singular point and quantum critical 
fluctuations restore the 2D classical Ising universality class close to $J_k^c$  at any finite $N$ even though the critical 
region itself shrinks. 
We also note that the development of a step-like scaling function helps to reduce the finite size effect and allows one to  
obtain a nice data collapse over a broader interval around $J_k^c$.

It is very natural to expect that the translationally invariant conduction electrons gas 
coupled  via a Kondo coupling $J_k>0$ to a dimerized lattice of magnetic  impurities 
will spontaneously reduce its symmetry by acquiring the same dimerization pattern as the local $f$ spins.
To confirm this expectation and quantify the degree of dimerization in the conduction electron layer, 
we measure the kinetic energy  dimer structure factor 
\begin{align}
	\mathcal{K}_c(\ve{q})  &  =
    \frac{1}{L} \sum_{\ve{i},\ve{j}}  e^{i \ve{q} \cdot \left( \ve{i} - \ve{j} \right) }  \times  \nonumber \\
  &   \left( \langle \hat{K}_{\ve{i},\ve{i}+\ve{1}} \hat{K}_{\ve{j},\ve{j}+\ve{1}} \rangle
     - \langle \hat{K}_{\ve{i},\ve{i}+\ve{1}} \rangle \langle \hat{K}_{\ve{j},\ve{j}+\ve{1} }
        \rangle \right),
\end{align}
 with $\hat{K}_{\ve{i},\ve{i}+\ve{1}} = 
    \ve{\hat{c}}^{\dagger}_{\ve{i}} \ve{\hat{c}}^{\phantom\dagger}_{\ve{i}+\ve{1}} + h.c.$
Figure~\ref{Fig:Rc} plots the corresponding correlation ratio 
\begin{equation}
        R_{c} =  1 - \frac{\mathcal{K}^{c}(\ve{Q} - \delta\ve{q}) }{\mathcal{K}^{c}(\ve{Q}) },
\label{eq:Rc}
\end{equation}
for $\ve{Q}=\pi$ as a function of $J_k/t$. On the one hand,  the observed increase of $R_c$ around $J_k^c$ is indicative 
of strong BOW fluctuations. Given the growth of $R_c$ with a system size $L$, we expect the onset of 
long range BOW order in the thermodynamic limit.
On the other hand,  by considering a finite spin gap $\Delta_s$ in the dimerized ground state of the bare SU($N\ge 4$) 
Heisenberg chain, one could argue in favor of a simple picture of a complete decoupling of both layers 
in the limit $J_k/\Delta_s\ll 1$.

\begin{figure*}[t!]
\includegraphics[width=0.23\textwidth]{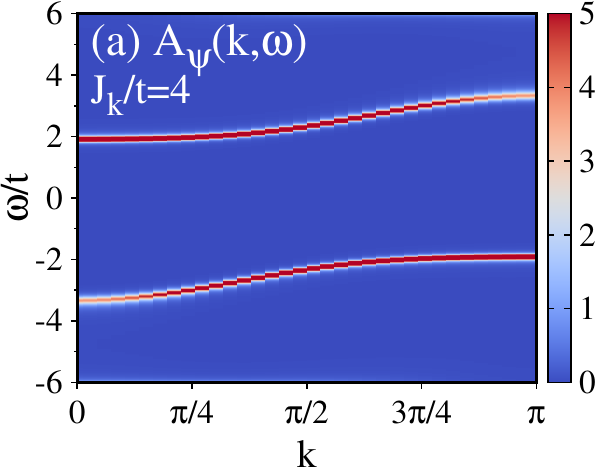}
\includegraphics[width=0.23\textwidth]{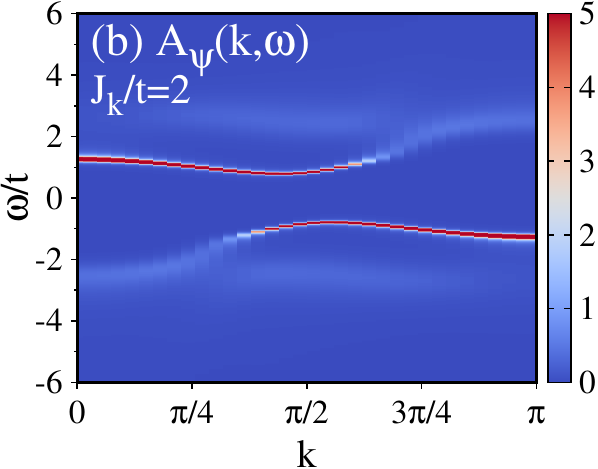}
\includegraphics[width=0.23\textwidth]{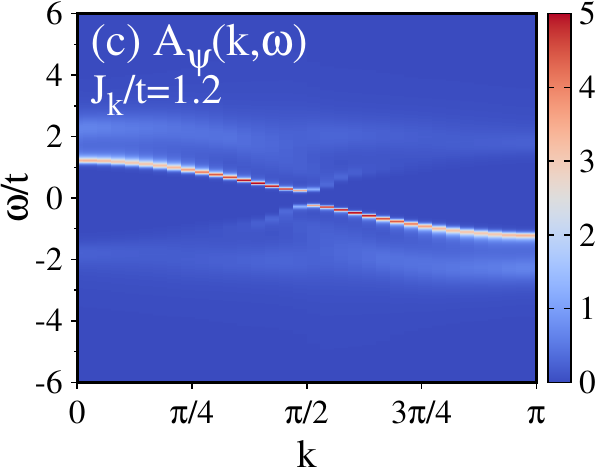}
\includegraphics[width=0.23\textwidth]{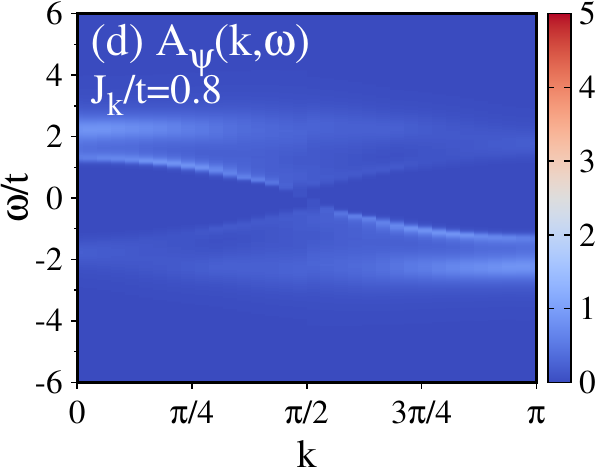}\\
\includegraphics[width=0.23\textwidth]{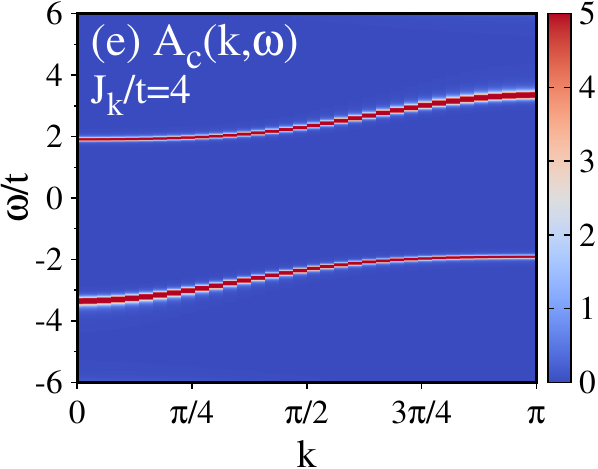}
\includegraphics[width=0.23\textwidth]{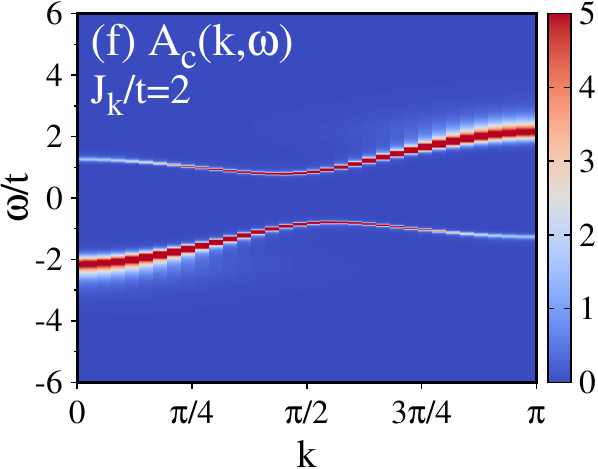}
\includegraphics[width=0.23\textwidth]{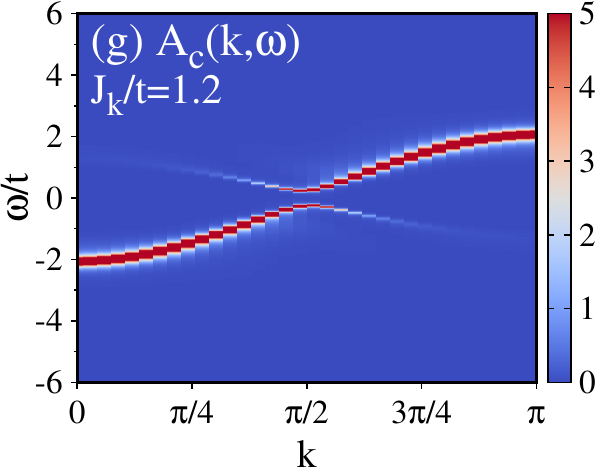}
\includegraphics[width=0.23\textwidth]{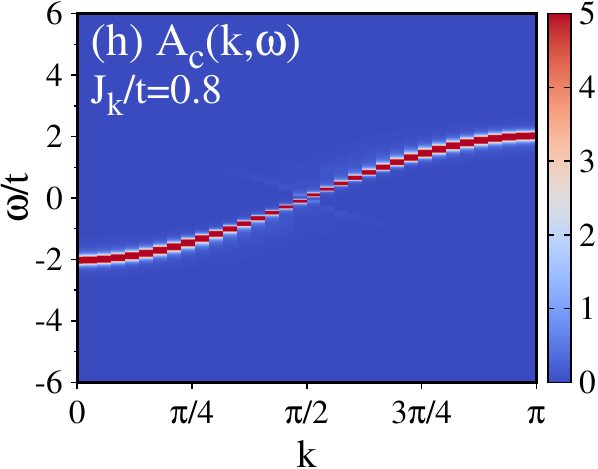}
	\caption{(a)-(d) Composite fermion $A_{\psi}(\ve{k},\omega)$ and (e)-(h) conduction electron $A_{c}(\ve{k},\omega)$  
	spectral functions of the $L=66$  SU(4) Kondo-Heisenberg chain for representative values of $J_k/t$; 
	the critical Kondo coupling $J_k^c/t = 0.987(2)$.} 
\label{Fig:AkN4kh}
\end{figure*}

To address this issue,  we plot in Fig.~\ref{Fig:Scf} the local spin-spin correlator 
\begin{equation}
S_{cf}=\frac{N}{(N^2-1)L}\sum_{\ve{i}} \sum_{a=1}^{N^2-1}\langle
        \ve{\hat{c}}^{\dagger}_{\ve{i}} \ve{T}^{a} \ve{\hat{c}}^{\phantom\dagger}_{\ve{i}}
         \cdot \ve{\hat{f}}^{\dagger}_{\ve{i}} \ve{T}^{a} \ve{\hat{f}}^{\phantom\dagger}_{\ve{i}} \rangle .
	\label{eq:Scf}
\end{equation}
Here   $T^{a}$   corresponds to the generators of  SU($N$)  with  the normalization condition:  
$ \text{Tr} \, T^{a} T^{b}  =  \frac{\delta_{a,b}}{2}$.
As is apparent,  this quantity remains finite at any $J_k>0$ such that the picture of a complete 
decoupling never holds. Thus we believe that  simulations performed on sufficiently long chains would reveal 
the signature of coexisting VBS and BOW orders as soon as $J_k>0$, see also Ref.~\cite{PhysRevB.109.014103}. 
Indeed, given that the long distance behavior of correlation functions determines the low energy physics, 
it is obvious that limited system sizes used in our QMC simulations act as the  energy  cutoff for long-wavelength 
dimer fluctuations. On the one hand, already our shortest $L=18$ chain is sufficient to produce the increase 
of $R_c$ around the quantum critical point $J_k^c$ where quantum fluctuations diverge and become 
scale invariant. On the other hand, on moving away from $J_k^c$ in the VBS phase while keeping the system 
size fixed, one inevitably arrives at the regime with effectively frozen $f$ spin dimers. This regime is 
captured by the large-$N$ theory and, as we have explicitely shown, fails to account for the BOW phase
arising from VBS order and erroneously leads to the breakdown of Kondo screening. 
This implies that to provide a definite answer concerning the fate of Kondo screening 
it is imperative to control first the energy scale of the BOW state.

In fact, Fig.~\ref{Fig:Rc} allows one to estimate the minimal value of $J_k/t$ above which system 
sizes $L$ available in our QMC simulations guarantee that the conduction electrons are subject to fluctuating  
$f$ spin dimers. Indeed, as a function of $L$, it is possible to overcome the initial decrease of $R_c$ for 
$J_k/t\ge 0.8$. In contrast, for smaller values of $J_k/t$ and up to our largest system sizes $L=130$, 
we are not able anymore to generate an increase of $R_c$ indicating that our results are biased by 
finite size effects.

Finally, it is noteworthy to point out a progressively steeper evolution of the local spin-spin correlation $S_{cf}$ 
across the transition point, see Figs.~\ref{Fig:Scf}(a)-\ref{Fig:Scf}(c). 
A similar tendency is also observed in Fig.~\ref{Fig:Ising} where the enlarged  SU($N$) symmetry of the model 
drives the onset of a step-like behavior of the scaling function $\mathcal{F}$.  
The similarity in the  response to the enlarged SU($N$) symmetry seen in $\mathcal{F}$ and $S_{cf}$ follows 
from the fact that the latter is directly proportional to the free energy derivative $\frac{\partial F}{\partial J_k}$. 
Hence, a discontinuity in $S_{cf}$ upon varying $J_k/t$  would signal a first order transition.  
The latter is expected on the basis of large-$N$ calculations to take place only in the limit of infinite degeneracy 
$N\to\infty$.

\subsubsection{\label{sec:dynam_single} Single particle spectra}

With restrictions on a minimal value of $J_k/t\simeq 0.8$ imposed by the available system sizes 
in QMC simulations in mind, we proceed to elucidate the fate of Kondo screening across the phase diagram 
in Fig.~\ref{Fig:PD}.

To this end, we calculate the evolution of the momentum resolved spectral function of the composite fermion 
$A_{\psi}(\ve{k},\omega) =  - \frac{1}{\pi} \text{Im}    G^{\text ret}_{\psi}(\k,\omega)$ with 
$ G^{\text ret}_{\psi}(\k,\omega)=
        -i \int^\infty_0 dt e^{i \omega t} \sum_{\sigma} 
        \big \langle \big\{ \hat{{\psi}}_{\k,\sigma}(t),\hat{ {\psi}}^\dagger_{\k,\sigma}(0) \big\}  \big\rangle $.
In the Kondo screened phase $A_\psi(\k,\omega)$ displays well defined 
quasiparticle poles whose dispersion relation follows that of the $f$ fermion in the large-$N$ approach~\cite{Danu21}.
In contrast, a continuum of excitations in $A_\psi(\k,\omega)$ signals the decay of a composite quasiparticle and points out 
the breakdown of a coherent lattice Kondo effect~\cite{Raczkowski22}. 
To identify the origin of dominant spectral features, we have equally calculated the electronic excitation spectrum 
for the conduction electrons $A_{c}(\ve{k},\omega) =  - \frac{1}{\pi} \text{Im} G^{\text ret}_{c}(\k,\omega)$. 


To begin with let us consider the SU(4) model. 
Figures~\ref{Fig:AkN4kh}(a)-\ref{Fig:AkN4kh}(c) [\ref{Fig:AkN4kh}(e)-\ref{Fig:AkN4kh}(g)] 
illustrate the evolution of $A_{\psi}(\ve{k},\omega)$ [$A_{c}(\ve{k},\omega)$], respectively,  upon 
going through the uniform Kondo screened phase. As a function of decreasing $J_k/t$,  there is a continuous change 
of the wavevector which determines the low energy dynamics from $\k=\pi$ to $\k=\pi$/2. Thus, the QMC data validate 
the findings within the large-$N$ approach accounting for both Kondo screening and the spinon description of the 
Kondo-Heisenberg chain, see Fig.~\ref{Fig:Ak_kh}(a)-\ref{Fig:Ak_kh}(c). 

This result is surprising and deserves more attention. The surprise comes from early QMC studies of a single-hole 
dynamics in the half-filled 2D Kondo-Hubbard model with an extra Hubbard interaction $U$ in the 
conduction band~\cite{PhysRevB.66.045103}.  Given that the low energy hole dynamics between the Kondo and Hubbard 
models is controlled by different wavevectors, it is not obvious how these conflicting tendencies can be 
accommodated in a single effective  electronic dispersion  when both interactions are brought together. 
It turns out, however, that irrespective of values of $U/t$, the lowest energy hole states occur 
for $J_k/t>0$  at $\ve{k}=(\pi,\pi)$ as in the pure Kondo lattice model. 
Te supremacy  of Kondo  and RKKY interactions in determining the low energy band structure of the uniform KI state 
can be nevertheless broken down by the Heisenberg interaction between the local moments by shifting the position of the minimal 
gap from $\k=\pi$ at large Kondo coupling $J_k/t=4$, see Fig.~\ref{Fig:AkN4kh}(a), to $\k=\pi$/2 at $J_k/t=1.2$ in Fig.~\ref{Fig:AkN4kh}(c). 
Importantly, this change does not require any extra spontaneous symmetry breaking. Hence,  a rigid shift of the hybridized bands upon doping
would lead to a Lifshitz transition where the Fermi surface topology in the heavy fermion liquid  changes from two  
to four points. This guarantees that the Luttinger volume remains preserved~\cite{PhysRevLett.79.1110}.

This piece of information has consequences for the interpretation of experimental data.   It implies that 
the form of the low energy band structure allows one to identify, at least in a 1D setup, 
the origin of the dominant magnetic interaction between the local $f$ spins: A direct Heisenberg exchange 
vs. the indirect RKKY interaction. It remains to be verified numerically if and how our result  carries over 
to higher dimensions.

We proceed now to discuss electronic spectra at our smallest value of $J_k/t\simeq 0.8$ in the ordered phase. 
As shown in Fig~\ref{Fig:AkN4kh}(d), a faint composite fermion band extending to the $\k=\pi$ point 
persists in the presence of fluctuating $f$ spin dimers. The latter are compatible with Kondo screening and trigger 
BOW fluctuations in the conduction electron layer. Both effects modify its tight binding cosine band 
which acquires a small gap at the Fermi level and develops backfolded  shadow bands tracking the composite 
fermion band, see Fig~\ref{Fig:AkN4kh}(h).
As we show latter on,  the gap opens up exponentially and thus it requires correspondingly large system sizes 
to simulate which renders the BOW phase intractable in the weak coupling limit $J_k/t<0.8$.

The above findings contrast with a large-$N$ outcome in Fig.~\ref{Fig:Ak_kh}(d). It displays a fully localized, i.e., 
flat, $f$ electron spectrum  specific to VBS order with frozen isolated  dimers  accompanied by the breakdown of 
Kondo screening.  As a consequence, the conduction electrons are fully decoupled from the localized spins and 
their dispersion relation  recovers a tight binding cosine form, see Fig.~\ref{Fig:Ak_kh}(h).


\begin{figure*}[t!]
\includegraphics[width=0.23\textwidth]{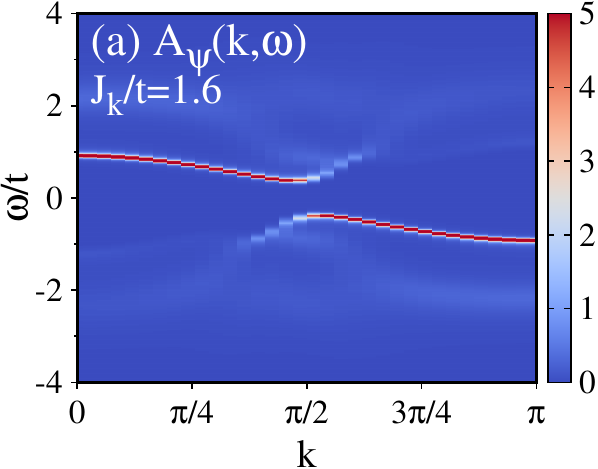}
\includegraphics[width=0.23\textwidth]{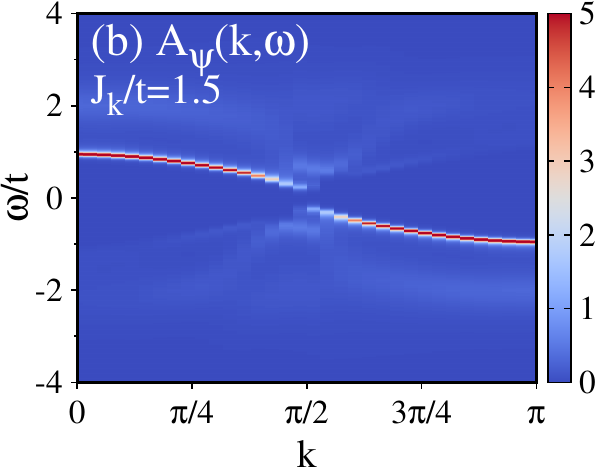}
\includegraphics[width=0.23\textwidth]{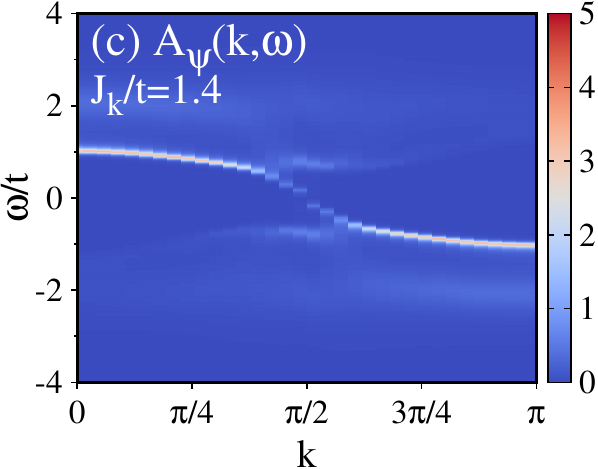}
\includegraphics[width=0.23\textwidth]{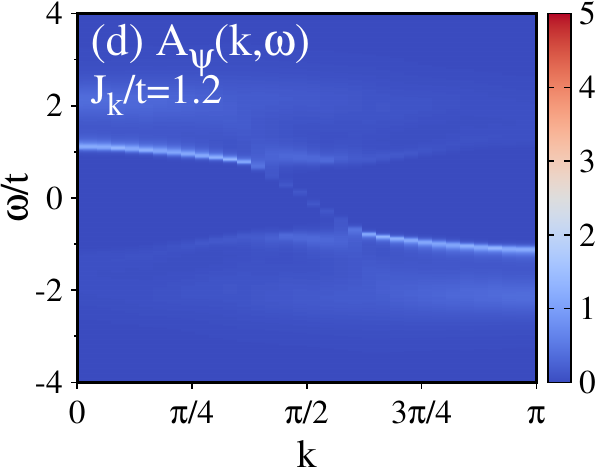}\\
\includegraphics[width=0.23\textwidth]{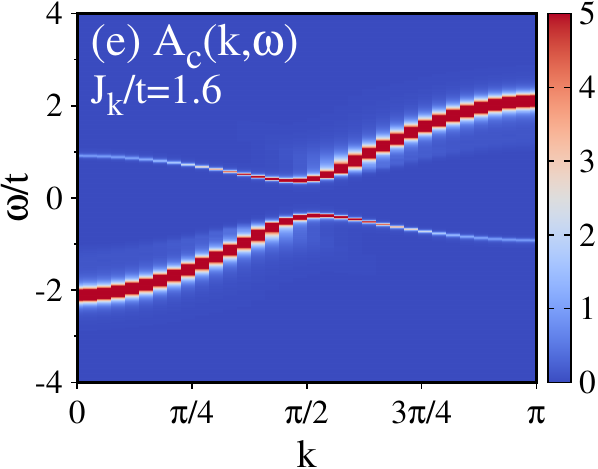}
\includegraphics[width=0.23\textwidth]{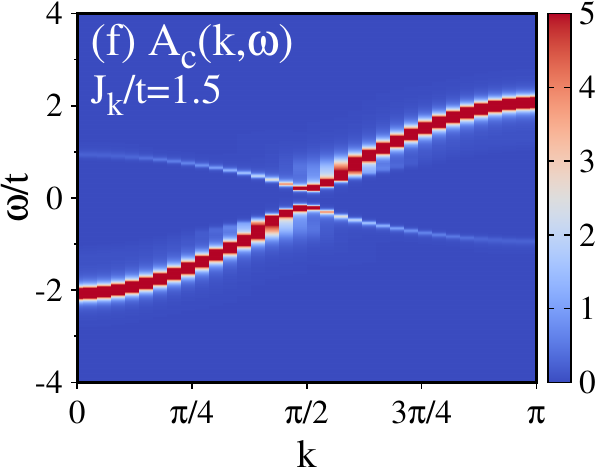}
\includegraphics[width=0.23\textwidth]{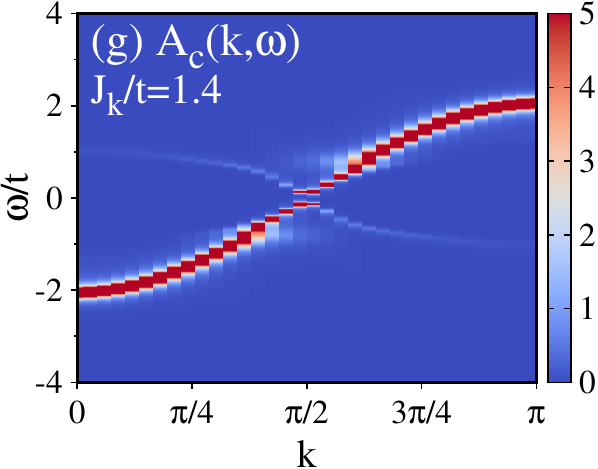}
\includegraphics[width=0.23\textwidth]{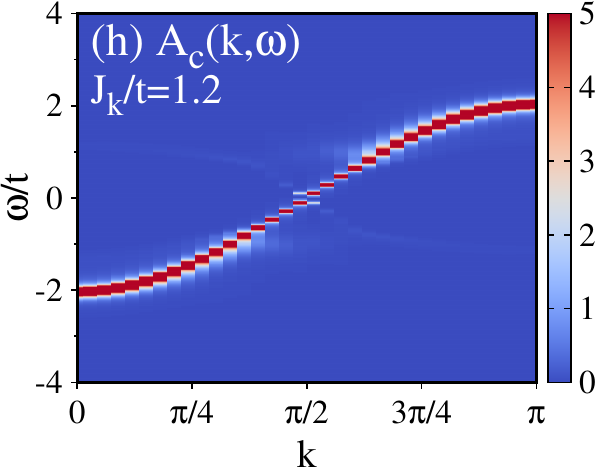}
        \caption{Evolution of (a)-(d) composite fermion $A_{\psi}(\ve{k},\omega)$ and (e)-(h) conduction electron $A_{c}(\ve{k},\omega)$
        spectral functions of the $L=66$ SU(8) Kondo-Heisenberg chain across the critical coupling $J_k^c/t=1.508(1)$.
         }
\label{Fig:AkN8kh}
\end{figure*}

\begin{figure*}[t!]
\includegraphics[width=0.23\textwidth]{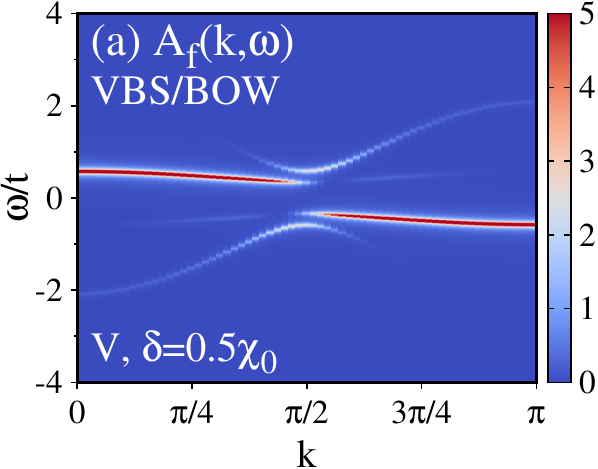}
\includegraphics[width=0.23\textwidth]{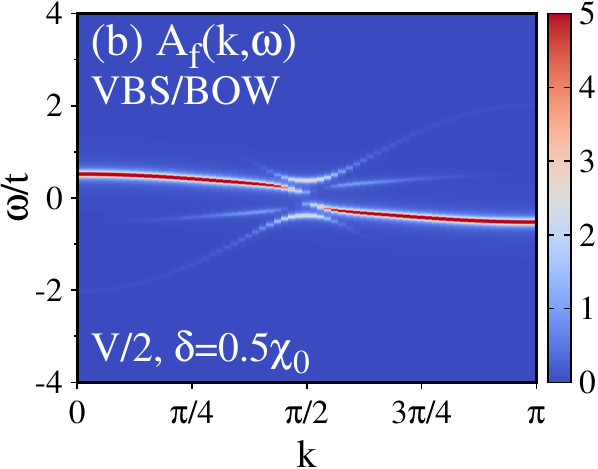}
\includegraphics[width=0.23\textwidth]{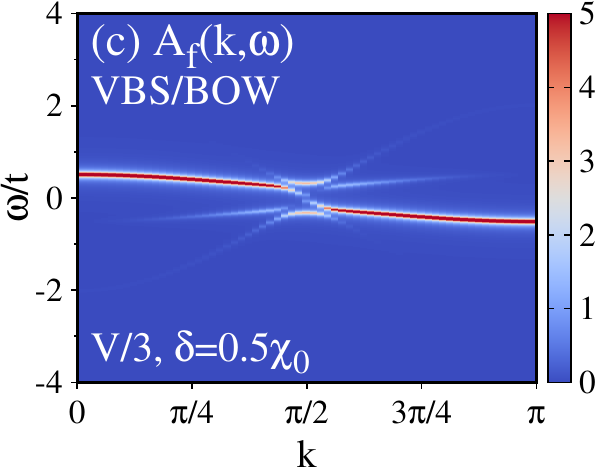}
\includegraphics[width=0.23\textwidth]{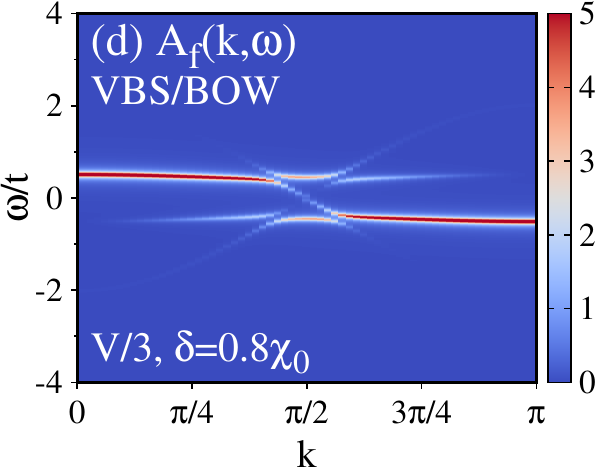}\\
\includegraphics[width=0.23\textwidth]{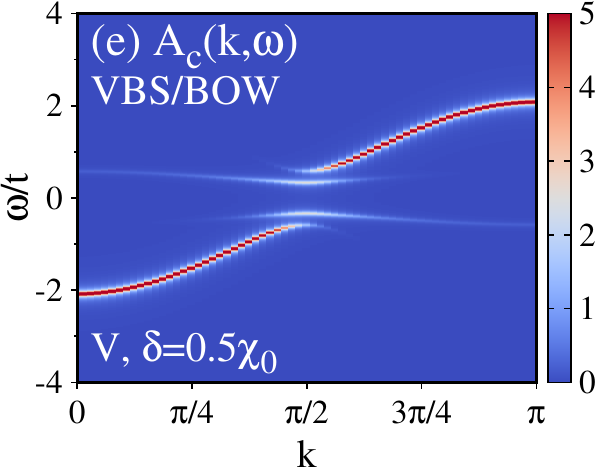}
\includegraphics[width=0.23\textwidth]{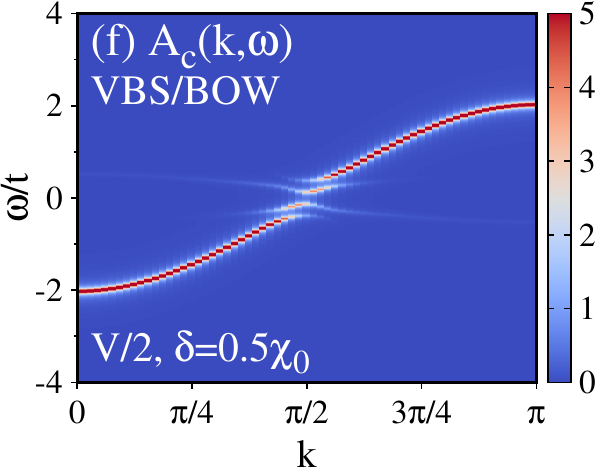}
\includegraphics[width=0.23\textwidth]{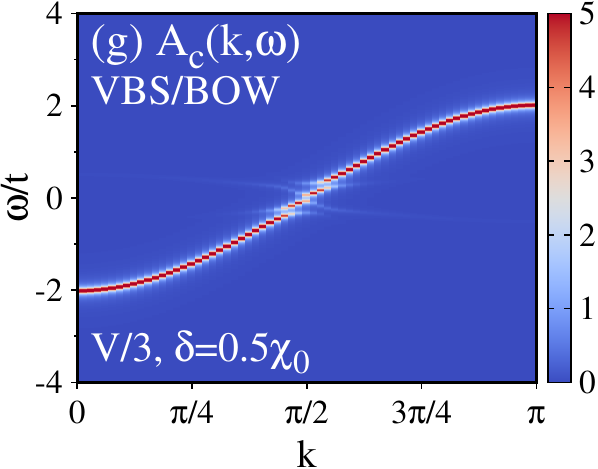}
\includegraphics[width=0.23\textwidth]{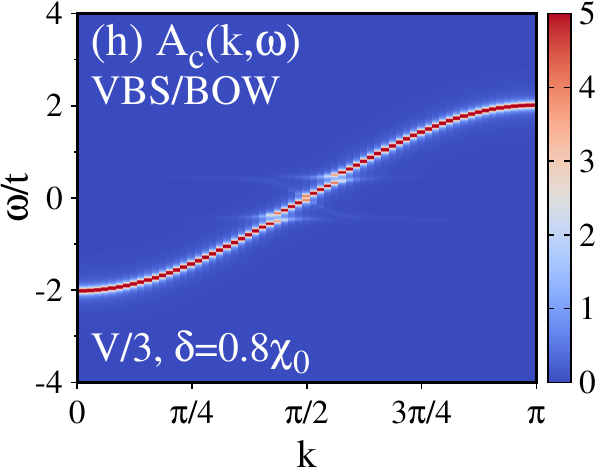}
        \caption{(a)-(d) $f$ fermion $A_{f}(\ve{k},\omega)$  and (e)-(h) conduction electron $A_{c}(\ve{k},\omega)$
        spectral functions as obtained from the large-$N$ approach accounting for both Kondo screening and spinon description
        of the Kondo-Heisenberg chain at $J_k^c/t=1.67$ and $J_h/t=1$.
	We assume here a two-site unit cell with an alternating between bonds $A$ and $B$ valence bond order parameter,
	i.e., $\chi_{A,B}=\chi_0\pm\delta$, and consider hybridization $V$ and dimerization $\delta$  amplitudes 
	as tuning parameters. In all panels $V/t=0.528$ and $\chi_0=-0.5$.
        }
\label{Fig:Ak_coex}
\end{figure*}

Unfortunately, a small critical coupling  $J_k^c/t\simeq 1$ of the SU(4) model leads to a low quasiparticle 
intensity and hampers a detailed analysis of spectral properties in the ordered phase. 
Given that an enhanced SU($N$) symmetry favors the dimerized ground state by shifting the critical coupling 
to larger values of $J_k/t$, one way around this problem is to consider a system with larger $N$. 

Thus we consider the SU(8) model for which  $J_k^c/t\simeq 1.5$. 
Here, our primary focus is on tracking the evolution of spectra in the vicinity of $J_k^c$, 
see Fig.~\ref{Fig:AkN8kh}.
Given a relatively high symmetry of the SU(8) model bringing the underlying physics closer to the large-$N$ limit, 
our evidence of the 2D Ising universality found in Sec.~\ref{sec:Ising} does not necessarily guarantee
persistent Kondo screening in the ordered phase. Indeed, VBS/BOW order alone is sufficient to gap out 
both the charge and spin degrees of freedom such that the same universality would also hold in the case of the 
full collapse of Kondo screening at the critical Kondo coupling.
Nevertheless, the observed continuous  evolution of  composite fermion quasiparticle at $\k=\pi$ 
confirms that quantum critical dimer fluctuations do not destroy Kondo screening. 
Moreover, the onset of long range VBS/BOW order enriches both the $A_{\psi}(\ve{k},\omega)$ and $A_{c}(\ve{k},\omega)$ 
spectral functions with extra features such that one can distinguish around $\k=\pi/2$  a total of four bands, 
see, e.g., Figs.~\ref{Fig:AkN8kh}(b) and ~\ref{Fig:AkN8kh}(g).  
 
To get some further insight into the physical origin of these bands, it is worthwhile to draw a comparison between the QMC data 
and large-$N$ modeling of spectral functions as accounted for by the Hamiltonian in Eq.~(\ref{LargeN_vbs.eq}). 
To this end, we focus on the critical value  $J_k^c/t=1.67$ and consider the hybridization $V$ and dimerization $\delta$  
amplitudes as tuning parameters while keeping fixed $\chi_0=-0.5$. 
We first impose a moderate dimerization $\delta=0.5\chi_0$ and weaken the Kondo screening amplitude from  
$V/t=-0.528$, through $V/2$, down to $V/3$, see Figs.~\ref{Fig:Ak_coex}(a)-\ref{Fig:Ak_coex}(c) with $A_{\psi}(\ve{k},\omega)$ 
and \ref{Fig:Ak_coex}(e)-\ref{Fig:Ak_coex}(g) with $A_{c}(\ve{k},\omega)$.
As can be seen, the combined effect of Kondo screening and broken translation symmetry gives rise to a four band 
electronic structure around $\k=\pi/2$.
The two inner weakly dispersive bands have nearly pure $f$ character while the two outer bands stem predominantly 
from the cosine band of $c$ electrons which acquired both a gap and a pronounced backfolding. 
The latter reflects a doubling of the unit cell due to a dimerized  kinetic energy of the conduction electrons 
emergent in response to the VBS pattern of $f$ spins.

Upon decreasing $V/t$, the inner bands in $A_{f}(\ve{k},\omega)$ bend towards each other while their backfolded segments 
lose the spectral weight around $\k=\pi/2$. Due to the total spectral sum rule, the weight reappears in the corresponding part of 
$A_{c}(\ve{k},\omega)$ forming the "missing" segments of the cosine band.  
Meanwhile, the lowest energy part of the outer bands in $A_{f}(\ve{k},\omega)$ flattens and gains the intensity. 
Both effects are further enhanced by a stronger dimerization $\delta=0.8\chi_0$, see  Fig.~\ref{Fig:Ak_coex}(d).
Hence, a starting point to understand the resultant spectrum $A_{f}(\ve{k},\omega)$ is a fully dimerized system 
with two flat bands shown in Fig.~\ref{Fig:Ak_kh}(d). A weak hybridization with the conduction electron band via the Kondo effect 
breaks each of the dimerized bands into outer and inner segments. 
The latter carries less spectral weight close to the $\k=\pi/2$ point and its backfolded part displays a strong modulation of 
spectral intensity so as to fulfill the total spectral sum rule. 
Likewise,  in the limit of  strong dimerization,  the cosine band in the conduction electron spectrum $A_{c}(\ve{k},\omega)$ 
is broken into four pronounced segments and is accompanied by backfolded shadow bands, see \ref{Fig:Ak_coex}(h).

By comparing Fig.~\ref{Fig:AkN8kh}(d) with Fig.~\ref{Fig:Ak_coex}(d) and Fig.~\ref{Fig:AkN8kh}(h) with Fig.~\ref{Fig:Ak_coex}(h),  
it becomes clear that the large-$N$ approach gives good overall account of the four band structure seen in both  
$A_{\psi}(\ve{k},\omega)$ and $A_{c}(\ve{k},\omega)$. 
Hence we trace the occurrence of these four bands back to the coexistence of Kondo screening and VBS/BOW order.   
Given that creating a Kondo singlet in the VBS phase  requires the breakup of a $f$ spin dimer,
it is tempting to interprete this coexistence as a partial Kondo screening along the imaginary time axis
--- at a given time $\tau$, a finite fraction of the local moments is paired into intersite  dimer singlets while 
the rest form Kondo singlets with conduction electron spins. Below we test this interpretation by looking at the evolution of spin excitation spectra.

\subsubsection{\label{sec:dynam_spin} Spin excitation spectra}

In the Lehmann representation, the dynamical spin structure factor $S( \ve{q}, \omega)$    reads
\begin{equation}\label{Eq:Lehmann}
  S ( \ve{q}, \omega)  \equiv    \pi   \sum_{ n}  | \langle  \Psi_n |  \hat{S}_{\ve{q}}  | \Psi_0 \rangle |^2
  \delta( E_n - E_0 - \omega )
\end{equation}
where $\hat{S}_{\ve{q}}=\tfrac{1}{\sqrt{L}}\sum_{\ve{i}}  e^{i \ve{q} \cdot \ve{i} } \hat{S}_{\ve{i}}$ and 
$ \hat{H} |\Psi_n \rangle = E_n |\Psi_n \rangle $.  
The SU($N$) version of the imaginary time  spin-spin correlator,
\begin{equation}\label{eq:Sq_tau}
	S(\ve{q},\tau)  =   \frac{1}{N^2-1} \sum_{\mu,\nu}  
	\langle  \Psi_0  | \hat{S}^{\mu}_{\ve{q},\nu}(\tau) \hat{S}^{\nu}_{-\ve{q},\mu} | \Psi_0 \rangle,
\end{equation}
where $\hat{S} ^{\mu}_{\ve{q},\nu}$ are Fourier transformed SU($N$) generators defined in Eq.~(\ref{generator}),  
is directly measured in QMC simulations and  related to the real frequency observable by
\begin{equation}
  S(\ve{q}, \tau) = \frac{1}{\pi}
  \int  \, {\rm d} \omega \,  e^{- \tau \omega} S(\ve{q},\omega).
\end{equation}
For the analytical continuation of the QMC data, we have used the ALF-2.0 implementation
\cite{ALF_v2} of the stochastic maximum entropy method~\cite{Sandvik98,Beach04a}.
In addition to $S(\ve{q},\omega)$ corresponding to the total spin, we have separately calculated 
the $f$ spin $S_f(\ve{q},\omega)$  and conduction electron spin $S_c(\ve{q},\omega)$ excitation spectra. 
Having both quantities at hand helps to clarify the origin of distinct features in $S(\ve{q},\omega)$.

\begin{figure}[t!]
\includegraphics[width=0.23\textwidth]{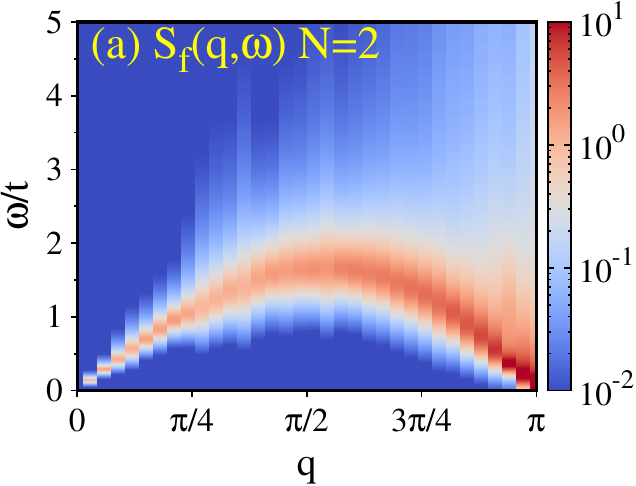}
\includegraphics[width=0.23\textwidth]{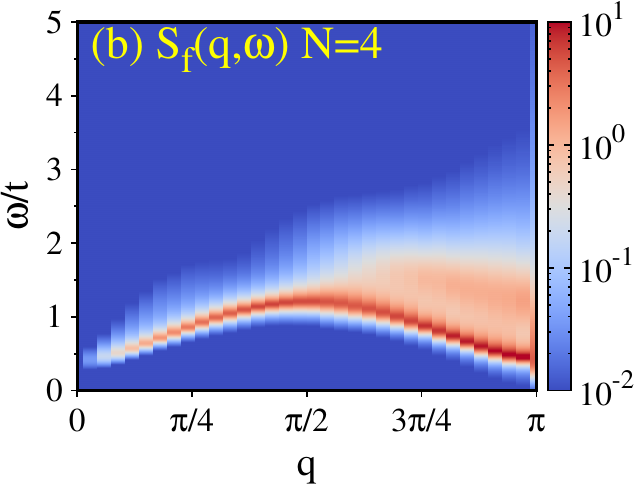}
\includegraphics[width=0.23\textwidth]{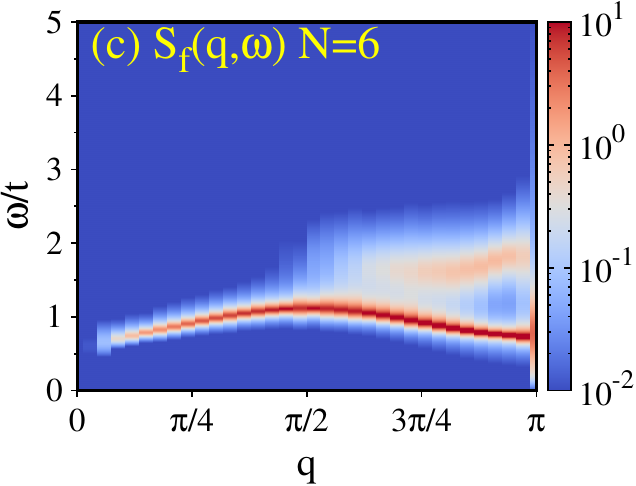}
\includegraphics[width=0.23\textwidth]{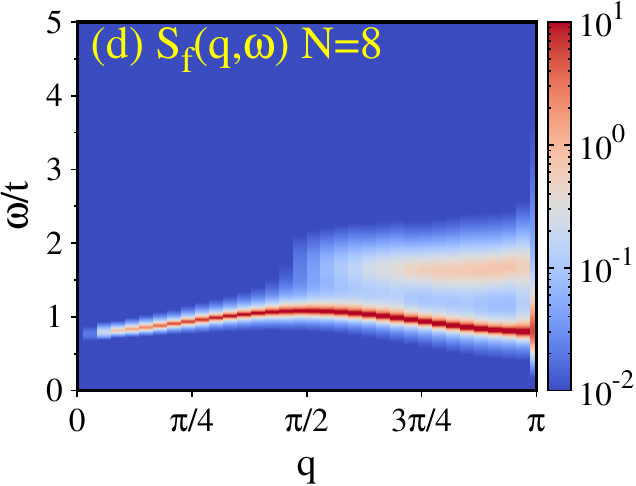}
	\caption{Dynamical spin structure factor $S_f(\ve{q},\omega)$ of the bare $L=66$ SU($N$) Heisenberg chain: 
	(a) $N=2$; (b) $N=4$; (c) $N=6$, and (d) $N=8$.
	}
\label{Fig:Sf_Heisen}
\end{figure}

The salient difference between the bare Heisenberg and Kondo chains concerns the nature of elementary spin excitations. 
In the spin-1/2 Heisenberg antiferromagnetic chain, the key elementary excitations are gapless spinons~\cite{Clo62,PhysRevLett.111.137205,Ronnow13}.
As shown in Fig.~\ref{Fig:Sf_Heisen},  a spontaneous  dimerization present in the  SU$(N\ge 4)$ version of the  Heisenberg chain  
gaps out the spin degrees of freedom but it clearly leaves spinons as the fundamental excitations above the spin gap, 
see also Ref.~\cite{PhysRevB.25.4925}.

In contrast,  the primary excitations of a bare Kondo chain   correspond to gapped triplons~\cite{Ueda97}. 
The latter have a composite character involving both the $f$ moments and conduction electron spins.  
Thus, the relative importance of Heisenberg versus Kondo interactions should be 
reflected in the evolution of spin excitation spectrum upon tuning the value of $J_k/t$. 

\begin{figure*}[t!]
\includegraphics[width=0.23\textwidth]{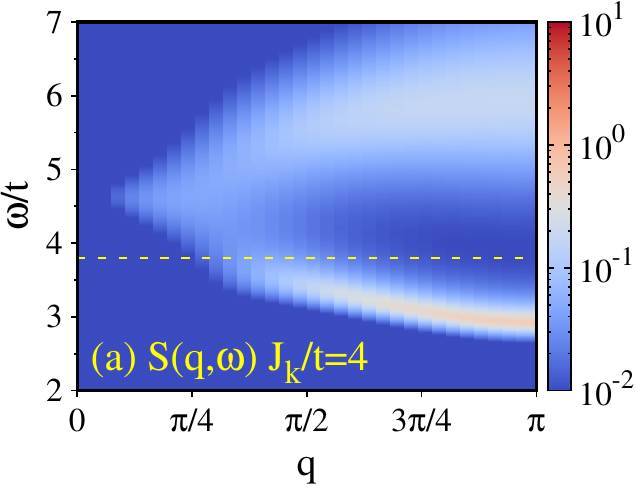}
\includegraphics[width=0.23\textwidth]{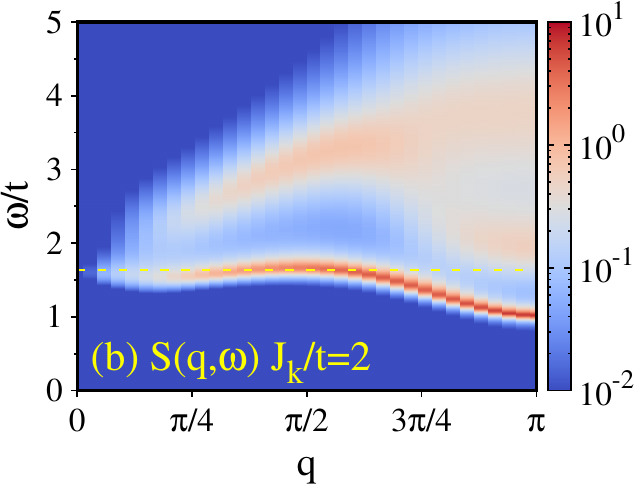}
\includegraphics[width=0.23\textwidth]{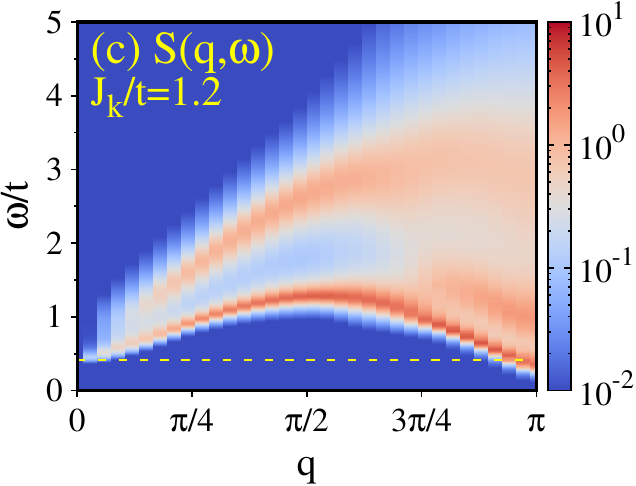}
\includegraphics[width=0.23\textwidth]{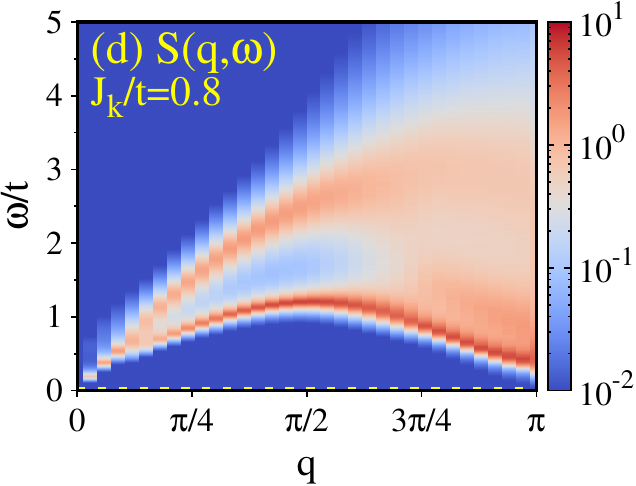}\\
\includegraphics[width=0.23\textwidth]{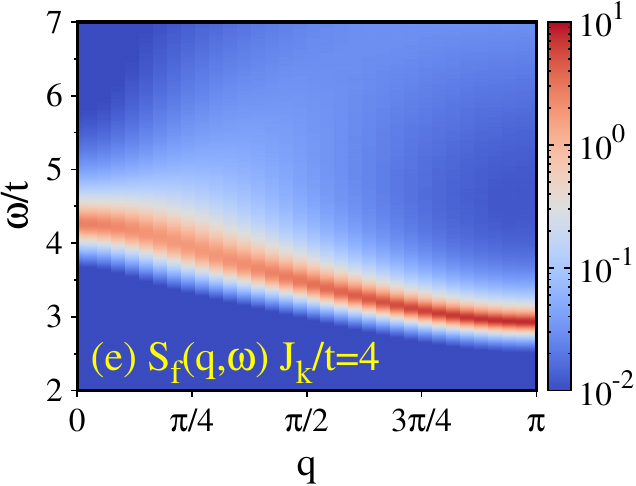}
\includegraphics[width=0.23\textwidth]{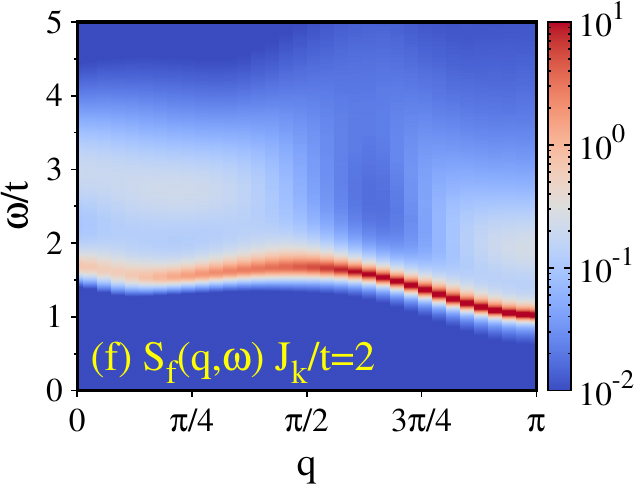}
\includegraphics[width=0.23\textwidth]{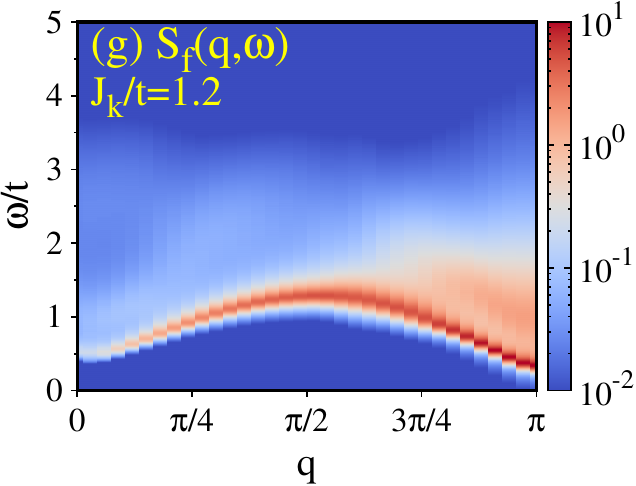}
\includegraphics[width=0.23\textwidth]{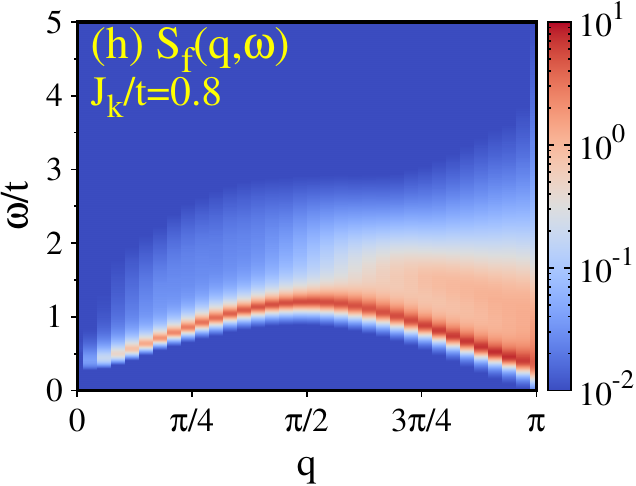}\\
\includegraphics[width=0.23\textwidth]{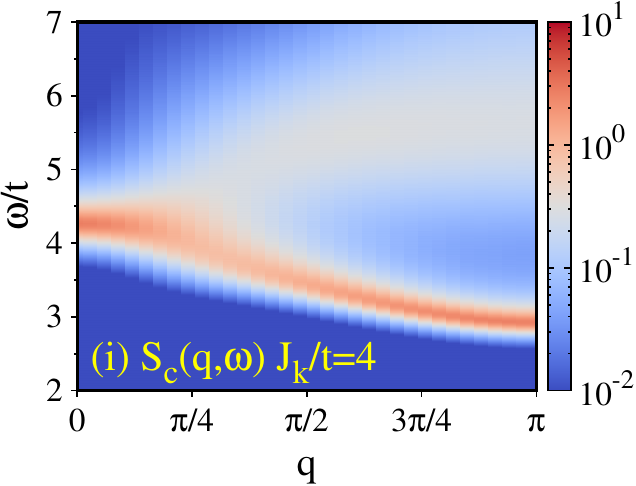}
\includegraphics[width=0.23\textwidth]{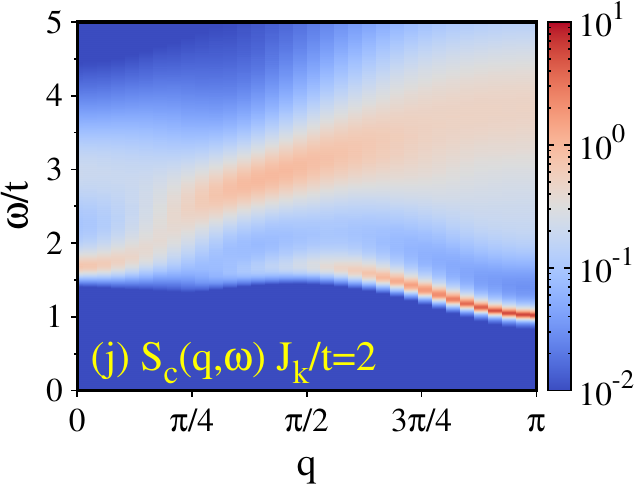}
\includegraphics[width=0.23\textwidth]{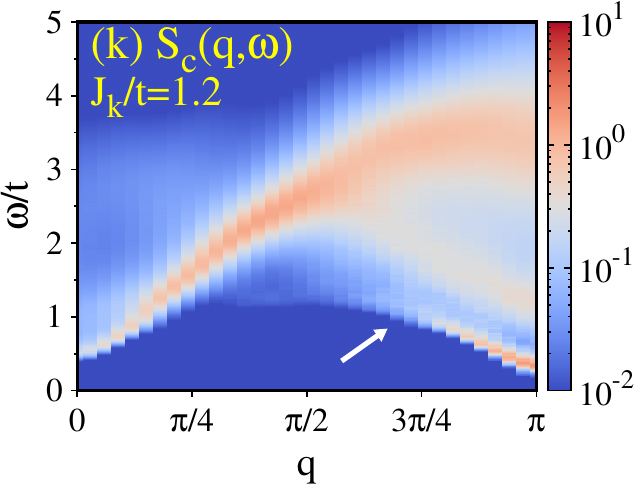}
\includegraphics[width=0.23\textwidth]{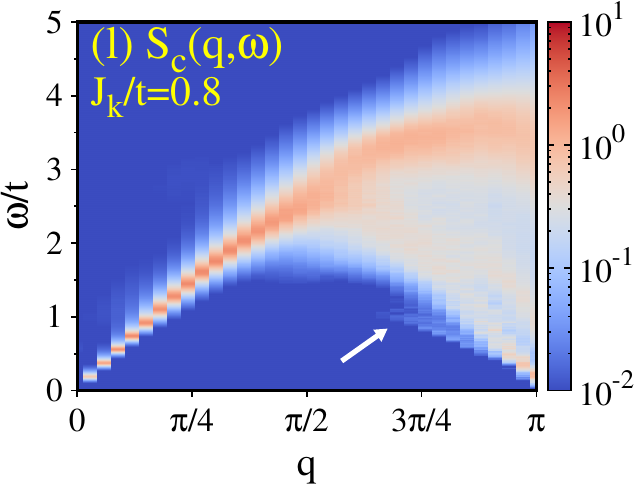}
	\caption{Dynamical spin structure factor of the $L=66$ SU(4) Kondo-Heisenberg chain for: 
	(a)-(d) total spin  $S(\ve{q},\omega)$; (e)-(h) $f$ spin $S_f(\ve{q},\omega)$, and (i)-(l) 
	conduction electron spin $S_c(\ve{q},\omega)$ for values of $J_k/t$ as in Fig.~\ref{Fig:AkN4kh}. 
	Dashed lines indicate the particle-hole continuum threshold; the arrows help to track the image of a remnant 
	triplon mode.}
\label{Fig:SqN4kh}
\end{figure*}

\begin{figure*}[t!]
\includegraphics[width=0.23\textwidth]{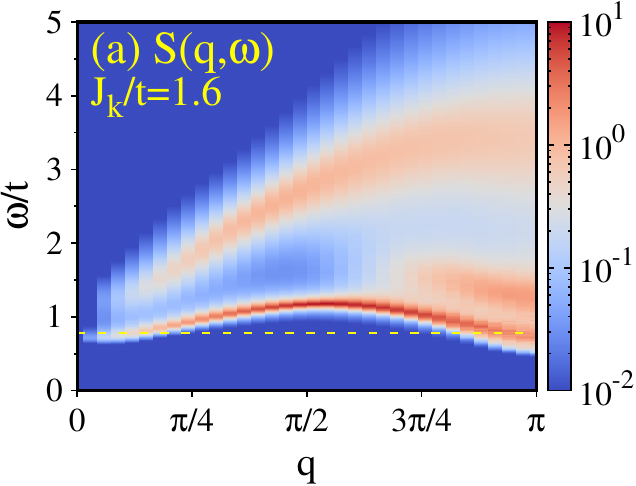}
\includegraphics[width=0.23\textwidth]{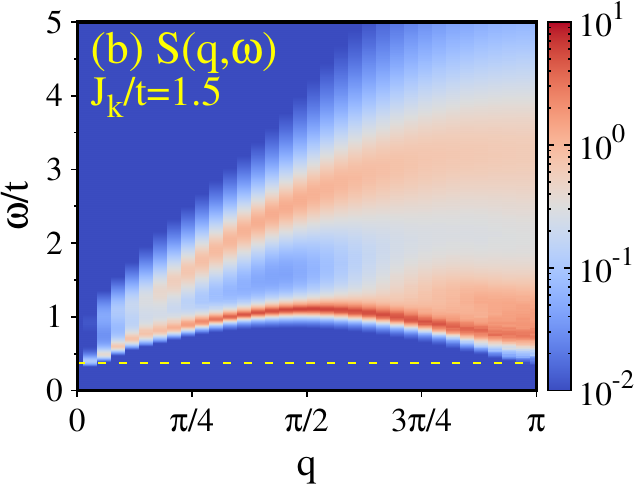}
\includegraphics[width=0.23\textwidth]{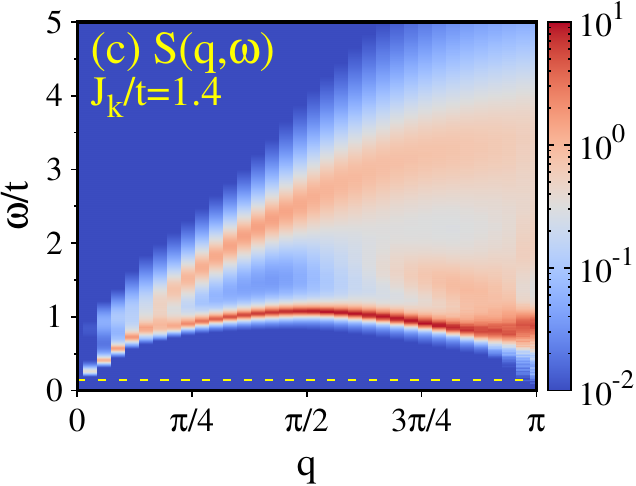}
\includegraphics[width=0.23\textwidth]{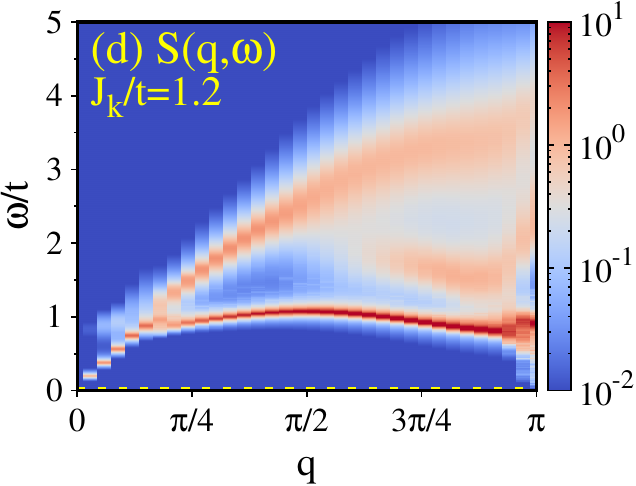}\\
\includegraphics[width=0.23\textwidth]{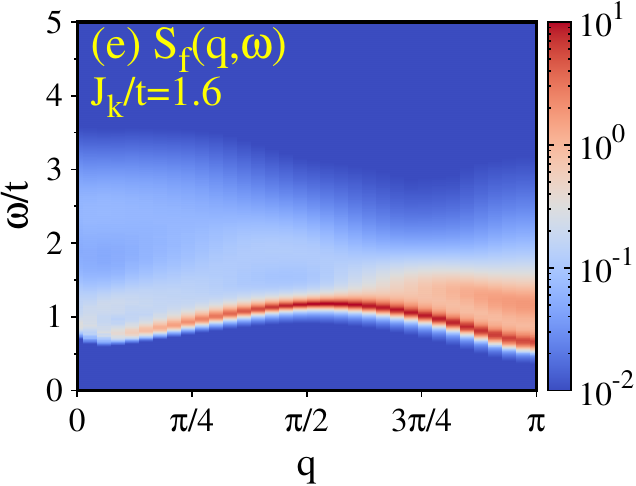}
\includegraphics[width=0.23\textwidth]{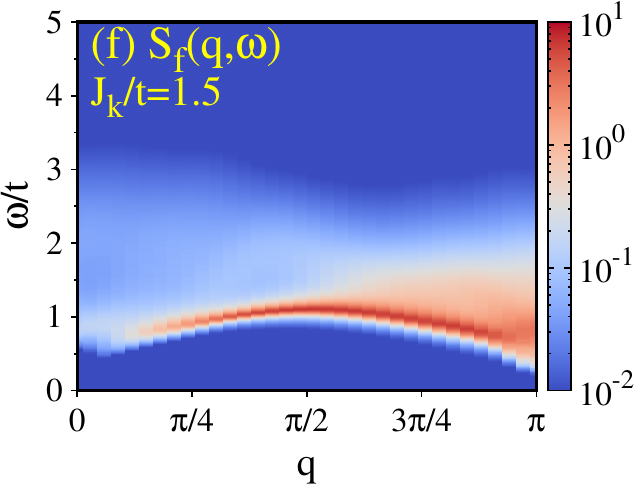}
\includegraphics[width=0.23\textwidth]{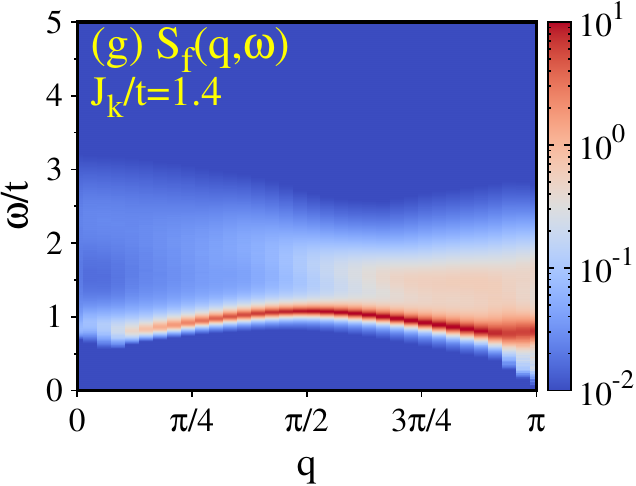}
\includegraphics[width=0.23\textwidth]{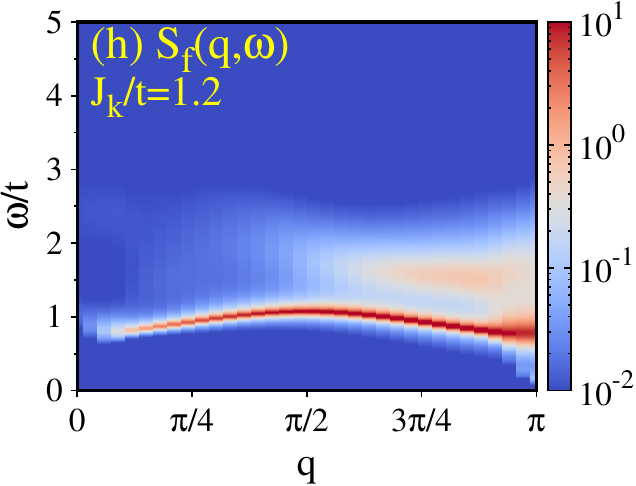}\\
\includegraphics[width=0.23\textwidth]{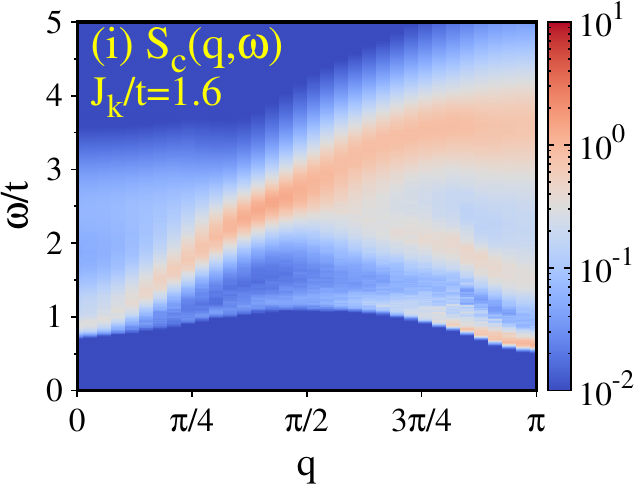}
\includegraphics[width=0.23\textwidth]{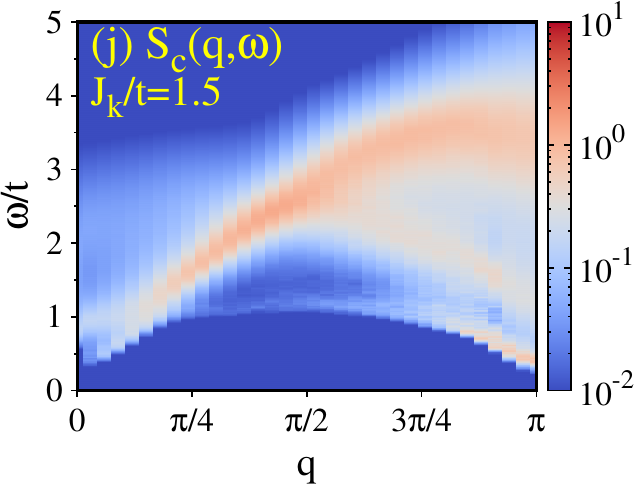}
\includegraphics[width=0.23\textwidth]{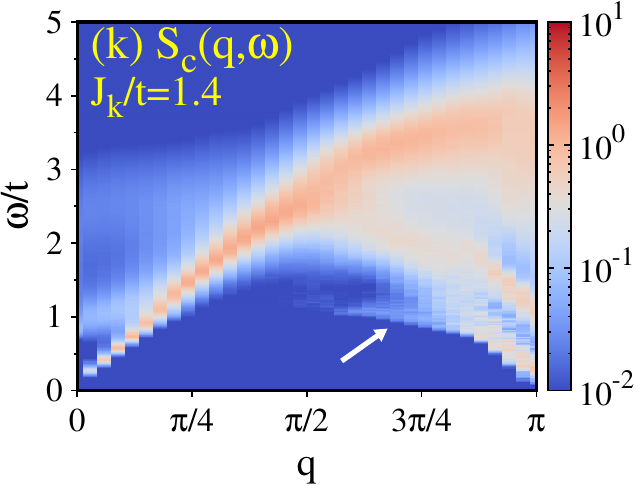}
\includegraphics[width=0.23\textwidth]{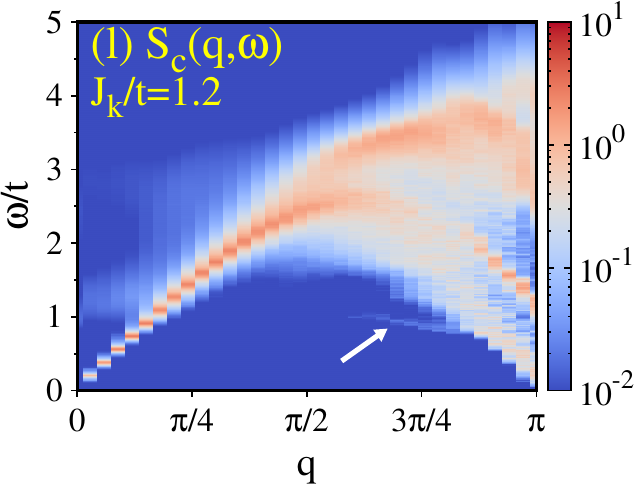}
	\caption{Dynamical spin structure factor of the $L=66$ SU(8) Kondo-Heisenberg chain for:
        (a)-(d) total spin  $S(\ve{q},\omega)$; (e)-(h) $f$ spin $S_f(\ve{q},\omega)$, and (i)-(l)
        conduction electron spin $S_c(\ve{q},\omega)$ for values of $J_k/t$ as in Fig.~\ref{Fig:AkN8kh}. 
	Dashed lines indicate the particle-hole continuum threshold; 
	the arrows help to track the image of a remnant triplon mode.
        }
\label{Fig:SqN8kh}
\end{figure*}

At large $J_k/t=4$, see Fig.~\ref{Fig:SqN4kh}(a), the dynamical (total) spin structure factor 
$S(\ve{q},\omega)$ is consistent with that of a pure Kondo chain and exhibits a low energy triplon 
mode corresponding to bound electron-hole pairs. 
The bound triplon shows a minimal gap at the $\q=\pi$ wavevector and becomes washed out upon merging 
in the particle-hole continuum whose energy threshold [dashed line in Figs.~\ref{Fig:SqN4kh}(a)-\ref{Fig:SqN4kh}(d)] 
is given by the two particle charge gap $2\Delta_{qp}$. 
Figures~\ref{Fig:SqN4kh}(e) and ~\ref{Fig:SqN4kh}(i) plot the corresponding $f$ spin $S_f(\ve{q},\omega)$   
and conduction electron spin $S_c(\ve{q},\omega)$ excitation spectra. As expected for the Kondo screened
phase~\cite{PhysRevB.57.R12659,PhysRevLett.81.4939,doi:10.1143/JPSJ.68.3138}, both quantities display 
the same support but carry different spectral weights. 

Reducing $J_k/t$  modifies the spectrum so as to reflect the emerging  Heisenberg spin physics.
It brings about (i) softening of the triplon mode in the long-wavelength limit $\q=0$ such that it gradually 
acquires a concave dispersion and (ii) a buildup of incoherent spectral weight around $\q=\pi$, 
just above the triplon mode, see Fig.~\ref{Fig:SqN4kh}(c). 
The latter is also seen in the $f$ spin spectrum $S_f(\ve{q},\omega)$ in Fig.~\ref{Fig:SqN4kh}(g). 
The enhanced spectral intensity concentrated close to the lower boundary of the spectrum is a good indicator of fractional 
spinon excitations~\cite{PhysRevB.71.020405,Raczkowski13,PhysRevB.94.205145}. Thus it is tempting to identify 
it as the two-spinon continuum. 
As shown in Fig.~\ref{Fig:SqN4kh}(k), the lowest edge  of the conduction electron spin spectrum 
$S_c(\ve{q},\omega)$ continues to display, albeit with strongly reduced weight around $\q=\pi/2$, 
the same support as the $f$ spin spectrum $S_f(\ve{q},\omega)$. 
On the other hand, a large majority of spectral weight in $S_c(\ve{q},\omega)$ above the charge gap 
is located on the upper boundary of the spectrum as expected for the particle-hole continuum of a nearly 
free conduction electron gas.

\begin{figure*}[t!]
\includegraphics[width=0.32\textwidth]{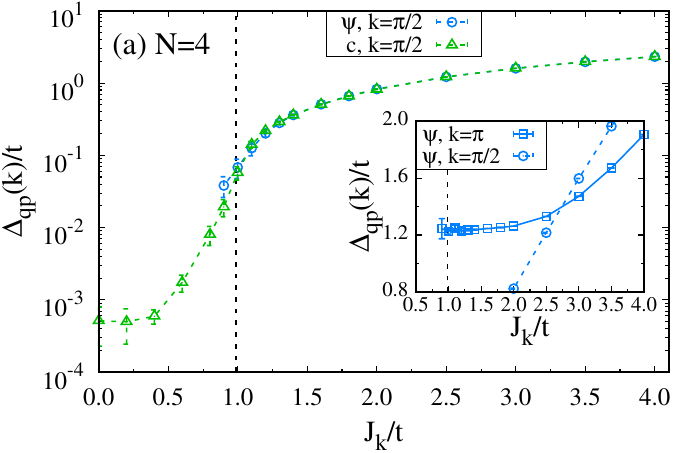}
\includegraphics[width=0.32\textwidth]{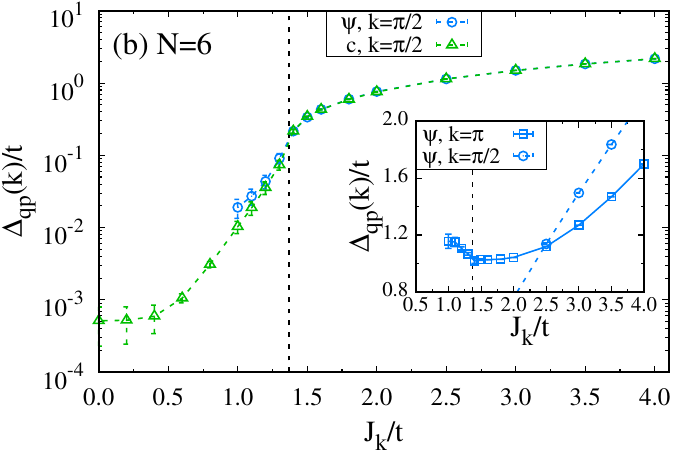}
\includegraphics[width=0.32\textwidth]{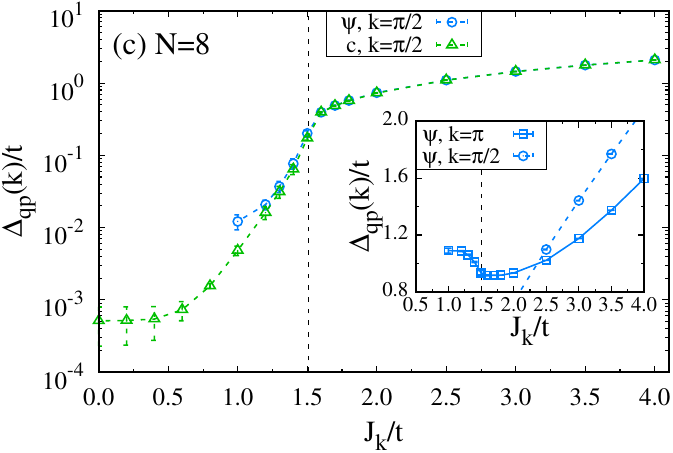} \\
\includegraphics[width=0.32\textwidth]{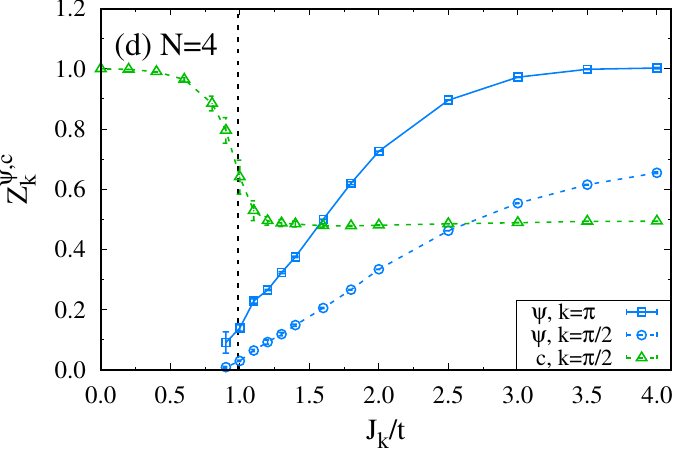}
\includegraphics[width=0.32\textwidth]{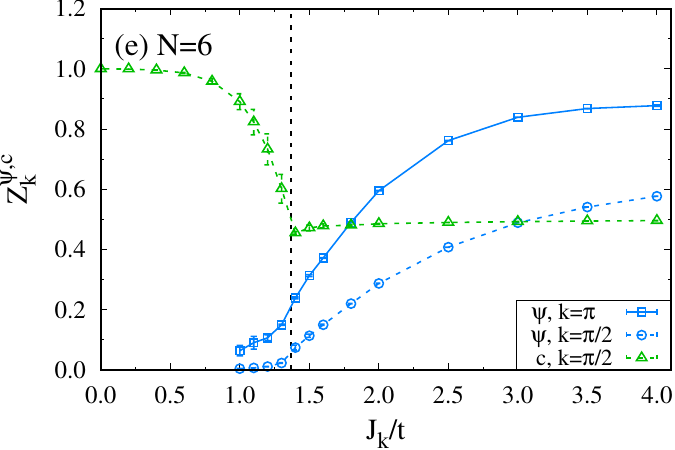}
\includegraphics[width=0.32\textwidth]{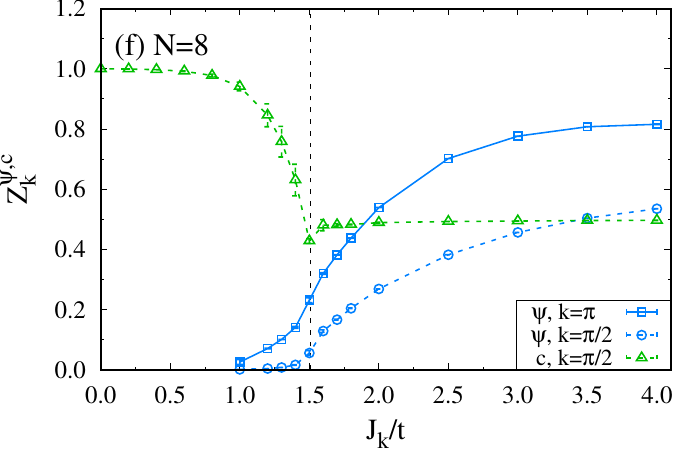} \\
\includegraphics[width=0.32\textwidth]{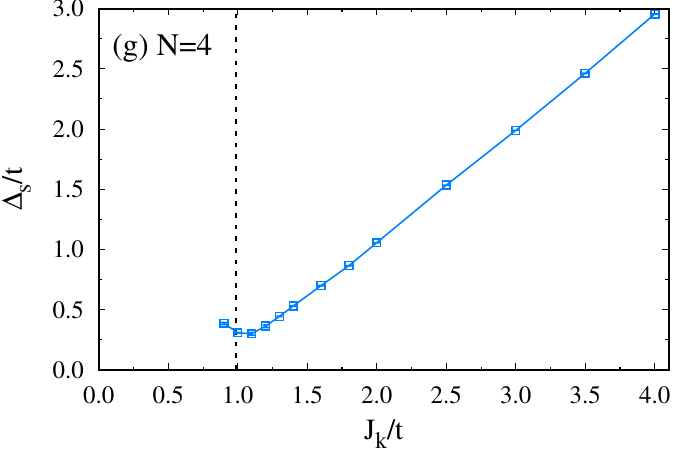} 
\includegraphics[width=0.32\textwidth]{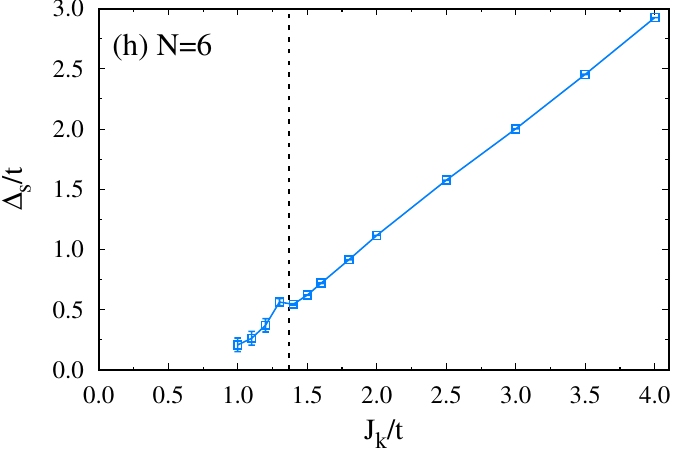} 
\includegraphics[width=0.32\textwidth]{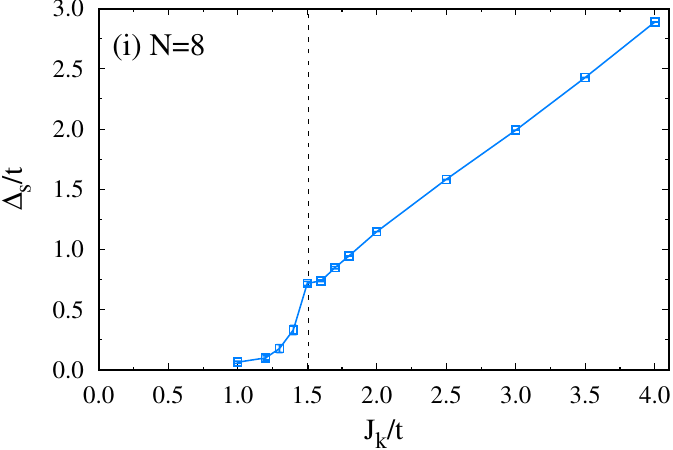} 
	\caption{ (a)-(c) Single particle gap $\Delta_{qp}(\ve{k})$  at $\ve{k}=\pi/2$ on a semilog scale 
	and (d)-(f) quasiparticle residue $Z^{\psi}_{\ve{k}}$  at $\ve{k}=\pi/2$ and $\ve{k}=\pi$  
	of the pole in the composite fermion Green's function  as a function of $J_k/t$ in the SU($N$) Kondo-Heisenberg 
	chain with $J_h/t=1$. Insets show a crossing between the quasiparticle gaps at $\ve{k}=\pi/2$ and $\ve{k}=\pi$.   
	To show an exponentially small gap at $\ve{k}=\pi/2$ in the BOW phase of the $c$ electron layer, 
	we also plot $\Delta_{qp}(\ve{k})$ together with the corresponding residue $Z^{c}_{\ve{k}}$  extracted  
	from the conduction electron Green's function. Panels (g)-(i) show the evolution of the spin gap 
	$\Delta_{s}(\ve{q})$ at $\ve{q}=\pi$ from the imaginary time  (total) spin correlation function.
	All the quantities are representative of the thermodynamic limit.
	Dashed lines indicate the respective critical Kondo couplings.}
\label{Fig:gaps_KHLM}
\end{figure*}

Figure~\ref{Fig:SqN4kh}(d) plots $S(\ve{q},\omega)$ at $J_k/t=0.8$ in the ordered phase.  
In the long-wavelength limit $\q=0$, $S(\ve{q},\omega)$ develops a nearly gapless linear mode. 
The latter is also resolved in the conduction electron spin spectrum  $S_c(\ve{q},\omega)$, 
see Fig.~\ref{Fig:SqN4kh}(l), being a consequence of the exponentially small single particle gap. 
The softening of conduction electron spin excitations is also seen at $\q=\pi$. 
Nevertheless, a fingerprint of electron-hole binding is still noticeable in $S_c(\ve{q},\omega)$ 
as a faint low energy image of the $f$ spin excitations emerging below the particle-hole continuum 
sufficiently away from $\q=\pi$, see the arrow in Fig.~\ref{Fig:SqN4kh}(l). 
These common spectral aspects shared by $S_f(\ve{q},\omega)$ and $S_c(\ve{q},\omega)$ signify 
a remnant Kondo screening~\cite{PhysRevB.57.R12659,PhysRevLett.81.4939,doi:10.1143/JPSJ.68.3138}. 
Meanwhile, a striking resemblance of the $f$ spin spectrum $S_f(\ve{q},\omega)$, see Fig.~\ref{Fig:SqN4kh}(h), 
to that of the corresponding isolated Heisenberg chain in Fig.~\ref{Fig:Sf_Heisen}(b) is very suggestive 
of the two-spinon excitations.

The coexistence of distinct spectral features brought about by Kondo and Heisenberg interactions across the phase 
transition is more tractable in the spin dynamics of the SU(8) case shown in Fig.~\ref{Fig:SqN8kh}.
In addition to the two-spinon continuum, other prominent features include a nearly gapless linear 
mode at $\q=0$ occurring together with a remnant triplon excitation at slightly higher frequency, see Figs.~\ref{Fig:SqN8kh}(c) 
and \ref{Fig:SqN8kh}(d). A faint image of the latter is also resolved in $S_c(\ve{q},\omega)$ both in the 
$\q\to 0$ and $\q\to \pi$ limits, see Figs.~\ref{Fig:SqN8kh}(k) and \ref{Fig:SqN8kh}(l).

\subsubsection{\label{sec:gaps} Spin and charge energy scales}

Let us summarize this section with a compilation of the $J_k/t$ dependence of the relevant energy scales 
in Fig.~\ref{Fig:gaps_KHLM}.  All the quantities have been extrapolated to the thermodynamic limit based on QMC results 
on chains up to $L=130$ sites.

We begin with the single particle gap at $\k=\pi/2$ shown in Figs.~\ref{Fig:gaps_KHLM}(a)-\ref{Fig:gaps_KHLM}(c).  
This quantity can be extracted  by fitting the tail of the imaginary time Green’s function to the exponential form     
$G_{}(\ve{k},\tau) \stackrel{\tau \to \infty}{\to}  Z^{}_{\ve{k}}e^{-\Delta_{qp}(\ve{k}) \tau }$, 
where $Z^{}_{\ve{k}}$ is the quasiparticle residue of the doped hole at momentum $\ve{k}$ and frequency 
$\omega=-\Delta_{qp}$. To assess the quality of the data we consider both the composite fermion  $G_{\psi}(\ve{k},\tau)$ 
and conduction electron $G_{c}(\ve{k},\tau)$ Green’s functions. In the presence of Kondo screening they shall display 
the same asymptotic behavior, and indeed, a good data match is obtained (note the semilog scale) in the vicinity of 
$J_k^c$.  However,  a growing discrepancy occurs  upon going deeper into the ordered phase where  the low intensity 
of the composite quasiparticle [circles in Fig.~\ref{Fig:gaps_KHLM}(d)-\ref{Fig:gaps_KHLM}(f)] makes the analysis 
more difficult. Therefore, we are left with the conduction electron data which are consistent with an exponentially 
small gap. Thus accordingly larger systems sizes are indispensable to reliably track the onset of the gap at smaller 
values of $J_k/t$.


\begin{figure*}[t!]
\includegraphics[width=0.23\textwidth]{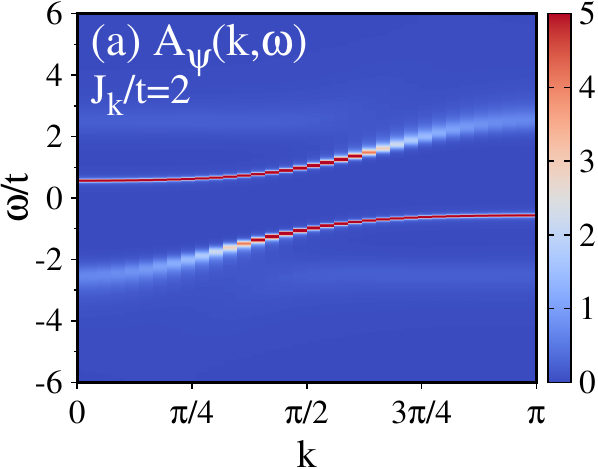}
\includegraphics[width=0.23\textwidth]{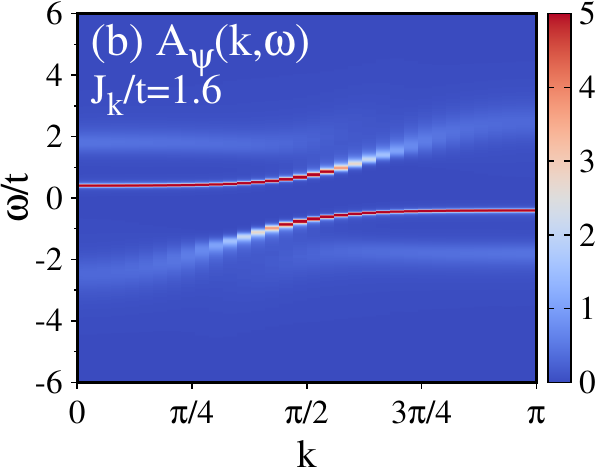}
\includegraphics[width=0.23\textwidth]{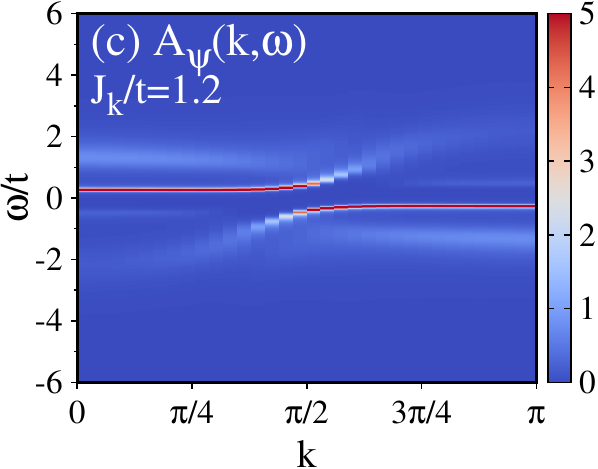}
\includegraphics[width=0.23\textwidth]{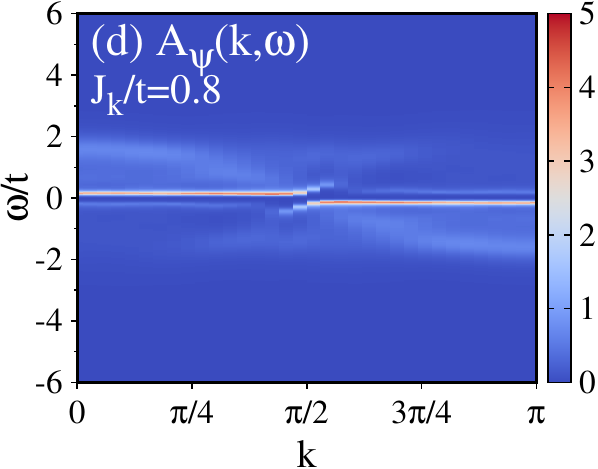}\\
\includegraphics[width=0.23\textwidth]{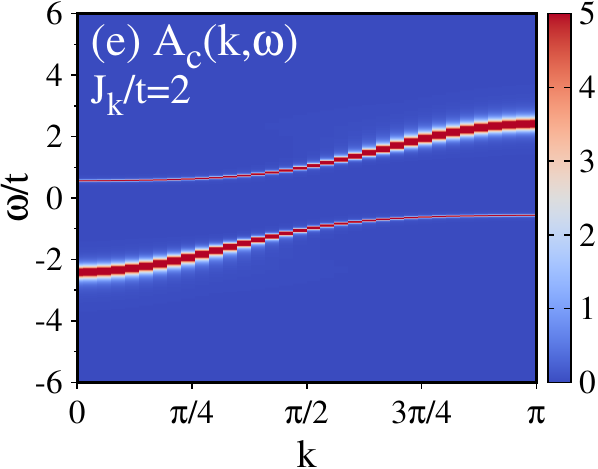}
\includegraphics[width=0.23\textwidth]{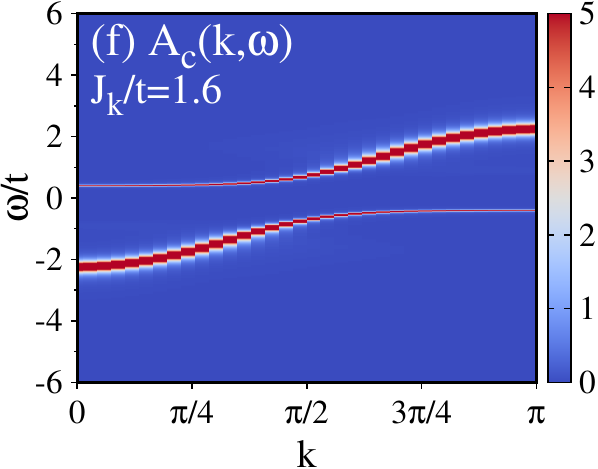}
\includegraphics[width=0.23\textwidth]{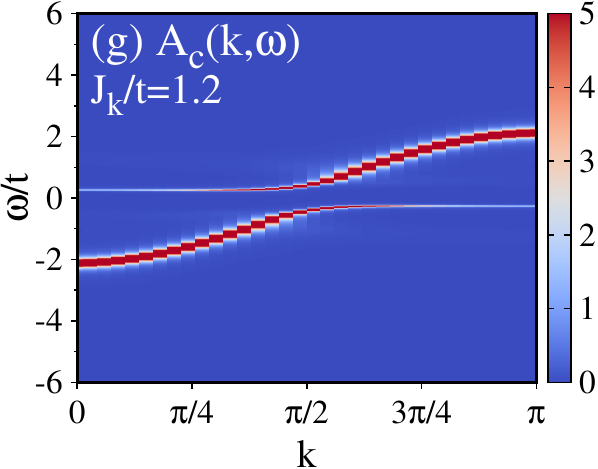}
\includegraphics[width=0.23\textwidth]{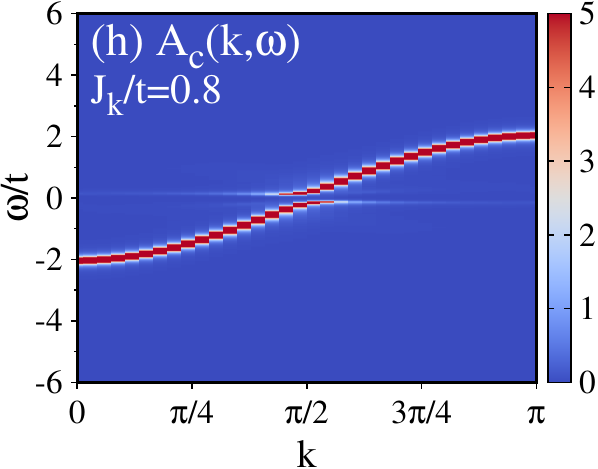}
	\caption{(a)-(d) Composite fermion $A_{\psi}(\ve{k},\omega)$  and (e)-(h) conduction electron $A_{c}(\ve{k},\omega)$  
	spectral functions of the bare ($J_h=0$) $L=66$ SU(4) Kondo chain for selected values of $J_k/t$. 
	The ground state in all panels corresponds to a uniform Kondo insulator.}
\label{Fig:AkN4}
\end{figure*}

\begin{figure}[t!]
\includegraphics[width=0.23\textwidth]{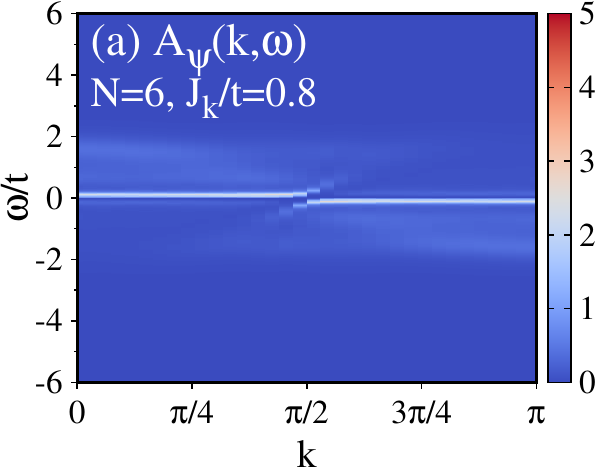}
\includegraphics[width=0.23\textwidth]{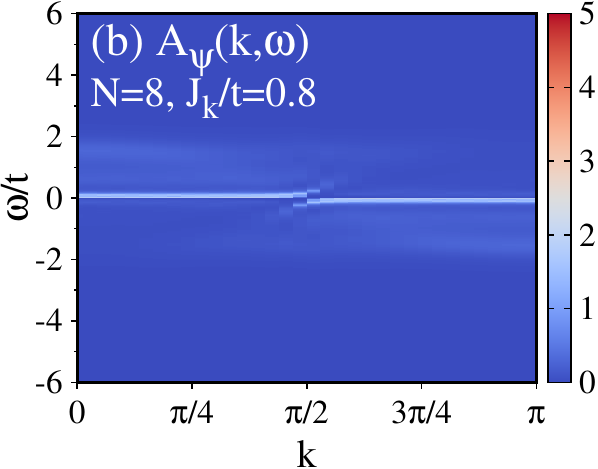}
        \caption{Composite fermion $A_{\psi}(\ve{k},\omega)$  spectral function of the $L=66$ 
	(a) SU(6) and (b) SU(8) Kondo chain with $J_k/t=0.8$. 
        }
\label{Fig:AkN68}
\end{figure}

To locate the change in the position of the minimal quasiparticle gap as a function of $J_k/t$,  
we have equally extracted the single particle gap at $\k=\pi$. As shown in the insets of 
Figs.~\ref{Fig:gaps_KHLM}(a)-\ref{Fig:gaps_KHLM}(c), the gap at $\k=\pi/2$ grows faster and eventually  
surpasses the one at $\k=\pi$ bringing  the electronic structure in line with that of a pure Kondo chain. 
We also note the hardening of the gap at $\k=\pi$ on the ordered side of $J_k^c$. We attribute it to the 
combined effect of dimerization and the persistence of Kondo screening. The latter gives rise to a smooth evolution 
of composite fermion residue $Z^{\psi}_{\ve{k}=\pi}$  [squares in Figs.~\ref{Fig:gaps_KHLM}(d)-\ref{Fig:gaps_KHLM}(f)] 
across the phase transition for each value of $N$.
In contrast, the evolution of conduction electron residue $Z^{c}_{\ve{k}=\pi/2}$ 
[triangles in Figs.~\ref{Fig:gaps_KHLM}(d)-\ref{Fig:gaps_KHLM}(f)] across $J_k^c$ is highly nontrivial. 
On the one hand, it quickly saturates for $J_k > J_k^c$ at 1/2 implying the equal distribution of spectral weight between the 
two bands of the Kondo insulator. In the  $J_k=0$ limit, it reaches the value of unity as expected for the free electron gas.  
On the other hand, for $N>4$ the data quality allows us to resolve a small drop in $Z^{c}_{\ve{k}=\pi/2}$ 
in the vicinity of $J_k^c$ on the ordered side. It is consistent with the redistribution of spectral weight 
so as to reflect the emergent four band structure, cf. Fig.~\ref{Fig:AkN8kh}(g). This effect is quickly masked 
by the overall growth of the $c$ electron weight driven by the weakening of Kondo screening.

Finally,  Figs.~\ref{Fig:gaps_KHLM}(g)-\ref{Fig:gaps_KHLM}(i) plot the evolution of spin gap 
$\Delta_{s}(\ve{q}=\pi)$ obtained from the asymptotic behavior for large $\tau t\gg 1$ of the  
imaginary time spin-spin correlation function $S_{}(\ve{q},\tau) \propto  e^{-\Delta_{s}(\ve{q}) \tau }$.
As is evident, the spin gap in the disordered phase  scales as $J_k$ including the $N=4$ case with a relatively 
small $J_k^c/t\simeq 1$. This linear behavior indicates that the assumed Heisenberg coupling $J_h/t=1$ automatically 
places the system in the strong coupling limit $J_k/t\gg 1$ of the bare Kondo chain where $\Delta_s\propto J_k$~\cite{Ueda97}. 
Notably, the critical coupling $J_k^c$  seems to be marked
by a weak dip. The latter is superseded by a rapid softening of $\Delta_s$ in the ordered phase. 
These findings are in accord with the observed smearing of the triplon mode in the SU(8) spin excitation spectrum, 
see Figs.~\ref{Fig:SqN8kh}(b) versus \ref{Fig:SqN8kh}(c).


\subsection{\label{sec:KLM} SU($\pmb{N}$) Kondo chain}

Having explored the physics of the SU($N$) Kondo-Heisenberg model, we turn now to a pure ($J_h=0$) SU($N$) Kondo chain. 
Given that already  the SU(4) spin rotational symmetry of the Heisenberg chain is sufficient to stabilize 
a dimerized ground state~\cite{Paramekanti07,PhysRevB.100.085103},  
the key question to address is whether the bare SU($N$) Kondo chain spontaneously dimerizes too, i.e.,  
can the low dimensionality of the RKKY interaction alone produce long range dimer order?  
 
On the one hand, in Sec.~\ref{sec:KHLM} we found that the critical magnetic interaction required to trigger 
dimer order in the SU(4) Kondo-Heisenberg chain seems to scale linearly with $J_k$. Since the RKKY interaction 
scales for small Kondo couplings  as $\frac{J_k^2}{N}$~\cite{Raczkowski20}, one may anticipate the absence of dimer order 
in the  SU(4) Kondo chain. 
On the other hand, a higher symmetry of fermion flavors shifts the critical Kondo coupling $J_k^c$ to larger values and 
increases the domain of stability of the VBS phase, see the phase diagram in Fig.~\ref{Fig:PD}. 
Thus, the question of a spontaneous dimerization  still remains an open issue in a generic SU($N$) case.

We begin our study with the evolution of composite fermion $A_{\psi}(\ve{k},\omega)$ and 
conduction electron $A_{c}(\ve{k},\omega)$ spectral functions of the SU(4) Kondo chain shown in Fig.~\ref{Fig:AkN4}.
While the main spectral features are quantitatively reproduced  within a simple large-$N$ approximation 
accounting for Kondo screening in the translationally invariant Kondo chain, see Fig.~\ref{Fig:Ak_klm}, 
one observes that extra features in $A_{\psi}(\ve{k},\omega)$, i.e., shadow bands become more pronounced 
at small values of $J_k/t$.
As argued in Ref.~\cite{Danu21}, they reflect antiferromagnetic spin fluctuations driven by the RKKY interaction, and they 
correspond to the image of the $c$ electron spectral function shifted by the antiferromagnetic wavevector $\ve{Q}=\pi$.
Since the RKKY energy scale is suppressed as $1/N$, the shadow bands shall lose their intensity with increasing $N$ bringing 
the overall spectrum, in the absence of enhanced spin dimer correlations, in line with the large-$N$ outcome in 
Fig.~\ref{Fig:Ak_klm}. We confirm this point explicitly in Fig.~\ref{Fig:AkN68} by plotting $A_\psi(\k,\omega)$ 
at our smallest value $J_k/t=0.8$ for the $N=6$ and $N=8$ case. The apparent similarity between the QMC data and the 
corresponding large-$N$ result in Fig.~\ref{Fig:Ak_klm}(d) suggests the absence of significant dimer-dimer correlations.

\begin{figure*}[t!]
\includegraphics[width=0.23\textwidth]{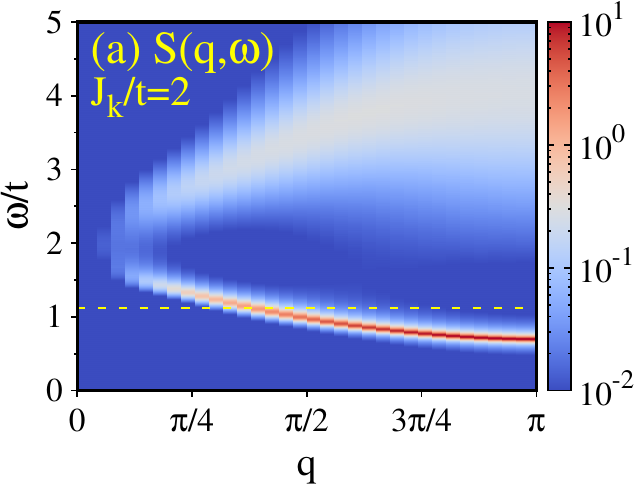}
\includegraphics[width=0.23\textwidth]{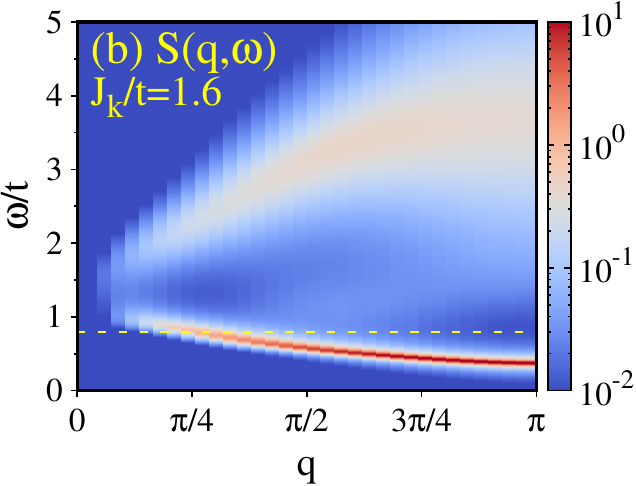}
\includegraphics[width=0.23\textwidth]{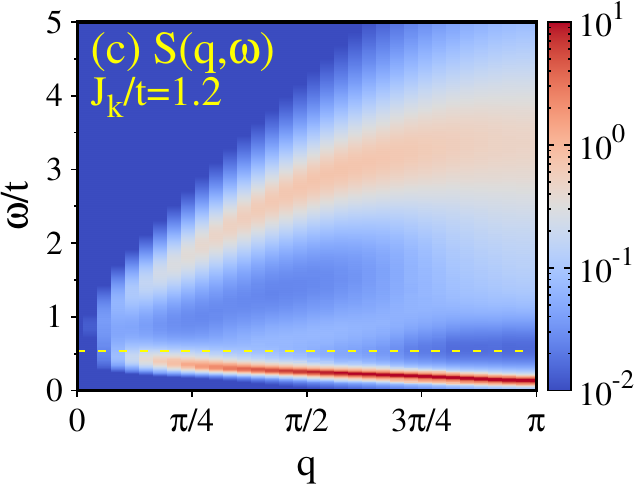}
\includegraphics[width=0.23\textwidth]{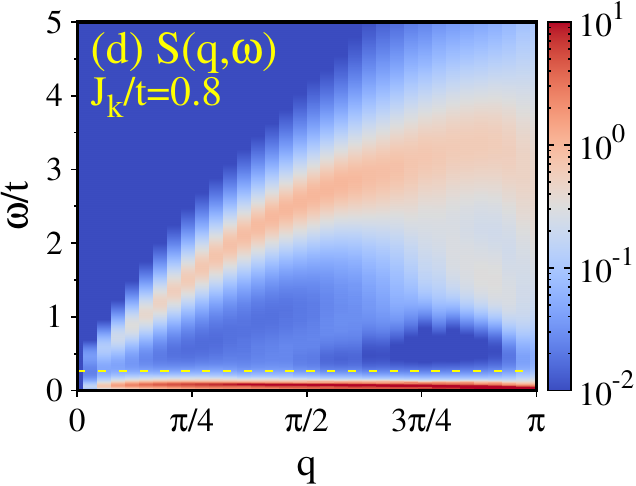}\\
\includegraphics[width=0.23\textwidth]{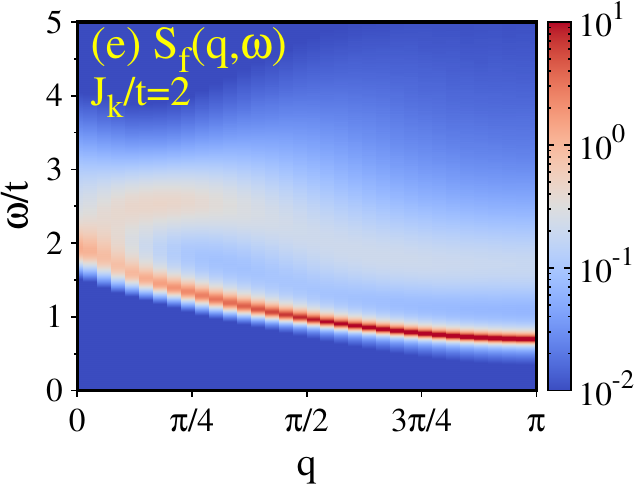}
\includegraphics[width=0.23\textwidth]{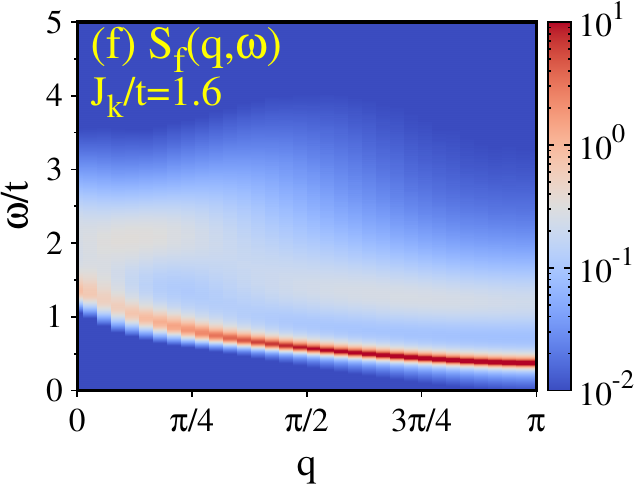}
\includegraphics[width=0.23\textwidth]{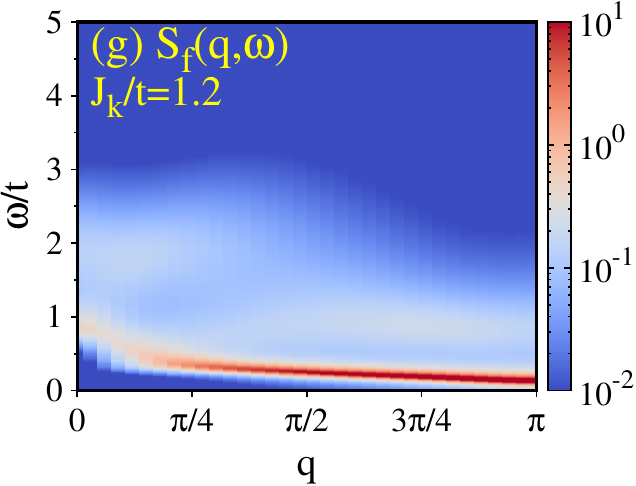}
\includegraphics[width=0.23\textwidth]{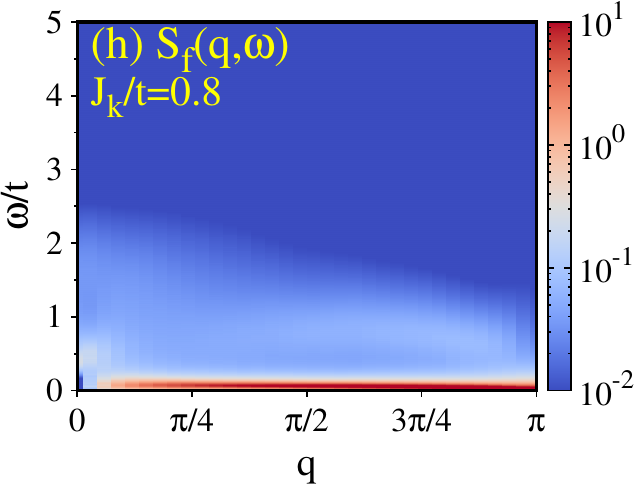}\\
\includegraphics[width=0.23\textwidth]{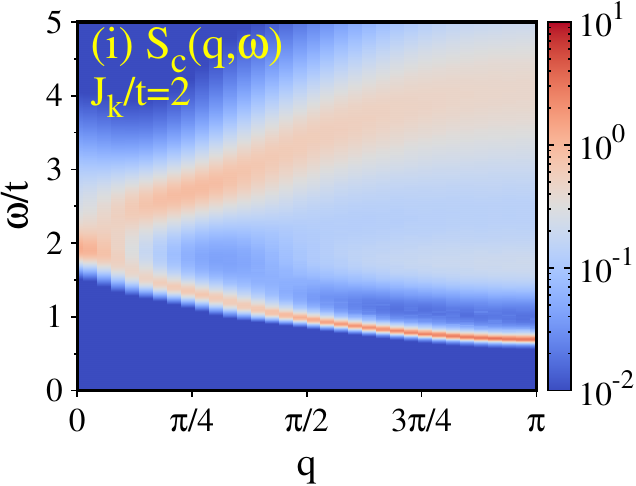}
\includegraphics[width=0.23\textwidth]{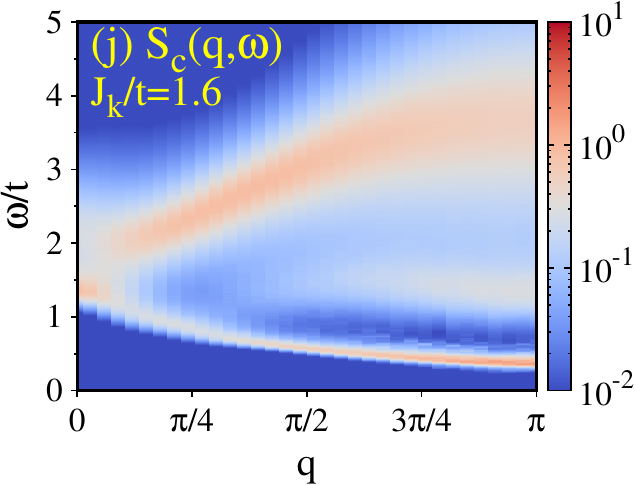}
\includegraphics[width=0.23\textwidth]{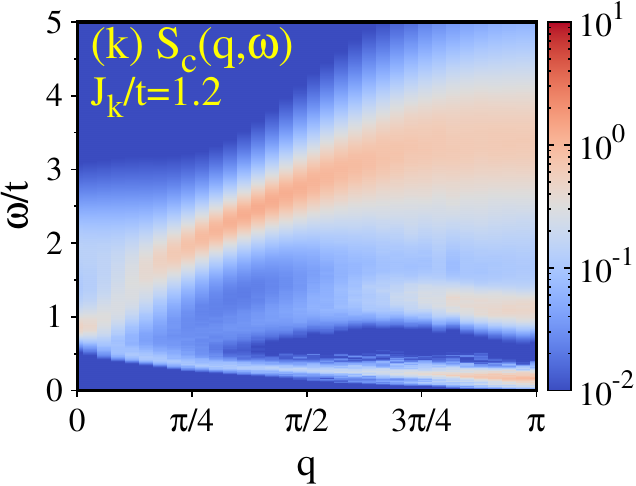}
\includegraphics[width=0.23\textwidth]{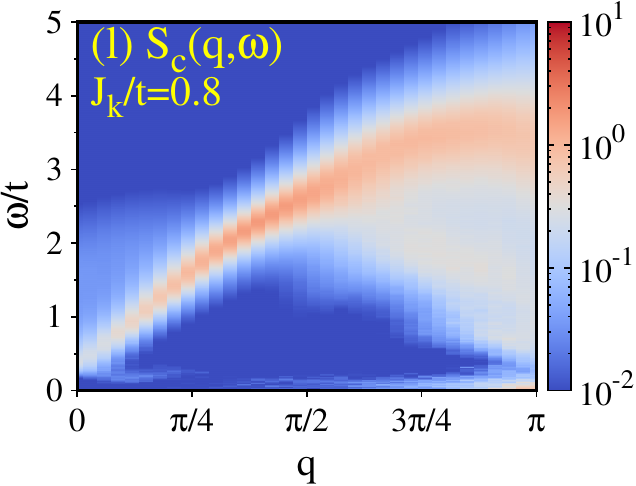}
        \caption{Dynamical spin structure factor  of the $L=66$ SU(4) Kondo chain for: 
	 (a)-(d) total spin  $S(\ve{q},\omega)$;
         (e)-(h) $f$ spin $S_f(\ve{q},\omega)$, and (i)-(l) conduction electron spin $S_c(\ve{q},\omega)$.
	Dashed lines indicate the particle-hole continuum threshold. 
	Parameters as in Fig.~\ref{Fig:AkN4}.} 
\label{Fig:SqN4}
\end{figure*}

\begin{figure*}[t!]
\includegraphics[width=0.23\textwidth]{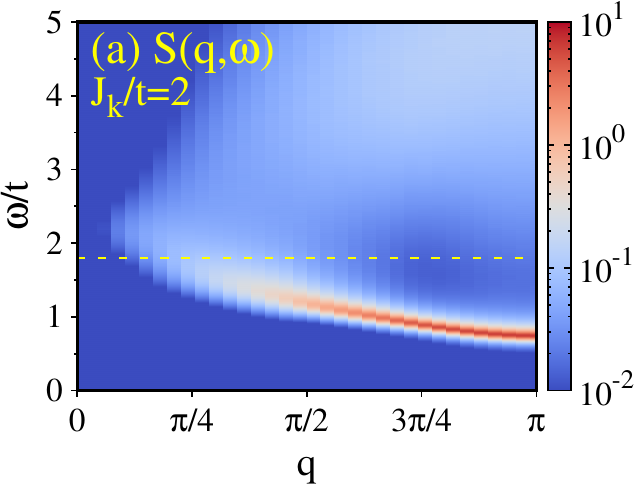}
\includegraphics[width=0.23\textwidth]{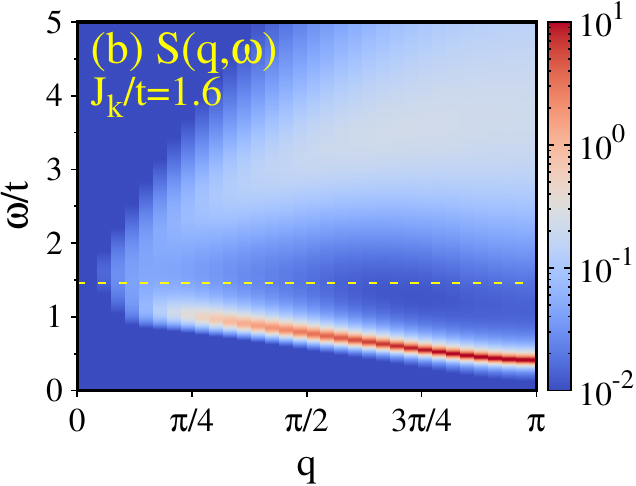}
\includegraphics[width=0.23\textwidth]{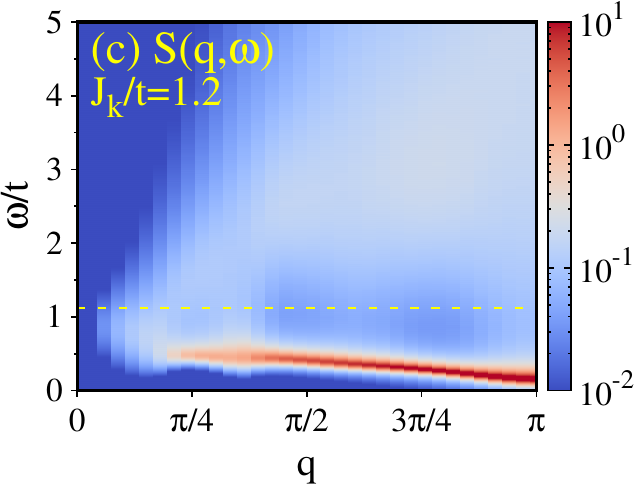}
\includegraphics[width=0.23\textwidth]{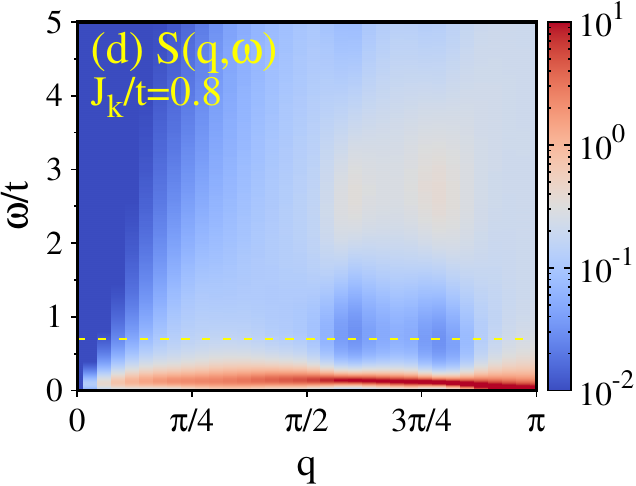}\\
\includegraphics[width=0.23\textwidth]{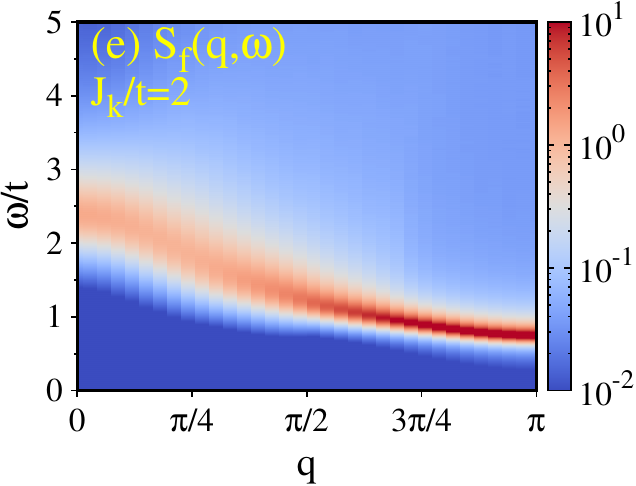}
\includegraphics[width=0.23\textwidth]{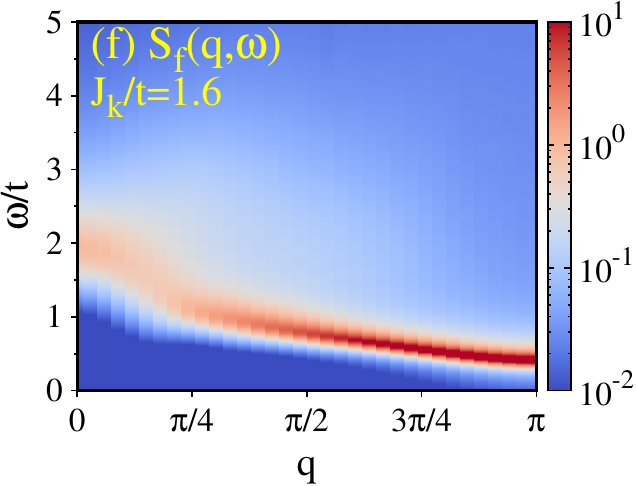}
\includegraphics[width=0.23\textwidth]{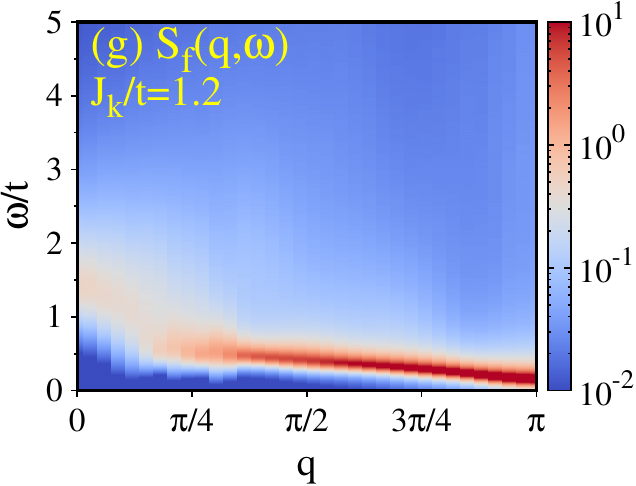}
\includegraphics[width=0.23\textwidth]{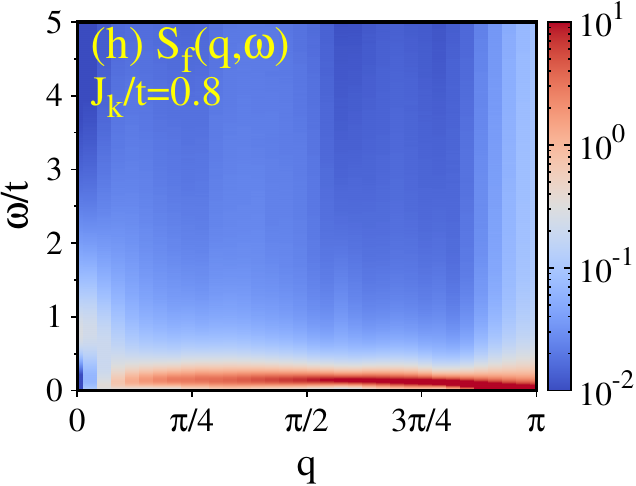}\\
\includegraphics[width=0.23\textwidth]{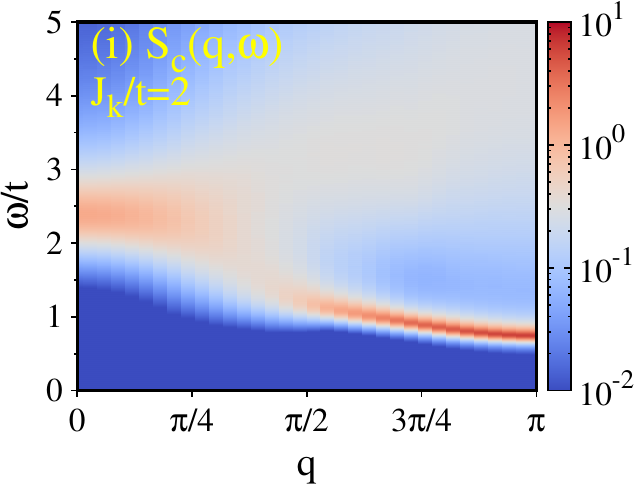}
\includegraphics[width=0.23\textwidth]{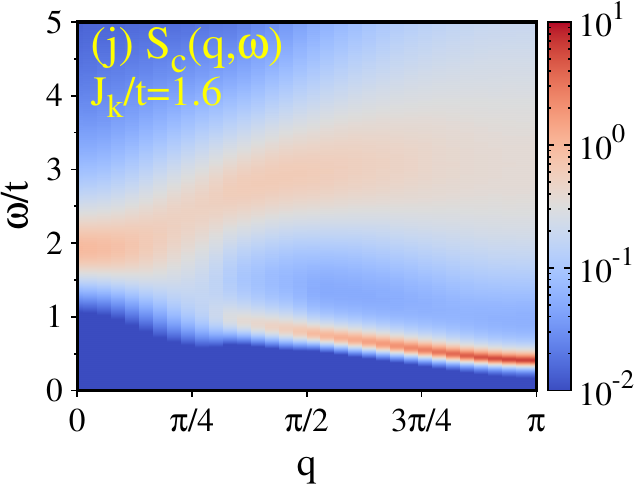}
\includegraphics[width=0.23\textwidth]{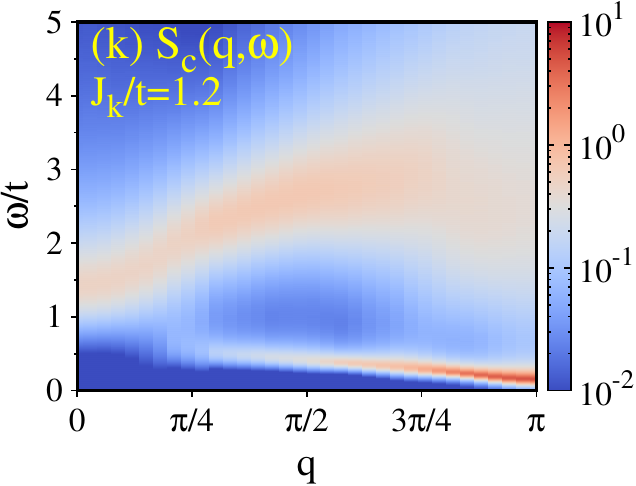}
\includegraphics[width=0.23\textwidth]{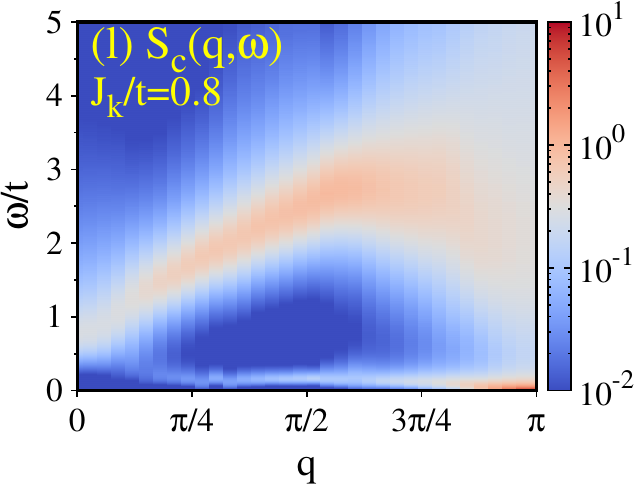}
        \caption{Same as in Fig.~\ref{Fig:SqN4} but for the SU(2) Kondo chain.}
\label{Fig:SqN2}
\end{figure*}

\begin{figure*}[t!]
\includegraphics[width=0.32\textwidth]{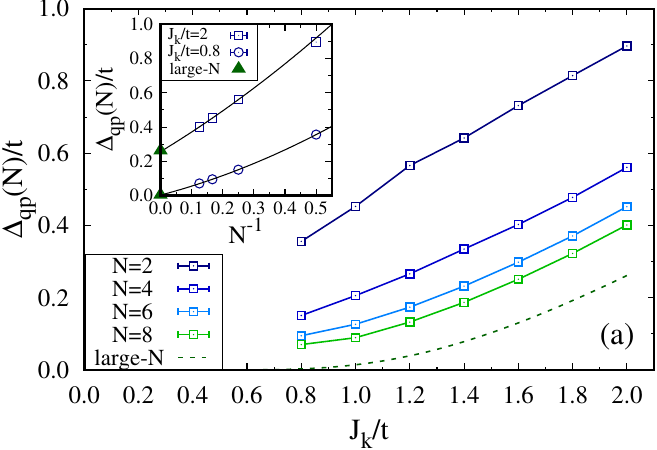}
\includegraphics[width=0.32\textwidth]{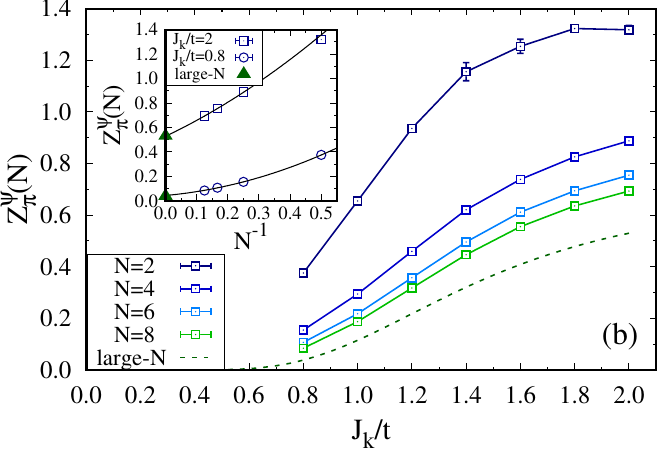}
\includegraphics[width=0.32\textwidth]{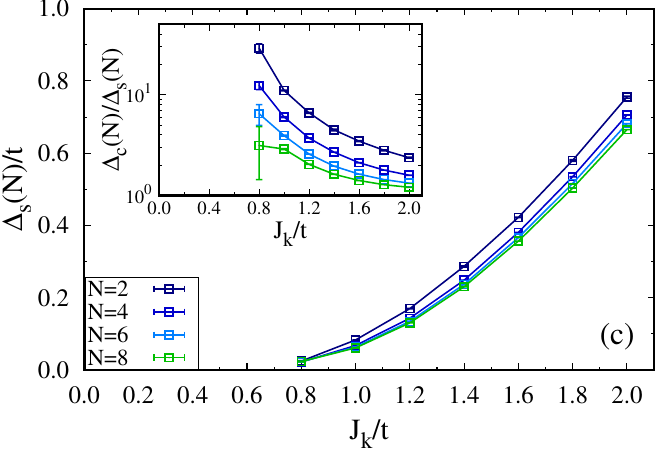}
	\caption{ (a) Single particle gap $\Delta_{qp}(N)$  at $\ve{k}=\pi$ and (b)
	the corresponding residue $Z^{\psi}_{\ve{k}=\pi}(N)$  of the pole in the composite fermion Green's function
	as a function of $J_k/t$ in the SU($N$) Kondo chain. Dashed lines correspond to the large-$N$ value 
	of $\Delta_{qp}$ and   $Z^{\psi}_{\ve{k}=\pi}$. Insets show second order polynomial fits to the QMC data  
	at $J_k/t=2$ and $J_k/t=0.8$ to extract the quasiparticle gap and residue in the $N\to\infty$ limit.
        Within error bars, the latter match the large-$N$ values (triangles). 
	Panel (c) shows the evolution of the spin gap $\Delta_{s}(N)$ at $\ve{q}=\pi$ extracted from 
	the imaginary time displaced  (total) spin correlation function and the inset shows 
	the ratio between the charge $\Delta_c($N$)=2\Delta_{qp}($N$)$ and spin gap on a semilog scale.
	All the quantities have been extrapolated to the thermodynamic limit, see Appendix~\ref{app:fss}.}
\label{Fig:gaps_KLM}
\end{figure*}

To complement our analysis of dynamical properties of the SU(4) Kondo chain, we plot in Fig.~\ref{Fig:SqN4} 
the corresponding spin excitation spectra. Irrespective of the value of $J_k/t$, the dynamical spin structure 
factor $S(\ve{q},\omega)$ displays the same main spectral features: 
(i) a low energy triplon mode with a minimal gap at the antiferromagnetic wavevector $\Q=\pi$, and 
(ii) particle-hole excitations of nearly free conduction electrons above the charge gap, 
see Figs.~\ref{Fig:SqN4}(a)-\ref{Fig:SqN4}(d). 
   
As is apparent,  Kondo screening unites the individual dynamics of the local $f$ and conduction electron spins 
such that both $S_f(\ve{q},\omega)$ and $S_c(\ve{q},\omega)$  spectra have the same support but differ in terms 
of the distribution of spectral intensity. 
In fact, the dynamical coupling between local and itinerant spin degrees of freedom is not restricted to the 
lowest energy triplon mode but the fingerprint of the particle-hole continuum in $S_f(\ve{q},\omega)$ extends over 
higher frequencies matching the energy window where sharp composite fermion bands occur 
in $A_{\psi}(\ve{k},\omega)$, see Figs.~\ref{Fig:AkN4}(a)-\ref{Fig:AkN4}(d). 

The observed flattening of the triplon mode for small $J_k/t$  stems from the slowing down of the spin velocity $v_s$ determined 
by the effective RKKY coupling $J_{\textrm{RKKY}}\propto \tfrac{J_k^2}{N}$.  
While the occurrence of the triplon mode induces the rearrangement of spectral intensity distribution in the particle-hole continuum, 
an extremely soft flat triplon in Fig.~\ref{Fig:SqN4}(d) is found below the charge gap across the whole Brillouin zone. 
It allows one to reveal a typical free-electron-like continuum delimited around $\q=\pi$ by a lower and upper boundary. 
The latter carries  most of the spectral weight and hence the overall form of both 
$S(\ve{q},\omega)$ and $S_c(\ve{q},\omega)$ above the triplon mode can be well approximated using a convolution of the single 
particle Green's function, i.e., particle-hole "bubble" diagram. 

Anticipating that the validity of the "bubble" diagram  might be specific to the enhanced SU(4) spin symmetry of the model,  
we clarify this issue by plotting in Fig.~\ref{Fig:SqN2} spin excitation spectra of the conventional SU(2) Kondo chain.  
As can be seen, spectral intensity in $S(\ve{q},\omega)$  does not rearrange so as to recover the 
free-electron-like continuum upon approaching the weak coupling, see Figs.~\ref{Fig:SqN2}(c) and \ref{Fig:SqN2}(d). 
Moreover, the "bubble" diagram cannot account for a high frequency part of $S_c(\ve{q},\omega)$  with the 
upper branch blurred close to $\q=\pi$ into broad featureless excitations, see Fig.~\ref{Fig:SqN2}(l). 
These findings underscore  the essential role of vertex correction effects in the spin dynamics of the SU(2) model. 

Importantly, regardless of the number of fermion flavors and down to our smallest $J_k/t=0.8$, we find no clear evidence 
of the two-spinon continuum in $S_f(\ve{q},\omega)$, see Figs.~\ref{Fig:SqN4}(h) and \ref{Fig:SqN2}(h),  
in contrast to the $f$ spin dynamics of the SU($N$) Kondo-Heisenberg chain shown in Figs.~\ref{Fig:SqN4kh}(h) and 
\ref{Fig:SqN8kh}(h).

As for the triplon mode, a more pronounced piling up of the spectral weight at $\q=\pi$ seen in $S_c(\ve{q},\omega)$ of the 
$N=2$ model, is consistent with a larger, with respect to the $N=4$ case,  effective RKKY coupling 
$J_{\textrm{RKKY}}\propto \tfrac{J_k^2}{N}$ producing stronger antiferromagnetic spin fluctuations.

\begin{figure*}[t!]
\includegraphics[width=0.32\textwidth]{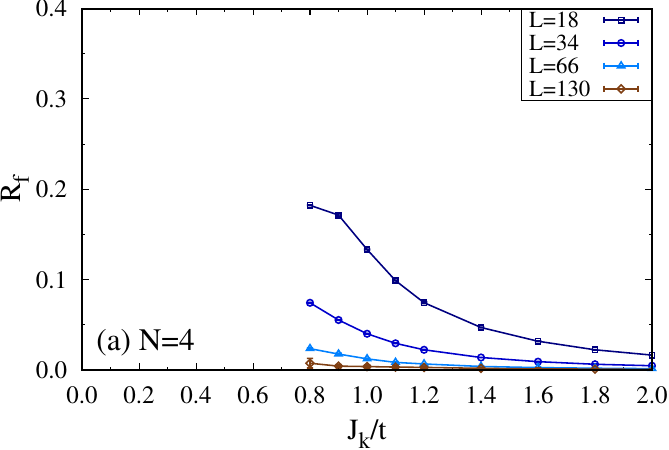}
\includegraphics[width=0.32\textwidth]{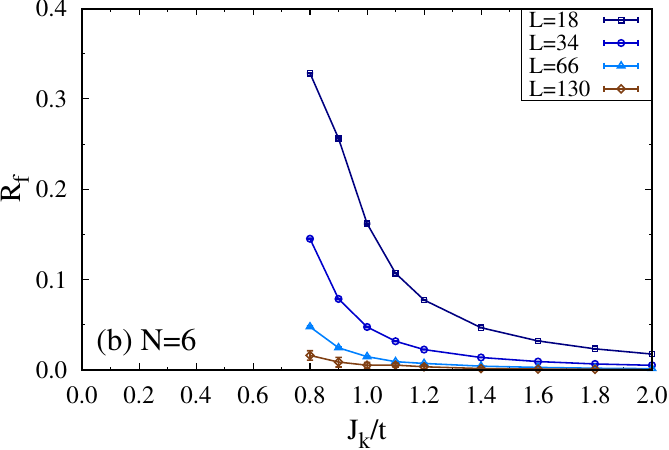}
\includegraphics[width=0.32\textwidth]{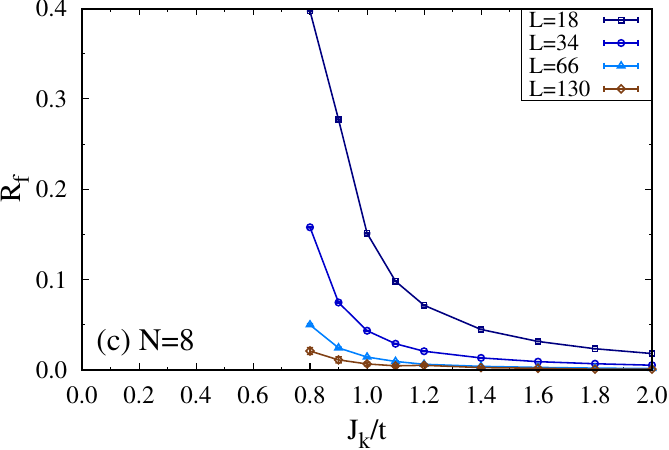}
	\caption{Correlation ratio $R_f$ from the structure factor of $f$ spin dimer
	correlations in the bare SU($N$) Kondo chain: (a) $N=4$; (b) $N=6$, and (c) $N=8$.}
\label{Fig:RN}
\end{figure*}

Next, we examine the impact of an enlarged SU($N$) symmetry of the model on reducing the separation of charge and spin energy 
scales specific to the KI state. To this end, we plot in Fig.~\ref{Fig:gaps_KLM} a single particle gap $\Delta_{qp}(N,\ve{k})$ 
at $\ve{k}=\pi$, the corresponding quasiparticle  residue $Z^{\psi}_{\ve{k}}(N)$, as well as the spin gap 
$\Delta_{s}(N,\ve{q})$ at $\ve{q}=\pi$ as a function of $J_k/t$.  
Following the same fitting procedure as in Sec.~\ref{sec:gaps}, we have extracted finite size estimates of these 
quantities from the asymptotic form of imaginary time composite fermion Green's $G_{\psi}(\k,\tau)$ and spin-spin 
correlation $S(\q,\tau)$  functions at large imaginary time $\tau$. The data in Fig.~\ref{Fig:gaps_KLM} correspond
the thermodynamic limit based on QMC results on chains up to $L=130$ sites. 
As shown in Fig.~\ref{Fig:gaps_KLM}(a), the $N=2$ quasiparticle gap clearly stands out and tracks $J_k$ in the weak coupling 
limit. This behavior bears a striking similarity to a linear dependence of $\Delta_{qp}$ in the magnetically ordered phase 
of the particle-hole symmetric 2D Kondo lattice 
model~\cite{Assaad99a,Capponi00,PhysRevB.82.245105,PhysRevB.96.155119,PhysRevB.98.245125}. 
Long range antiferromagnetic order does not occur in one dimension resulting in a finite spin gap 
$\Delta_s\propto e^{-t/J_k}$ in the weak coupling limit. 
This tiny gap inhibits reliable QMC simulations below $J_k/t=0.8$:  The extrapolated value of $\Delta_s$ 
at $J_k/t=0.7$ is not distinguishable within error bars from zero, spuriously indicating onset of long range magnetic order, 
see Appendix~\ref{app:fss}. 
Due to the exponentially small spin gap, the antiferromagnetic spin correlation length 
$\xi_s\propto \frac{v_s}{\Delta_s}$  is much longer than the conduction electron coherence length 
$\xi_c\propto \frac{t}{\Delta_{qp}}$. 
As a consequence, the $c$ electrons are effectively subject to a quasistatic staggered local moment which accounts 
for $\Delta_{qp}\propto J_k$ as in the magnetically ordered phase~\cite{Ueda97}.   
The observed asymptotic reduction of $\Delta_{qp}$ to its large-$N$ limit (dashed line) is driven by the vanishing  of  vertex  corrections 
brought by the RKKY energy scale $J_{\textrm{RKKY}}\propto \tfrac{J_k^2}{N}$.

Furthermore, since the composite fermion operator directly provides the measure of hybridization, 
${\hat{\psi}}^{\dagger}_{\i,\sigma}\propto {\hat{f}}^{\dagger}_{\i,\sigma} V$~\cite{Danu21},  
one would expect the quasiparticle residue of the composite fermion $Z^{\psi}_{\k}$ to track $V^2 Z_{\k}^{f}$, where 
$Z_{\k}^{f}$  is a large-$N$ coherence factor in Eq.~(\ref{QP_largeN}).
However, as quantified in Fig.~\ref{Fig:gaps_KLM}(b), $Z^{\psi}_{\ve{k}=\pi}$ for the $N=2$ case does not follow the 
expected exponentially small Kondo scale of the large-$N$ approach (dashed line).  
Our data obtained in the absence of long range  antiferromagnetic order  indicate that the observed  enhancement of 
spectral  weight is unrelated to the Fermi surface nesting driven magnetism. Instead, we believe that it is 
an intrinsic property of the composite fermion operator in the presence of strong antiferromagnetic spin fluctuations 
~\cite{Ueda97,SciPostPhys.14.6.166,PhysRevB.105.155134,PhysRevB.108.195116}.

Notably, we were able to recover  within error bars the large-$N$ values of both the single particle gap and quasiparticle weight 
by extrapolating the QMC data at $J_k/t=2$ and $J_k/t=0.8$ to the $N\to\infty$ limit, 
see the inset of Figs.~\ref{Fig:gaps_KLM}(a) and ~\ref{Fig:gaps_KLM}(b), 
confirming that the large-$N$ theory is a correct saddle point of the SU(2) Kondo chain.
In addition, to quantify the strength of correlation effects in the KI state using a single measure~\cite{PhysRevB.47.12451},
we plot in the inset of Fig.~\ref{Fig:gaps_KLM}(c) the ratio between the charge gap $\Delta_c=2\Delta_{qp}$ and 
the spin gap $\Delta_s$.   
As is evident, up to our largest value of $N=8$ this quantity is always larger than 1 and gets strongly enhanced 
in the weak-coupling limit. It marks a difference with respect to a conventional band insulator where the trivial 
splitting between bonding and antibonding bands produces  identical charge and spin excitation 
gaps~\cite{PhysRevB.80.155116,PhysRevB.106.125116}.

Finally, in order to directly confirm that the enlarged SU($N$) symmetry of the RKKY interaction  does not lead to 
the enhancement of staggered valence bond fluctuations, we plot in Fig.~\ref{Fig:RN} the correlation ratio $R_f$ given 
by Eq.~(\ref{eq:R}).
Even though larger $N$ leads on our shortest $L=18$ chain to a systematic increase of $R_f$ in the small $J_k/t$ region, 
the observed \textit{downscaling} of $R_f$ for increasingly larger system sizes implies the translational invariance 
of the ground state in the thermodynamic limit.

\section{\label{sec:summary} Summary and outlook}

The main motivation of this work was to investigate dynamical signatures of the formation of VBS order and its interplay 
with Kondo screening  in a model system accessible to sign free QMC simulations. 
For this reason, we have considered a 1D SU($N$) Kondo-Heisenberg lattice model in the fully antisymmetric self-adjoint 
representation.  We have been able to  determine the phase diagram for the case of a half-filled conduction electron 
band,  and we showed that in the $N\ge 4$ case there is a quantum phase transition between the Kondo insulating phase and the 
VBS/BOW state, which is in the universality class of the 2D classical Ising model. 

While making  a definite statement about Kondo breakdown at small Kondo couplings $J_k/t$  within the dimerized 
phase would  require us to simulate  exponentially  large lattices, our data extend the list of known examples 
(albeit with magnetic order) where the order-disorder  transition  and the breakdown of heavy  quasiparticles  
defined by the loss of a pole  in the  composite  fermion  Green's  function are detached. 
Thus, the coexistence of Kondo screening and long range order is not restricted to  spontaneous breaking  of a continuous 
SU($N$) spin rotational symmetry of the Hamiltonian and to the resultant emergence of  Goldstone bosons (spin waves), 
but it also happens in the case of discrete (translational) symmetry breaking by long range dimer correlations.  

The   U(1) gauge theory  formulation  of the Kondo lattice model in Sec.~\ref{sec:Model},  introduces   three  particles, 
the  bare  fermion,  $\hat{c}^{\dagger}_{\ve{i},\sigma}$,  the  composite  fermion,  $\hat{\Psi}^{\dagger}_{\ve{i},\sigma}$,  
as  well as the gauge  fermion,  $\hat{f}^{\dagger}_{\ve{i},\sigma}$.   The  bare  fermion and  the  composite  fermion have  
the quantum number of  the electron:  globally  conserved  charge and spin.   The $f$ fermion carries a  locally  conserved 
gauge  charge  and  globally  conserved spin  degrees of  freedom  but no electric  charge.  The composite  and bare  fermions  
can  delocalize,   and both will participate  in the Luttinger count.  On the other hand, the  $f$ fermions cannot  hybridize  
with  the  conduction electrons and  composite  fermions  due  to  the mismatch of  quantum numbers.   
In Monte Carlo  simulations, we can  investigate  the dynamics  of  the   bare  and   composite  fermions by computing  
the corresponding  spectral  function.
In contrast,  understanding the  dynamics  of the  $f$ fermions is  more  delicate since the  single particle  propagation involves  a  gauge  string. 
We  can, however,  extract the dynamical spin  structure  factor  $S_f(\ve{q},\omega)$  for the $f$ spins  and  by  comparing it  
to   that  of the isolated  SU($N$) Heisenberg chain, we can assess if  the  $f$ fermions are  present  as possibly  \textit{fractionalized} particles 
in  one  part  of  the spectrum.

\begin{figure*}[t!]
\includegraphics[width=0.32\textwidth]{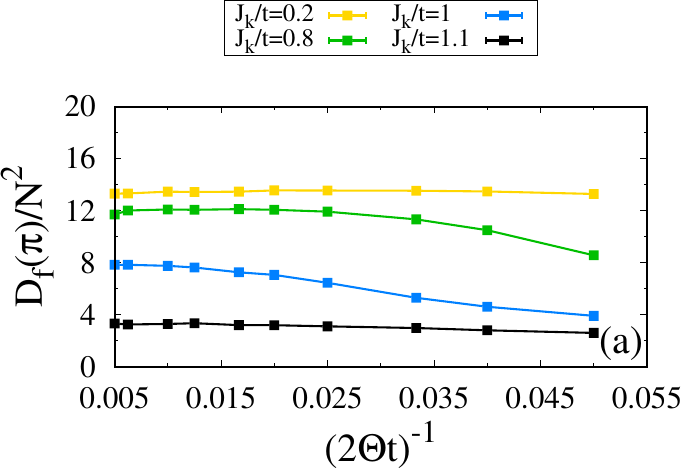}  
\includegraphics[width=0.32\textwidth]{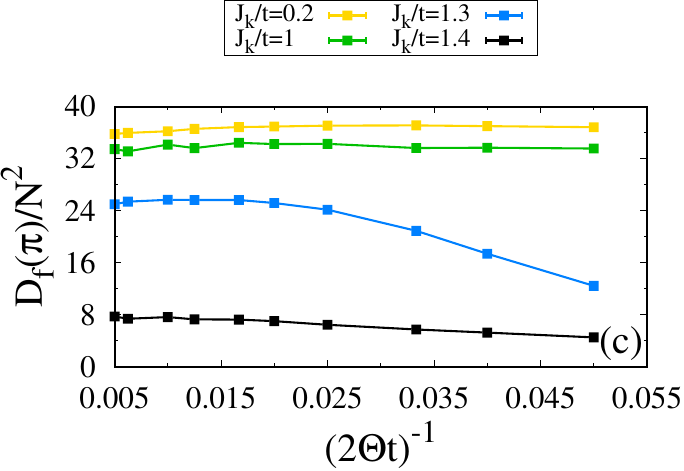}
\includegraphics[width=0.32\textwidth]{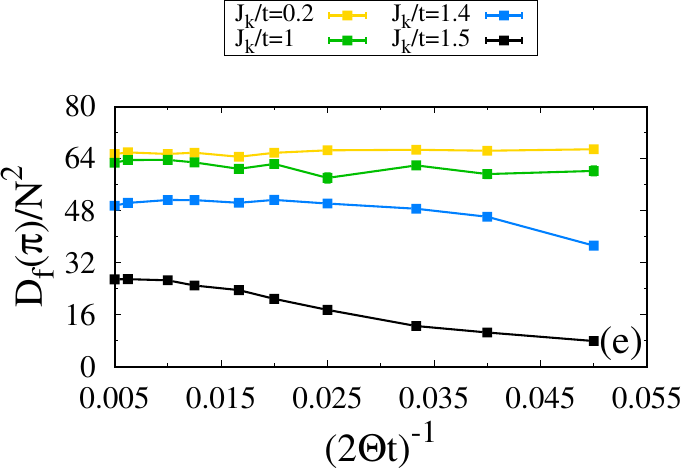}\\
\includegraphics[width=0.32\textwidth]{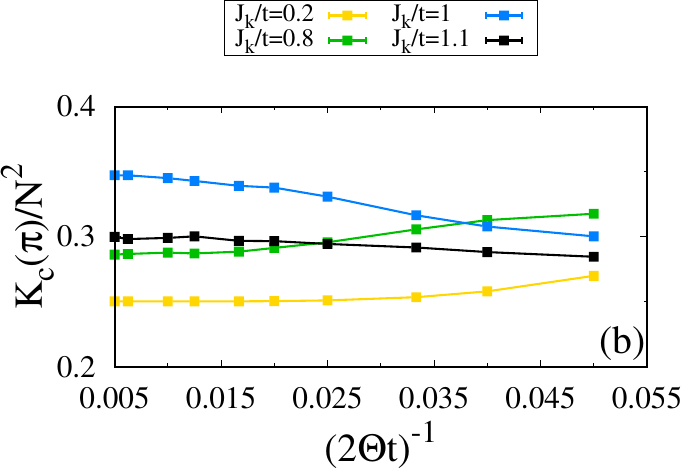}  
\includegraphics[width=0.32\textwidth]{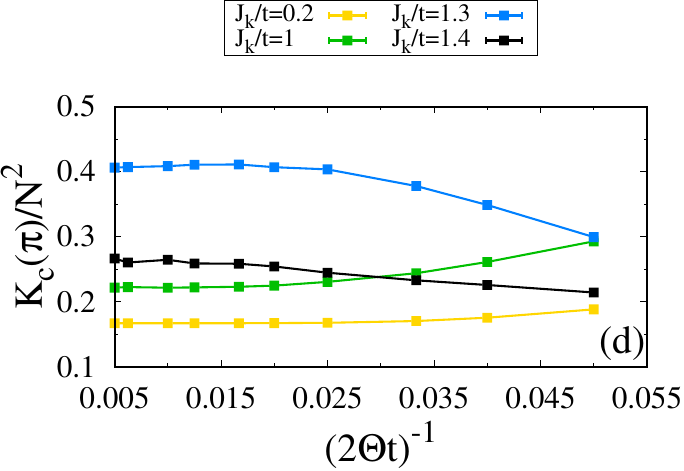}
\includegraphics[width=0.32\textwidth]{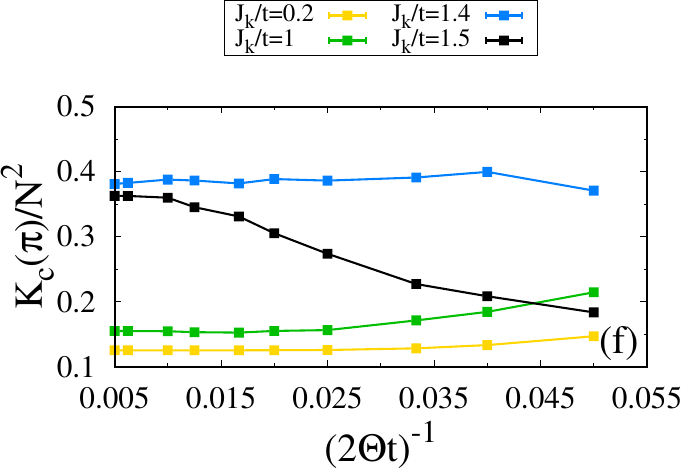}
	 \caption{Convergence of the $f$ spin dimer $D_f(\ve{q}=\pi)/N^2$ (top)  
	 and kinetic energy dimer $K_c(\ve{q}=\pi)/N^2$  (bottom) structure factors 
	 at representative values of $J_k/t$ as a function of the projection parameter $\Theta t$ for the 
	 $L=66$ SU($N$) Kondo-Heisenberg chain with $J_h/t=1$: (a),(b) $N=4$;  (c),(d) $N=6$,  and  (e),(f) $N=8$.
	 }
\label{Fig:KH_Theta}
\end{figure*}

Let   us  now  interpret  aspects  of  our  data  in  terms of  the above,  and concentrate on the SU(4) case.   The  Kondo screened phase  
at  large  $J_k/t=4$ can  be  understood  solely  in  terms of  the hybridization of  the composite and   bare   electrons, 
see Figs.~\ref{Fig:AkN4kh}(a) and \ref{Fig:AkN4kh}(e).  
The  triplon mode  observed  in the   spin  dynamical structure  factor  can be  well  accounted    for   by  considering   vertex  corrections  
to the particle-hole  excitations  that   form   a    bound  state  below  the particle-hole  continuum, 
see  Figs.~\ref{Fig:SqN4kh}(a),\ref{Fig:SqN4kh}(e), and~\ref{Fig:SqN4kh}(i).  
At  our smallest  value of  $J_k/t=0.8$  in the  VBS/BOW phase,   the  situation is  more  complicated   and  better  understood  
by taking the limit of a  Kondo breakdown.  That is,   solely  in terms of  the  $f$ fermions   and  $c$ electrons.  
The  dynamical spectral  function  of  the  $f$ fermions  in  Fig.~\ref{Fig:SqN4kh}(h) shows  a  dimerization gap  and, above it,  a continuum. 
On the one hand, the  similarity  to  the pure  SU(4)  Heisenberg chain in Fig.~\ref{Fig:Sf_Heisen}(b)  is  very evident and suggests 
the interpretation in terms of a two-spinon  continuum.  
On the other hand, given that the charge gap at $J_k/t=0.8$ is lower that the dominant features, one could argue that one cannot really distinguish 
between Landau damping and spinons. 
However, owing  to  the  different  quantum  numbers of  the  $f$ fermions and $c$ electrons,   the  two-spinon  continuum  cannot couple  to  
the particle-hole  continuum of  the conduction electrons. This line of arguing supports  our discussion in Sec.~\ref{sec:KHLM}  
based on the presence of fractionalized particles.  
Furthermore, a nonvanishing  local spin-spin correlation function $S_{cf}$ in Fig.~\ref{Fig:Scf} implies an ultraviolet  coupling between 
the spin chain and  conduction  electrons. Clearly, since the  spin chain  dimerizes, the  conduction electrons
will be subject  to a  dimerization potential. As a  consequence,   a  gap  opens  at  $\ve{k}=\pi/2$  in the  conduction electron spectral  
function and backfolding of  the  band  is  apparent in Fig.~\ref{Fig:AkN4kh}(h).

We can contrast the above results with those in Sec.~\ref{sec:KLM} for the bare  SU(4) Kondo chain which hosts 
solely the Kondo insulating phase. As a consequence of the hybridization of the composite and bare fermions, 
the spin excitation spectrum displays a more conventional structure with the low energy triplon mode and particle-hole 
excitations of nearly free conduction electrons above the charge gap, see Fig.~\ref{Fig:SqN4}.


The starting point of our study corresponds to a dimerized ground state of the SU($N$) Heisenberg chain. 
A complementary avenue is to consider an SU($N$) Kondo-Hubbard chain with an additional 
Hubbard interaction between the conduction electrons: 
In the limit $J_k=0$, there is evidence that a pure SU($N>2$) Hubbard chain at half filling 
hosts a dimerized ground state~\cite{PhysRevB.60.2299,PhysRevB.71.205108,PhysRevB.75.155108}. 
Introducing Kondo coupling $J_k$ to a lattice of magnetic impurities provides another possibility to 
study the effect of quantum critical dimer fluctuations on Kondo screening.

Along these lines, an enlarged symmetry of fermion flavors can pave the way to study the interplay 
of VBS order and Kondo screening in higher dimensions: It is known that the SU(6) Hubbard model 
on the square lattice exhibits at $U/t\simeq 13.3$  a phase transition between long range antiferromagnetic order 
and a VBS state~\cite{Wang19a}. 
Hence, of particular interest is the question of what happens in the vicinity of this 
quantum critical point in the presence of finite Kondo coupling $J_k$. Indeed, in the parameter range 
where Kondo screening, magnetism, and VBS order strongly compete, solutions with larger unit cells, e.g., 
Kondo stripes~\cite{PhysRevLett.100.236403,PhysRevB.96.115158} are conceivable. 
We hope to address these new research directions in the near future.

\begin{acknowledgments} 
The authors gratefully acknowledge the Gauss Centre for Supercomputing e.V. (www.gauss-centre.eu) for funding this project 
by providing computing time through the John von Neumann Institute for Computing (NIC) on the GCS Supercomputer 
JUWELS~\cite{JUWELS}  at J\"ulich Supercomputing Centre (JSC).
M.R. is funded by the Deutsche Forschungsgemeinschaft (DFG, German Research Foundation), Project No. 332790403.
F.F.A.    acknowledges financial support from the DFG under the grant AS 120/16-1 (Project No. 493886309) 
that is part of the collaborative research project SFB Q-M\&S funded by the Austrian Science Fund (FWF) F 86 
as well as   the W\"urzburg-Dresden Cluster of Excellence on Complexity and Topology in Quantum Matter 
ct.qmat (EXC 2147, Project No. 390858490).
\end{acknowledgments}

\appendix* 

\section{\label{app:data} Supplemental QMC data}

\begin{figure*}[t!]
\includegraphics[width=0.24\textwidth]{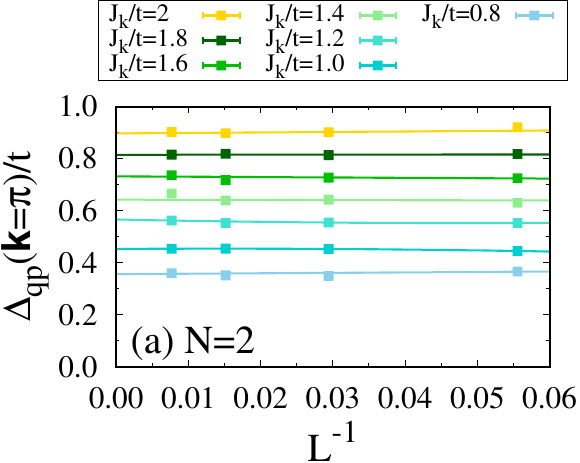}
\includegraphics[width=0.24\textwidth]{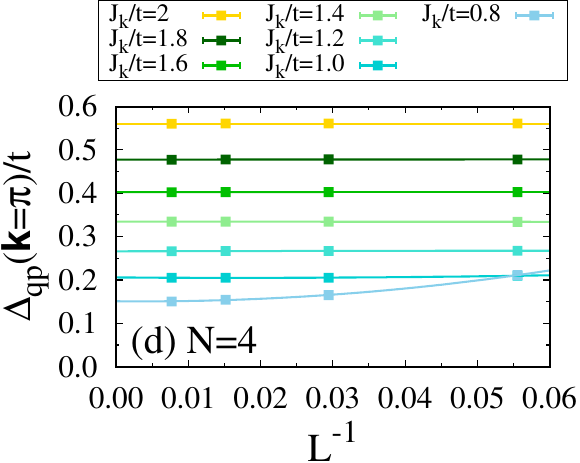}
\includegraphics[width=0.24\textwidth]{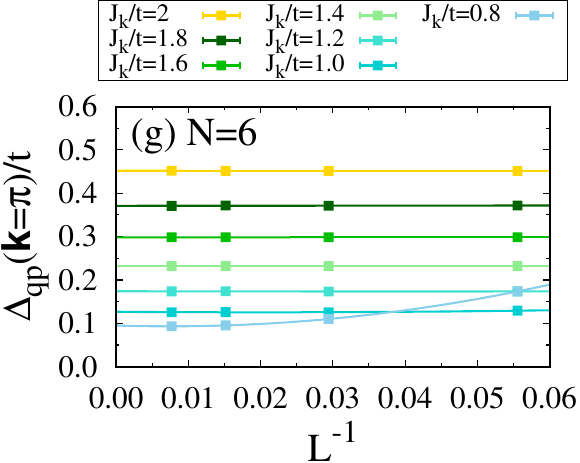}
\includegraphics[width=0.24\textwidth]{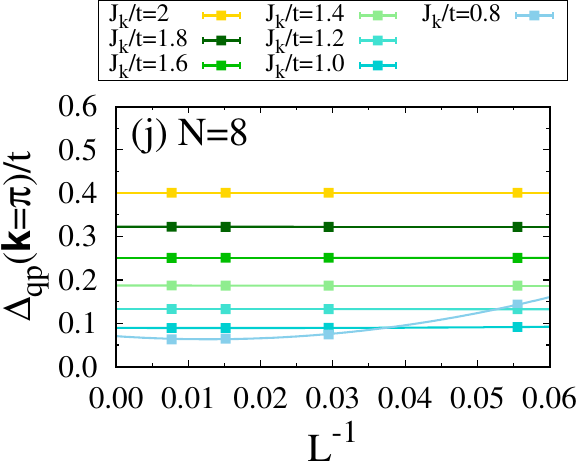}\\
\includegraphics[width=0.24\textwidth]{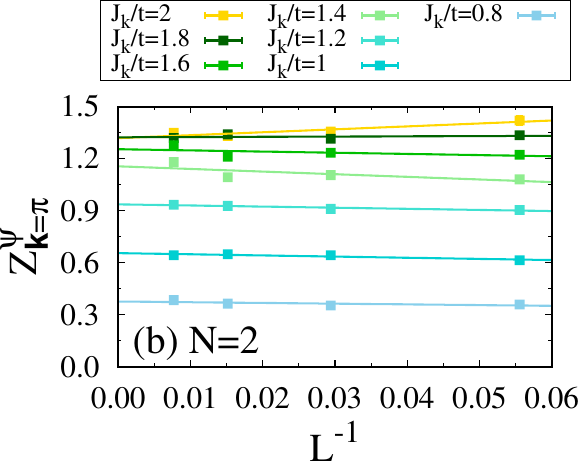}
\includegraphics[width=0.24\textwidth]{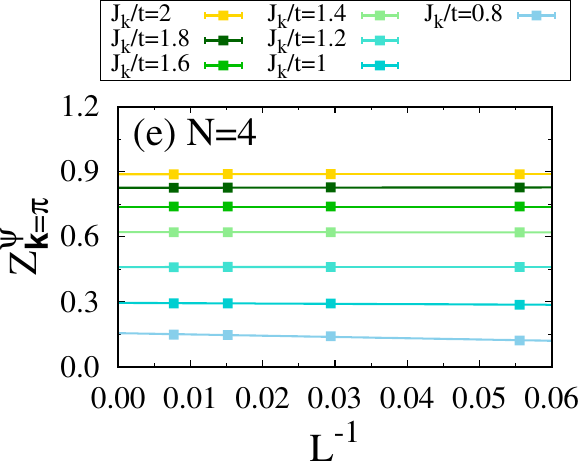}
\includegraphics[width=0.24\textwidth]{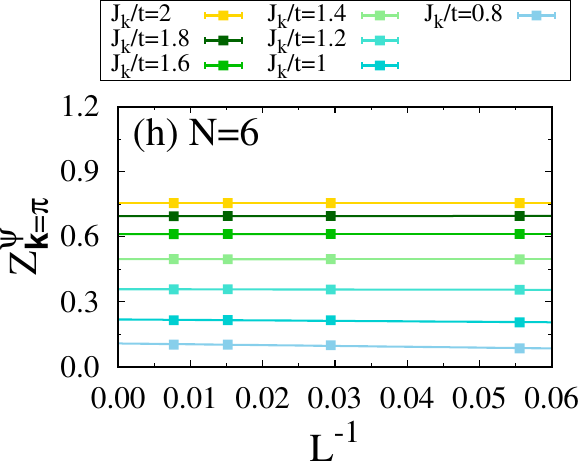}
\includegraphics[width=0.24\textwidth]{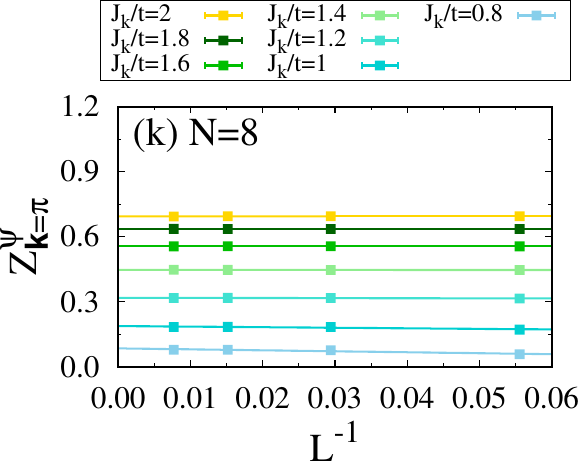}\\
\includegraphics[width=0.24\textwidth]{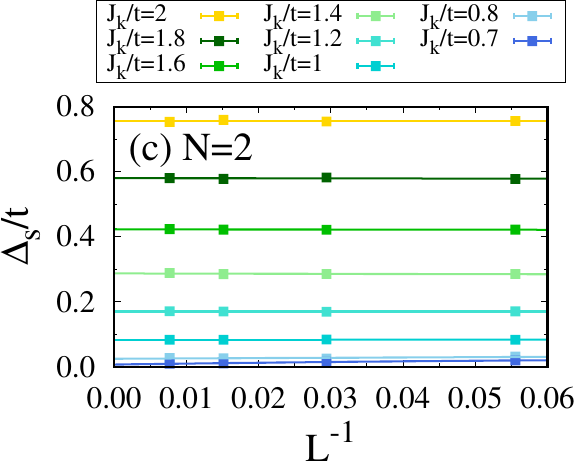}
\includegraphics[width=0.24\textwidth]{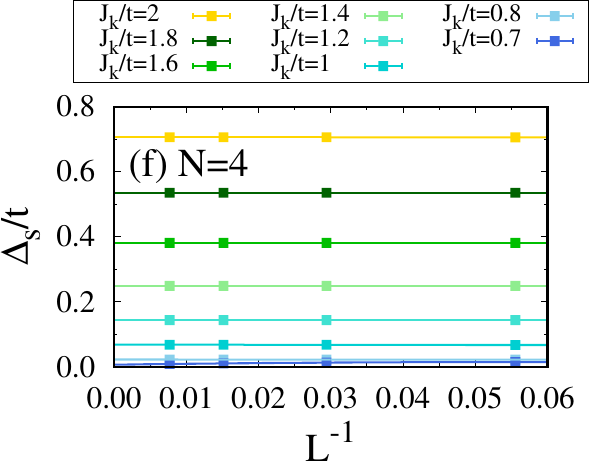}
\includegraphics[width=0.24\textwidth]{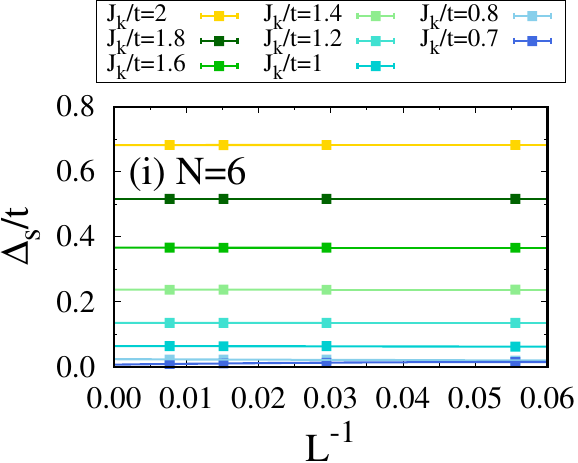}
\includegraphics[width=0.24\textwidth]{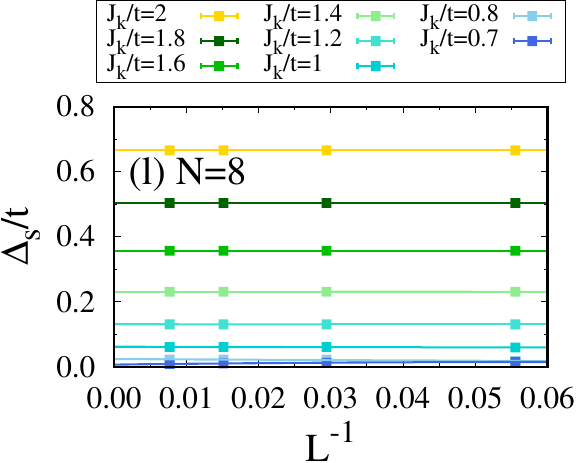}
         \caption{Finite size extrapolation of the single particle gap $\Delta_{qp}$ at $\ve{k}=\pi$ (top),
         the corresponding  residue $Z^{\psi}_{\ve{k}=\pi}$  (middle) of the pole in the composite fermion Green's
         function, and the spin gap $\Delta_s$ at $\ve{q}=\pi$ (bottom) for representative values of $J_k/t$ in
         the SU($N$) Kondo chain. Solid lines are linear and quadratic polynomial fits  to the QMC data:
         (a)-(c) $N=2$;  (d)-(f) $N=4$;  (g)-(i) $N=6$, and (j)-(l) $N=8$. At $J_k/t=0.7$, the extrapolated value of
         $\Delta_s$ is not distinguishable  within error bars  from zero.}
\label{Fig:KLM_L}
\end{figure*}

Here we provide further details about the QMC simulation results  discussed  in the main text. 

\subsection{\label{app:theta} Effect of a finite projection parameter $\Theta$  } 

In the adopted projective version of the QMC  algorithm, the expectation value of a physical observable operator $\hat{O}$
is obtained using 
\begin{equation}
  \frac{ \langle  \Psi_0 | \hat{O} |  \Psi_0 \rangle  }{ \langle  \Psi_0  |  \Psi_0 \rangle  }  =  
  \lim_{\Theta \rightarrow \infty} 
  \frac{ \langle  \Psi_\text{T} | e^{-\Theta \hat{\mathcal{H}}  } \hat{O} e^{-\Theta \hat{\mathcal{H}}   } |  \Psi_\text{T}\rangle  }
  { \langle  \Psi_\text{T}  | e^{-2\Theta \hat{\mathcal{H}} }  |  \Psi_\text{T} \rangle  }
\end{equation}
where $|\Psi_0 \rangle$ is the ground state and $\Theta$ is a projection parameter chosen to be large enough 
to ensure that the trial wave function $| \Psi_\text{T}\rangle$ is projected to the ground state.  
In practice, the ground state expectation value of the observable is obtained by searching for its convergent 
behavior as a function of $\Theta$.  
We illustrate this in Fig.~\ref{Fig:KH_Theta} which plots $f$ spin dimer $D_f(\ve{q}=\pi)$  and kinetic energy dimer 
$K_c(\ve{q}=\pi)$  structure factors from QMC simulations of the $L=66$ SU($N$) Kondo-Heisenberg chain 
for representative values of Kondo coupling $J_k/t$. 
From these plots, we find that $2\Theta t=40$ is sufficient in most cases to guarantee convergence of both quantities 
except for the quantum critical region where the ground state projection is slower and requires up to $2\Theta t=200$.    
Thus, the data for $K_c(\ve{q}=\pi)$ clarify that the weakening of  bond order wave fluctuations  in the valence 
bond solid phase, see  Fig.~\ref{Fig:Rc} of the main text, is not an artifact of the finite value of $\Theta$ but it rather 
stems from the finite size.

\subsection{\label{app:fss} Finite size scaling  }

Figure~\ref{Fig:KLM_L} shows examples of finite size scaling of the single particle gap $\Delta_{qp}$ at $\ve{k}=\pi$,  
the corresponding  residue $Z^{\psi}_{\ve{k}=\pi}$  of the pole in the composite fermion Green's function, and the 
spin gap $\Delta_s$ at $\ve{q}=\pi$ in  the SU($N$) Kondo chain. Based on these scalings, we restricted our analysis 
in Sec.~\ref{sec:KLM} to $J_k/t\ge 0.8$:  The extrapolated value of $\Delta_s$ at $J_k/t=0.7$ is not distinguishable 
within error bars from zero, spuriously indicating the onset of long range magnetic order. Thus, properly addressing 
the weak coupling limit requires the simulation of chains with more than $L=130$ sites used in our study. 
The scaling results led us to Fig.~\ref{Fig:gaps_KLM} in the main text.

\bibliographystyle{bibstyle}
\bibliography{marcin}

 \end{document}